%% file: RANA.tex
\documentclass[10pt, final, conference, a4paper, twocolumn]{IEEEtran}
\IEEEoverridecommandlockouts

\input{prae-flo-ieee.tex}

\def\BibTeX{{\rm B\kern-.05em{\sc i\kern-.025em b}\kern-.08em
    T\kern-.1667em\lower.7ex\hbox{E}\kern-.125emX}}

\renewcommand{\enquote}[1]{`#1'}

\newcommand{\PSPACE}{\textsc{PSpace}\xspace}
\newcommand{\EXPSPACE}{\textsc{ExpSpace}\xspace}

\title{Alternating Nominal Automata\\ with Name Allocation}

\author{
  \IEEEauthorblockN{Florian Frank\IEEEauthorrefmark{1}, Daniel Hausmann\IEEEauthorrefmark{2}, Stefan Milius\IEEEauthorrefmark{1}, Lutz Schröder\IEEEauthorrefmark{1}, Henning Urbat\IEEEauthorrefmark{1}}
  \IEEEauthorblockA{
    \IEEEauthorrefmark{1}
    \textit{Friedrich-Alexander-Universit\"at Erlangen-N\"urnberg}\\
    $\{\text{florian.ff.frank},\text{stefan.milius},\text{lutz.schroeder},\text{henning.urbat}\}$@fau.de}
  \IEEEauthorblockA{\IEEEauthorrefmark{2}\textit{University of Liverpool}\\
    d.hausmann@liverpool.ac.uk}

\thanks{Florian Frank acknowledges support by the Deutsche Forschungsgemeinschaft (DFG) as part of the Research and Training Group 2475 ``Cybercrime and Forensic Computing'' (393541319/GRK2475/2-2024). Daniel Hausmann acknowledges
  suppport by the ERC Consolidator grant D-SynMA (No. 772459) and by the EPSRC through grant EP/Z003121/1.
  Stefan Milius acknowledges support by the Deutsche Forschungsgemeinschaft (DFG) -- project number 419850228.
  Lutz Schröder  acknowledges support by the Deutsche Forschungsgemeinschaft (DFG) -- project number 517924115.
  Henning Urbat acknowledges support by the Deutsche Forschungsgemeinschaft (DFG) -- project number 470467389.}
}

\makeatletter
\hypersetup{
  pdfauthor={Florian Frank, Daniel Hausmann, Stefan Milius, Lutz Schröder and Henning Urbat},
  pdftitle={Alternating Nominal Automata with Name Allocation},
  hidelinks,final,
}
\makeatother

\usepackage{balance}

\begin{document}

\FXRegisterAuthor{ls}{als}{LS}%
\FXRegisterAuthor{dh}{adh}{DH}%
\FXRegisterAuthor{sm}{asm}{SM}%
\FXRegisterAuthor{hu}{ahu}{HU}%
\FXRegisterAuthor{ff}{aff}{FF}%

\maketitle

\begin{abstract}
  Formal languages over infinite alphabets serve as abstractions of
  structures and processes carrying data. Automata models over
  infinite alphabets, such as classical register automata or,
  equivalently, nominal orbit-finite automata, tend to have
  computationally hard or even undecidable reasoning problems unless
  stringent restrictions are imposed on either the power of control or
  the number of registers. This has been shown to be ameliorated in
  automata models with name allocation such as regular
  nondeterministic nominal automata, which allow for deciding language
  inclusion in elementary complexity even with unboundedly many
  registers while retaining a reasonable level of expressiveness. In
  the present work, we demonstrate that elementary complexity survives
  under extending the power of control to \emph{alternation}: We
  introduce \emph{regular alternating nominal automata (RANAs)}, and
  show that their non-emptiness and inclusion problems have elementary
  complexity even when the number of registers is unbounded. Moreover,
  we show that RANAs allow for nearly complete de-alternation,
  specifically de-alternation up to a single deadlocked universal
  state. As a corollary to our results, we improve the complexity of
  model checking for a flavour of \muBar, a fixed-point logic with name
  allocation over finite data words, by one exponential level.
\end{abstract}
\section{Introduction}\label{sec:intro}

Automata models over infinite alphabets, such as the classical register
automata~\cite{KaminskiFrancez94}, or automata over nominal sets~\cite{Pitts2013} such as
nondeterministic orbit-finite automata (NOFA)~\cite{BojanczykEA14}, have received long-standing
interest.  Infinite alphabets may be understood as representing data, for example, nonces in
cryptographic protocols~\cite{KurtzEA07}, data values in XML documents~\cite{NevenEA04}, object
identities~\cite{GrigoreEA13}, or parameters of method calls~\cite{HowarEA19}.  Generally, a
challenge posed by the use of infinite alphabets is that decision problems such as emptiness or
inclusion checking are often either undecidable or computationally very hard, even under severe
restrictions (see the related work section).  Schröder et al.~\cite{skmw17} have introduced
\emph{regular nondeterministic nominal automata} (RNNA), which on one hand extend NOFAs by name
allocating transitions but on the other hand impose finite (rather than orbit-finite)
branching; name-allocating transitions may be thought of as storing letters read from the input
in a register. Despite that restriction, RNNAs retain a reasonable level of expressivity. In
particular, they feature unrestricted nondeterminism and hence can express languages that are
not acceptable by deterministic or unambiguous register automata such as `some letter occurs
twice'; moreover, they allow an unbounded number of registers, and as such accept languages
that are not acceptable by, for instance, alternating 1-register
automata~\cite{skmw17}. Nevertheless, inclusion checking of RNNAs is in parametrized \PSPACE
(the parameter being the number of registers), in sharp contrast to the computational hardness
of other non-deterministic register automaton models.

In the present work, we move from nondeterminism to fully fledged \emph{alternation}; as in the
classical setting, this allows for a more succinct presentation of automata. We introduce
\emph{regular alternating nominal automata} (RANA), a natural common generalization of
classical alternating finite automata~\cite{cks81} and RNNAs, and we demonstrate that the
pleasant algorithmic properties of RNNAs are retained when moving to RANAs. In
infinite-alphabet automata models, expressiveness typically increases strictly with the power
of control (e.g.~extending deterministic to nondeterministic and further to alternating
computation), and indeed RANAs are strictly more expressive than RNNAs. Remarkably, however, we
show (\Cref{thm:corrdealt}) that RANAs can be de-alternated to so-called \emph{extended RNNAs},
with (only) doubly exponential blowup. Extended RNNAs, introduced in work on nominal temporal
logics~\cite{hms21}, extend the nondeterministic RNNA model only by $\top$-states,
i.e.~deadlocked universal states. No similar result appears to be known for alternating
register automata. Furthermore, we show that full de-alternation of RANAs to RNNAs is possible
under a specific form of data language semantics. As a corollary, under this semantics RANAs
are expressively equivalent to a subclass of non-guessing NOFAs~\cite{BojanczykEA14} and
register automata with nondeterministic reassignment~\cite{BojanczykEA14,KaminskiZeitlin10}.

In addition, we show that emptiness and inclusion checking (relevant for
model checking) of RANAs
are decidable in elementary complexity. More precisely, RANAs, like
RNNAs, support two different forms of freshness semantics for letters
read from the input word: \emph{global freshness} (as found, for
instance, in session automata~\cite{BolligEA14}) requires letters to
be distinct from all previous letters in the input, while \emph{local
  freshness} requires distinctness from letters currently stored in
the registers (as in register automata). Under global freshness,
emptiness and inclusion of RANAs are decidable in exponential space, and in
fact in parametrized polynomial space with the number of registers as
the parameter
(\cref{thm:decneRANA,thm:decincl}),
while under local freshness, these problems are decidable in doubly
exponential (and in parametrized singly exponential),
space. Notably, these results hold without any bound on the number of
registers or restrictions on alternation.

Both de-alternation and our decidability results hinge on
the notion of name dropping~\cite{skmw17}, which we adapt to RANAs:
Every RANA has an equivalent \emph{name-dropping} modification
(\Cref{thm:relAAnd}) in which all transitions may
nondeterministically lose any number of names from the support of states
(analogous to losing register contents in a register automaton).%

Our results impact on model checking in the finite-word linear-time nominal
fixed-point logic \muBar~\cite{hms21}: While the original algorithm for model checking \muBar
over RNNAs was based on a direct translation from \muBar to extended
RNNAs, we factor this translation through a simple translation from
\muBar to RANAs. For local freshness, our method then proceeds by
de-alternating the obtained RANA, yielding the same complexity as
before. Contrastingly, model checking under global freshness can be
performed directly by inclusion checking of RANAs, and hence is in
parametrized \PSPACE by our above-mentioned results; this improves on the
original bound~\cite{hms21} by one exponential level.
The translation above also retains the complexity of checking satisfiability 
of \muBar{} for both local and global freshness. %

In summary, our contributions are the following:
\begin{itemize}
\item We introduce regular alternating nominal automata (RANA), an
  alternating automata model for bar languages (\Cref{sec:rana}) that
  allows unrestricted alternation and unboundedly many registers.%
  \item We show their correspondence to the linear-time nominal
    temporal logic \muBar~\cite{hms21} (\Cref{sec:logic}),
    improving the complexity of model checking under global freshness.
  \item We prove that RANAs can be %
    de-alternated to extended RNNAs (\Cref{sec:dealt}), which are
    purely nondeterministic except for a single deadlocked universal
    state. %
  \item We show that checking non-emptiness and inclusion of RANAs are decidable in
    \EXPSPACE, in sharp contrast to other alternating or even just nondeterministic
    register-based models (\Cref{sec:decincl}).
\end{itemize}

\IEEEpubidadjcol{}
\paragraph*{Related Work}%
The expressive power of automata models over infinite alphabets
generally increases with the power of control
(deterministic/nondeterministic/alternating)~\cite{KlinEA21}. In
deterministic models, language inclusion can often be decided in
reasonable complexity; this remains true for unambiguous register
automata~\cite{MottetQuaas19,Colcombet15,bfkm24}. For nondeterministic
register automata and the equivalent nondeterministic orbit-finite
automata~\cite{BojanczykEA14}, emptiness is decidable but inclusion is
undecidable unless one restricts to at most two
registers~\cite{KaminskiFrancez94} (one register in terms of the definition
given by Demri and Lazi{\'c}~\cite{DemriLazic09}). Similarly, language inclusion
(equivalently emptiness) of alternating register automata is
undecidable unless one restricts to at most one register, and even
then is not primitive recursive~\cite{DemriLazic09}. Safety
alternating automata with one register (which are not closed under
complement) have \EXPSPACE-complete non-emptiness checking,
while inclusion is decidable but not primitive
recursive~\cite{Lazic11}. Automata models for infinite alphabets outside
the register paradigm include data walking automata~\cite{ManuelEA16},
whose inclusion problem is decidable even under nondeterminism but at
least as hard as reachability in Petri nets (equivalently vector
addition systems)~\cite{ColcombetManuel15}, as well as the highly
expressive data automata~\cite{BojanczykEA11}, whose nonemptiness
problem is decidable but, again, at least as hard as Petri net
reachability. By recent results, Petri net reachability is
not elementary~\cite{CzerwinskiEA21}, namely
Ackermann-complete~\cite{LerouxSchmitz19,Leroux22,CzerwinskiOrlikowski22}.

As mentioned above, our results relate to work on the
nominal fixed-point logic \muBar~\cite{hms21}. Our de-alternation
construction partly builds on ideas used in that work; on the other
hand, both the central positivization construction
(\Cref{lem:constr02}) and our algorithmic treatment of inclusion
checking by reduction to alternating finite automata are new. %

\section{Preliminaries}\label{sec:prelims}

\paragraph*{Nominal Sets}
The theory of nominal sets~\cite{Pitts2013} offers a convenient framework for dealing with names, and related concepts such as binding and freshness. For our purpose, names play the role of data.
In the following, we briefly recall some terminology and basic facts.

For the rest of the paper, we fix a countably infinite set~$\names$ of \emph{names} (or \emph{data values}). A \emph{finite permutation of $\names$} is a bijective map $\pi\colon \names\to\names$ such that $\pi(a) = a$ for all but finitely many $a\in \names$.
We denote by $\Perm(\names)$ the group of all \emph{finite permutations}, with multiplication given by composition.
The group $\Perm(\names)$ is generated by the \emph{transpositions} $\makecycle{a, b}$ for $a \neq b \in \names$.
Recall that $\makecycle{a, b}$ swaps $a$ and $b$, while fixing all $c\in \names\setminus \set{a, b}$.
A \emph{$\Perm(\names)$-set} is a set~$X$ equipped with a group action $\cdot\colon \Perm(\names) \times X \to X$,
denoted by $(\pi,x)\mapsto \pi\cdot x$. A subset $S\seq \names$ \emph{supports} the element $x\in X$ if $\pi \cdot x = x$ for every
$\pi\in\Perm(\names)$ such that $\pi(a) = a$ for all $a \in S$. A \emph{nominal set} is a $\Perm(\names)$-set $X$ such that every element $x\in X$ has a finite support.
This implies that $x$ has a least finite support, denoted by $\supp x \seq \names$. A name $a \in \names$ is \emph{fresh} for~$x$, denoted $a \fresh x$, if $a \notin \supp x$. 

Informally, we think of a nominal set $X$ as a set of syntactic objects (e.g.~words, trees,
$\lambda$-terms), and of $\supp x\seq \At$ as the (finite) set of names (freely) occurring in $x\in X$.

An element $x \in X$ is \emph{equivariant} if $\supp(x) = \emptyset$, while a subset $X$ of a nominal set $Y$ is \emph{equivariant} if $\pi\cdot x\in X$ for all $x\in X$ and $\pi\in\Perm(\names)$.
The finite subsets of a nominal set~$X$ form a nominal set with the expected group action: $\pi \cdot Y = \setw{\pi \cdot y}{y \in Y}$ for every finite $Y \subseteq X$.
A map $f\colon X \to Y$ between nominal sets is \emph{equivariant} if $f(\pi \cdot x) = \pi \cdot f(x)$ for all $x \in X$
and $\pi \in \Perm(\names)$, which implies $\supp f(x) \subseteq \supp
x$ for all $x \in X$.

We write $X\times Y$ for the cartesian product of nominal sets $X,Y$ with coordinatewise action and $\coprod_{i\in I} X_i = \setw{(i,x)}{i\in I,\, x\in X_i}$ for the coproduct (disjoint union) with the action inherited from the $X_i$.
Given a nominal set $X$ equipped with an equivariant equivalence relation $\sim$, we write $\quotient{X}{\sim}$ for the nominal quotient set under the group action $\pi \cdot [x]_\sim = [\pi\cdot x]_\sim$.

A vital role in the theory of nominal sets is played by \emph{abstraction sets}, which provide semantics for binding mechanisms~\cite{gp99}.
Given a nominal set $X$, we define the equivariant equivalence relation $\approx$ on $\names \times X$ by $\maketuple{a,x} \approx \maketuple{b,y}$ iff $\makecycle{a,c} \cdot x=\makecycle{b,c} \cdot y$ for some, or equivalently all, names $c$ that are fresh for $a, b, x$ and $y$.
The abstraction set $\Abstr{X}$ is the quotient set $\quotient{(\names\times X)}{\approx}$. The $\approx$-equivalence
class of $\maketuple{a,x} \in \names \times X$ is denoted by $\braket{a}x$.
We think of $\approx$ as an abstract notion of $\alpha$-equivalence and of $\braket{a}x$ as binding the name $a$ in $x$.
Indeed, we have $\supp(\braket{a}x) = \supp x \setminus \set{a}$ (while $\supp\maketuple{a,x} =
\set{a} \cup \supp x$), as expected. %

Given a nominal set $X$, the \emph{orbit} of an element $x\in X$ is the set  $\setw{\pi\cdot x}{\pi \in \Perm(\names)}$.
The orbits form a partition of~$X$.
A nominal set is \emph{orbit-finite} if it has finitely many orbits. For every finite set $S\seq\names$, an orbit-finite nominal set contains only finitely many elements supported by $S$. The \emph{degree} of an orbit-finite nominal set $X$ is  $\deg(X) = \max_{x\in X} |\supp(x)|$. 

\begin{example}
  The set $\names$ with the $\Perm(\names)$-action defined by $\pi\cdot a = \pi(a)$ is a nominal set, as is the set $\Ats$ of finite words over $\names$ with $\pi\cdot w=\pi(a_1)\cdots\pi(a_n)$ for $w=a_1\cdots a_n$.
  Then $\supp(a_1\cdots a_n)=\{a_1,\ldots,a_n\}$. The set~$\Ats$ has infinitely many orbits;
  its equivariant subsets $\names^n$ (words of a fixed length $n$) are orbit-finite.
  For instance, $\names^2$ has the two orbits $\{aa: a\in \names\}$ and $\{ab: a\neq b\in \names\}$.
    The equivariant subset $\At^{\#n}\seq \At^n$ given by $\names^{\#n} = \setw{a_1\cdots a_n}{a_i\neq a_j
    \text{ for $i\neq j$}}$ has a single orbit. Both $\names^{\# n}$ and $\names^n$ have degree $n$.
\end{example}

\paragraph*{Data and Bar Languages}

We will use the set $\names$ of names as the data domain, and work with words over $\At$ where a bar symbol (\enquote{$\midmid$}) might precede names, which indicates that the next letter is bound until the end of the word.
We will call these words \emph{bar strings} \cite{skmw17}.
Intuitively, a bar string can be seen as a pattern that determines the way letters are read from the input: an occurrence of $\newletter{a}$ corresponds to reading a letter from the input and binding this letter to the name $a$, while an undecorated occurrence of $a$ means that the letter $a$ occurs literally in the input.
Bound names can be \emph{renamed}, giving rise to a notion of $\alpha$-equivalence of bar strings.
However, the new name must be \emph{fresh}, i.e.~cannot occur freely in the scope of the binding.
For instance, in $ba\newletter{b}ab$ the $\midmid$ binds the letter $b$ in $\newletter{b}ab$.
The string $ba\newletter{b}ab$ therefore is $\alpha$-equivalent to $ba\newletter{c}ac$, but not to $ba\newletter{a}aa$, since $a$ occurs freely in $\newletter{b}ab$.
We say that the latter renaming of $b$ to $a$ is \emph{blocked}.
Formal definitions are:%

\begin{defn}
  We put $\barNames = \names \cup \setw{\newletter{a}}{a \in \names}$ and refer to elements $\newletter{a}$ of $\barNames$ as \emph{bar names}, and to elements $a \in \names$ as \emph{plain names}.
  A \emph{bar string} is a finite word $w = \alpha_1 \cdots \alpha_n \in \barAs$, with \emph{length} $\abs{w} = n$.
  The empty string is denoted by $\varepsilon$.
  We turn $\barNames$ into a nominal set by letter-wise action of $\Perm(\names)$ given by $\pi \cdot a = \pi(a)$ and $\pi \cdot \newletter{a} = \newletter{\pi(a)}$.
  We define \emph{$\alpha$-equivalence} on bar strings to be the equivalence relation generated by $w\newletter{a}v \alphaequiv w\newletter{b}u$ if $\braket{a}v = \braket{b}u$ in $\Abstr{\barAs}$.
  We write $[w]_\alpha$ for the $\alpha$-equivalence class of $w$.
  A name $a$ is \emph{free} in a bar string~$w$ if there is an occurrence of the plain letter $a$ in $w$ that is to the left of the first occurrence (if any) of the bar letter~$\newletter{a}$.
  We write $\FN(w)$ for the set of free names in $w$.
  A bar string~$w$ is \emph{closed} if $\FN(w) = \emptyset$.
  It is \emph{clean} if all bar names $\newletter{a}$ in $w$ are pairwise
    distinct, and for all bar names $\newletter{a}$ in $w$ one has $a \notin \FN(w)$. For a set $S \seq \names$ of names, we put $\bars(S) =
    \setw{w \in \barAs}{\FN(w) \seq S}$.
\end{defn}

\begin{example} For $w = ba\newletter{b}ab$, we have $\FN(w)=\{a,b\}$, and $w \alphaequiv ba\newletter{c}ac$ for all $c\neq a$.
\end{example}
 We work with three different types of languages:

\begin{defn} A \emph{data} / \emph{literal} / \emph{bar language} is, respectively, a subset of $\Ats$, $\barAs$, $\quotient{\barAs}{\alphaequiv}$.
    A bar language is \emph{closed} if it is an equivariant subset of $\quotient{\barAs}{\alphaequiv}$. 
\end{defn}
Every bar language $L \seq \barAs/{\alphaequiv}$ can be converted into a data language by interpreting name binding as reading either \emph{globally fresh} letters (letters not read so far) or \emph{locally fresh} letters (letters not currently stored in memory).
These two interpretations arise from two disciplines of $\alpha$-renaming as known from $\lambda$-calculus~\cite{barendregt85}, with \emph{global freshness} corresponding to a discipline of \emph{clean} renaming where bound names are never shadowed, and \emph{local freshness} corresponds to an unrestricted naming discipline that allows shadowing. Formally, let $\ub(w)\in \Ats$ emerge from $w\in \barAs$ by erasing all bars; for instance, $\ub(ba\newletter{b}ab)=babab$. We put
\begin{equation}
  \begin{aligned}
    N(L) & = \setw{\ub(w)}{\text{$[w]_\alpha \in L$ and $w$ is
    clean}}; \\
    D(L) & = \setw{\ub(w)}{[w]_\alpha \in L}.
  \end{aligned}\label{eq:ND}
\end{equation}

We see that $N$ yields the global freshness interpretation of $L$, while $D$ yields local freshness.

Note that the global freshness operator $N$ is injective on closed bar
languages, more precisely preserves and reflects inclusion, mainly
because $\ub$ is injective on closed clean bar
strings~\cite[Lemma~A.3]{skmw17}.

\section{Regular Alternating Nominal Automata}\label{sec:rana}
We next introduce our automaton model, the regular alternating nominal automaton (RANA),
as a nominalization of classical alternating finite automata (AFA)~\cite{cks81}. %
We shall consider three different flavours of RANAs, which are subsequently shown equivalent. 
Let us first set up the types of boolean formulae that will feature in transitions:

\begin{defn}[Boolean formulae]\label{defn:boolform}
Let $X$ be a set of \emph{atoms}.
  \begin{enumerate}%
    \item Let $\negbool(X)$ denote
      the set of \emph{Boolean formulae} over $X$ as defined by
      the grammar $\varphi,\psi ::= \top\,\vert\,\bot\,\vert\,x\,\vert\,
      \neg\varphi\,\vert\,\varphi\vee\psi\,\vert\,\varphi\wedge\psi$ where $x\in X$. 
    \item Denote by $\posbool(X)$ the subset of
      $\negbool(X)$ given by all \emph{positive}
      Boolean formulae over $X$, i.e.~Boolean formulae that
      do not contain any negation.
    \item Put $\dualbool(X) = \posbool(X \cup X_\mathsf{d})$,
        where $X_\mathsf{d} = \setw{\dual{x}}{x \in X}$ is a copy of $X$.
    \end{enumerate}
If $X$ is a nominal set, we regard $\negbool(X)$, $\posbool(X)$, $\dualbool(X)$ as nominal sets with the group action extending the action of $X$, e.g.~$\pi\cdot (x\wedge \neg \dual{y}) = (\pi\cdot x)\wedge \neg \dual{(\pi\cdot y)}$.  The least support $\supp(\varphi)$ of a formula $\varphi$ is the union of all $\supp(x)$, where $x\in X$ is an atom occurring in $\phi$. 
\end{defn}
\begin{defn}[RANA]\label{defn:rana}
  For a nominal set $Q$, let $X_Q$ denote the nominal set $1 + \names \times Q + \Abstr{Q}$.
  \begin{enumerate}
    \item A \emph{regular alternating nominal automaton} (RANA) is a triple $A = \maketuple{Q,
        \delta, q_0}$ consisting of
      orbit-finite nominal set $Q$ of \emph{states}, an equivariant \emph{initial state} $q_0 \in Q$ and an equivariant
      \emph{transition function} $\delta \colon Q \to \negbool(X_Q)$.
    \item A RANA is \emph{positive} if its transition function $\delta$ restricts to positive formulae ($\posbool$) in
      the codomain; we overload notation and write $\delta\colon Q \to \posbool(X_Q)$ for the restriction again.
    \item An \emph{explicit-dual RANA} is defined like a RANA but with an equivariant transition function of type
      $\delta\colon Q \to \dualbool(X_Q)$.
    \item An \emph{extended regular nondeterministic nominal automaton} (ERNNA) is a positive RANA in which none of the
      transition formulae uses a conjunction ($\land$).
  \end{enumerate}
In each case, the \emph{degree} of $A$ is the degree of the state set $Q$. 
\end{defn}

\begin{rem} \label{rem:equivRNNA}
  ERNNAs have been introduced by Hausmann et al.~\cite{hms21}. They further restrict to \emph{regular nondeterministic nominal automata} (RNNA), introduced by Schröder et al.~\cite{skmw17}, by disallowing $\top$ in transition formulae.
\end{rem}

\noindent Note that in contrast to classical AFAs, a state in our
model has a \emph{single} transition formula mentioning letters
together with states, rather than one formula for every individual
letter in the alphabet. This simplifies the presentation for automata
over infinite alphabets.

\begin{notation}
  We denote the unique atom in the set $1$ by $\epsilon$, as it
  represents the behaviour on the empty word. Moreover, we denote the
  atoms $\maketuple{a,q}$, $\braket{a}q$, $\dual{\maketuple{a,q}}$,
  $\dual{\left(\braket{a}q\right)}$ by $\Diamond_a q$,
  $\Diamond_{\scriptnew{a}} q$, $\Box_a q$, $\Box_{\scriptnew{a}} q$,
  respectively.  We refer to $\Diamond_a,\Diamond_{\newletter
    a},\Box_a,\Box_{\newletter a}$ as
  (\emph{diamond} / \emph{box}) \emph{modalities}, and to $\Diamond_a q$
  etc.\ as \emph{modal atoms}.  This alludes to the connection
  of RANAs to the logic \muBar introduced by Hausmann et
  al.~\cite{hms21}: The behaviour of automata on the four types of
  atoms will correspond to the semantics of the respective modalities
  in the logic. This is reflected in the following definition.
\end{notation}

\begin{defn}[Semantics of RANAs]\label{def:ranasem}
  Let $A = (Q,\delta,q_0)$ be a (positive/explicit-dual) RANA. Given a transition
  formula $\varphi \in \posbool(X) \cup \negbool(X) \cup \dualbool(X)$, where $X = 1 + \names \times Q + \Abstr{Q}$,
  and a bar string $w \in \barAs$, we define satisfaction $w \vDash \varphi$ recursively by
  the conventional clauses for Boolean connectives ($\top$, $\bot$,
  $\neg$, $\vee$, $\wedge$), and on atoms from $X$ by
  \begin{IEEEeqnarray}{lCl}\label{eq:valuation-transition}
    w \vDash \varepsilon & \iff & w = \varepsilon; \IEEEyesnumber*\label{eq:val-EPS}\\
    w \vDash \Diamond_a q & \iff & \exists v \in \barAs.\ w = av \text{ and } v \vDash \delta(q);\label{eq:val-PLAIN}\\
    w \vDash \Diamond_{\scriptnew{a}} q & \iff &
    \begin{IEEEeqnarraybox}[\relax][t][0.3\linewidth]{l} \exists v, v' \in \barAs, b, c \in \names, q' \in Q. \\
      w = \newletter{b}v \alphaequiv \newletter{c}v',\ \braket{a}q = \braket{c}q' \text{ and } \\
      v' \vDash \delta(q');
    \end{IEEEeqnarraybox} \label{eq:val-BOUND}\\%
    w \vDash \dual{\varepsilon} & \iff & w \neq \varepsilon; \IEEEnonumber \\
    w \vDash \Box_a q & \iff & \forall v \in \barAs.\ w = av \implies v \vDash \delta(q); \IEEEnonumber \\
    w \vDash \Box_{\scriptnew{a}} q & \iff & 
    \begin{IEEEeqnarraybox}[\relax][t][0.65\linewidth]{l}%
      \forall b \in \names, v \in \barAs, w = \newletter{b}v \implies \\
      \big(\exists v' \in \barAs, c \in \names, q' \in Q.\, w \alphaequiv \newletter{c}v',\\
      \braket{a}q = \braket{c}q' \text{ and } v' \vDash \delta(q')\big).
    \end{IEEEeqnarraybox}\label{eq:val-BARBOX}
  \end{IEEEeqnarray}
  A state $q \in Q$ \emph{accepts} a bar string $w \in \barAs$ if
  $w \vDash \delta(q)$. We define the \emph{literal language $L_0(A)$ accepted by $A$}
  \begin{IEEEeqnarray*}[\relax]{lCl}
    L_0(A) & = & \setw{w\in \bars(\emptyset)}{\text{$q_0$ accepts $w$}};
    \\
    \noalign{\noindent and the \emph{bar language $L_\alpha(Q)$ accepted by $A$}\vspace{\jot}}
    L_\alpha(A) & = & \setw{[w]_\alpha}{w\in \bars(\emptyset) \text{ and $q_0$ accepts $w$}}.
  \end{IEEEeqnarray*}
We stress that, by definition, only closed bar strings are admitted as inputs of the automaton. 
\end{defn}

\begin{example}\label{ex:barmu-rana}
Consider the literal language of all bar strings $\newletter{a}\newletter{b}w$  ($a\neq b\in \At$, $w\in \{a,b\}^\star$) where $a$ occurs in $w$ at every second position and $b$ occurs in $w$ at least once. It is accepted by a positive RANA with states $\{q_0,q_{\top}\}+\{q_1\}\times \names + \{q_2,\ol{q}_2, q_3\}\times \names^{\#2}$ and
transition formulae as follows:
\begin{IEEEeqnarray*}{rClcrCl}
  \delta(q_\top) & = & \top; & \quad & \delta(q_2(a,b)) & = & \varepsilon \vee \Diamond_a \ol{q}_2(a,b); \\ 
  \delta(q_0) & = & \Diamond_{\scriptnew a} q_1(a); & & \delta(q_3(a,b)) & = & \Diamond_a q_3(a,b) \vee \Diamond_b q_{\top};\\
  \delta(q_1(a)) & = & \mathrlap{\Diamond_{\scriptnew b}q_2(a,b) \wedge \Diamond_{\scriptnew b}q_3(a,b);} \\
  \delta(\ol{q}_2(a,b)) & = & \mathrlap{\varepsilon \vee \Diamond_a q_2(a,b) \vee \Diamond_b q_2(a,b).}
\end{IEEEeqnarray*}
  State $q_2(a,b)$ accepts strings $w \in \{a,b\}^\star$ of arbitrary length in which every other letter is an $a$,
  while state $q_3(a,b)$ accepts bar strings $w \in \barAs$ of the form $a^*bv$ (for $v \in \barAs$).
  Due to the conjunction in its transition formula, $q_1$ only accepts bar strings $\newletter{b}w$
  for $w \in \{a,b\}^\star$ where $b$ occurs in $w$ at least once and there is an $a$ at every other position in $w$.
\end{example}

\begin{example}\label{ex:succinct}
  Put $\pown{N}(\names) := \setw{S \seq \names}{\card{S} \leqslant N}$ for $N \geqslant 1$.
  The literal language $L_{N}$ of all closed bar strings
  $uvvw \in \barAs$ such that $w \in \barAs$, $u \in (\barNames\setminus\names)^*$, $v \in \namesN$ can be accepted by the following positive RANA
  $A = \maketuple{Q, \delta,\maketuple{0,\emptyset}}$:
  \begin{enumerate}
    \item $\textstyle Q = \set{q_\top} + \coprod_{i = 0}^{N} \pown{N}(\names) + \coprod_{i = 1}^N
      (\pown{N}(\names) \times \names)$,
      \item Transition formulae are given by ($N > i > 0$, $j > 1$):
      \begin{IEEEeqnarray*}{rCl}
        \delta(q_\top) & = & \delta(N, S) = \top; \\ %
        \delta(0, S) & = & 
        \begin{IEEEeqnarraybox}[\relax][t][0.8\linewidth]{l} 
          \textstyle\bigvee_{\sigma \in S} (\Diamond_\sigma\maketuple{1,S} \wedge \Diamond_\sigma\maketuple{N,S,\sigma}) \vee{} \\
          \textstyle\bigvee_{{S' \subseteq
          S \cup \set{a},\,\card{S'} \leqslant N}} \Diamond_{\scriptnew{a}}(0, S');
        \end{IEEEeqnarraybox}\\
        \delta(i, S) & = & \textstyle\bigvee_{\sigma \in S} (\Diamond_\sigma\maketuple{i + 1,S} \wedge \Diamond_\sigma\maketuple{N,S,\sigma}); \\
        \delta(1, S, \sigma) & = & \Diamond_\sigma q_\top; \\
        \delta(j, S, \sigma) & = & \textstyle\bigvee_{\alpha \in S} \Diamond_{\alpha} \maketuple{j - 1, S, \sigma} \vee \Diamond_\sigma\maketuple{j-1, S, \sigma}.
      \end{IEEEeqnarray*}
  \end{enumerate}
  The degree of $Q$ is $N + 1$ and its number of orbits is
  \[\textstyle
    1 + \sum_{i = 0}^{N} (N + 1) + \sum_{i = 1}^{N} 2 \cdot (N + 1)
    = 3 \cdot N^2 +  4 \cdot N + 2.
  \]
  Indeed, this follows from the fact that an orbit of the product
  $\pown{N}(\names) \times \names$ is completely determined by the size of the set in the first
  component together with the information whether the name in the second component is contained
  in that set or not. 
  One can show that every ERNNA accepting $L_N$ requires a number of orbits exponential in
  $N$. Intuitively, such an ERNNA has to keep track of all $N$ (possibly equal) letters~of
  the first substring $v$ to check the second copy of $v$ against those names. This suggests the need for
  a state set like $\namesN$, which has ex\-po\-nen\-tial\-ly many orbits. A
  formal argument, based on a nominal adaptation of the \emph{fooling set}
  technique~\cite{hk11} for proving lower bounds on the state complexity of nondeterministic
  finite automata, is found in \Cref{app:secRANA}.%

\end{example}

Further examples for data (and bar) languages recognized by RANAs are given later in~\cref{ex:barmutls}
with the connection to the nominal fixed-point logic $\muBar$.

\begin{rem}
  A nondeterministic orbit-finite automaton (NOFA)~\cite{BojanczykEA14} is
  \emph{non-guessing} (\emph{non-spontaneous} in~\cite{skmw17}) if
  for all transitions $q \xra{a} q'$ we have $\supp(q') \seq \supp(q) \cup \set{a}$
  and the initial state is equivariant. Schröder et al.~\cite{skmw17}
  showed that RNNAs under local freshness are expressively equivalent
  to a subclass of non-guessing NOFAs, so-called \emph{name-dropping} ones.
  Intuitively, a NOFA is name-dropping if transitions may behave lossy
  with respect to the state's support. Additionally, RNNAs under local freshness
  are expressively equivalent to name-dropping register automata with nondeterministic
  reassignment~\cite{KaminskiZeitlin10}.
  We show in~\Cref{sec:dealt}, that under local freshness RANAs are
  expressively equivalent to RNNAs and thus also to
  a subclass of non-guessing NOFAs and register automata, while retaining succinctness.
  This is motivated with~\Cref{ex:succinct}, which translates directly to
  local freshness.
\end{rem}

\begin{rem}\label{rem:rana:coalg}
  For readers familiar with coalgebras~\cite{Rutten00}, we note that (positive/explicit-dual) RANAs are precisely
  pointed orbit-finite coalgebras for the endofunctors $\RANAFunc_x$ 
  on $\Nom$, the category of nominal sets and equivariant maps, given by $\RANAFunc_x Q =
  \allbool_x(1 + \names \times Q + \Abstr{Q})$ for $x \in \set{+,\mathsf{n},\mathsf{d}}$
  on objects, and on morphisms $f\colon Q\to Q'$ by applying $f$ componentwise.
\end{rem}

\begin{rem}\label{rem:ranadefn}
  Since Boolean operators receive their standard
  semantics, one can transform transition formulae using standard
  equivalences, e.g.~into disjunctive normal form (DNF). Also,
  $\varepsilon \wedge \Diamond_\alpha q$ is equivalent to $\bot$, and
  $\varepsilon \wedge \Box_\alpha q$ is equivalent to $\varepsilon$;
  dually, $\neg\varepsilon\vee\Diamond_\alpha q$ is equivalent to
  $\neg\varepsilon$, and $\neg\varepsilon\vee\Box_\alpha q$ is
  equivalent to~$\top$. Moreover $\neg\varepsilon$ and
  $\dual{\varepsilon}$ are equivalent; we will prefer the former in
  the notation.
\end{rem}

\begin{rem}\label{ex:restrictedsemantics}
The above definitions involve a few subtleties that are worth pointing out:
  \begin{enumerate}
    \item\label{rem:ranadefn:6} Our alternating automata have no explicit final states.
      However, we may express final states as states whose transition formula, given in DNF, contains a disjunct $\varepsilon$.
      Our choice of encoding finality into transition formulae again comes from a logical point of view; using this semantics makes our modalities conform with their logical counterparts of diamonds~($\Diamond$) and boxes~($\Box$).
      Otherwise, diamonds would need to accept empty words depending on some finality predicate, a deviation from the intuitive meaning of the atoms. 
    \item Our use of a single initial state simplifies some proofs, but does not restrict the expressivity one expects to have with initial formulae.
      Indeed, just like for classical AFA, an initial formula $\sigma \in \posbool(Q)$ can be replaced with a newly added initial state $q_0$, with the transition
      formula $\delta(q_0)$ obtained from $\sigma$ by replacing each $q\in Q$ in $\sigma$ with $\delta(q)$.
    \item  The possibility of $\alpha$-renaming bar strings in the clauses for $\Diamond_{\scriptnew{a}} q'$ and $\Box_{\scriptnew{a}} q'$ may appear too permissive, especially when comparing RANAs
      to RNNAs~\cite{skmw17}. However, renaming is necessary, since otherwise negation is not $\alpha$-invariant. To see this, denote by $\rsem$ the restricted variant of $\models$ without renaming (see \Cref{rem:implrenW} below for a formal definition). Consider the RANA with 
      $Q = \set{q_0,\ \negstate{q}} + \set{q_1} \times \names + \set{q_2} \times \names^{\# 2}$ and $\delta$ defined
      as follows: $\delta(q_0) = \Diamond_{\scriptnew{a}} q_1(a)$, $\delta(\negstate{q}) = \neg\delta(q_0)$,
      $\delta(q_1(a)) = \Diamond_{\scriptnew{b}} q_2(a, b)$, $\delta(q_2(a, b)) = \varepsilon$. We see that $\newletter{a}\newletter{a} \not\rsem \delta(q_0)$, while $\newletter{a}\newletter{b} \rsem
      \delta(q_0)$. Therefore, $\newletter{a}\newletter{a} \rsem \delta(\negstate{q})$ and
      $\newletter{a}\newletter{b} \not\rsem \delta(\negstate{q})$. Hence the accepted bar languages of $\negstate{q}$
      and~$q_0$ are not disjoint, as both contain $[\newletter a \newletter a]_\alpha=[\newletter a \newletter b]_\alpha$. For~$\models$, this does not happen: Here, $\newletter{a}\newletter{a} \models \delta(q_0) =
      \Diamond_{\scriptnew{a}} q_1(a)$; with a choice of $v' = \newletter{b}$ we see that $v' \models
      \delta(q_1(a))$. 

      In~\Cref{subsec:name-dropping}, we will show that every RANA has a \emph{name-dropping modification} that accepts the same bar language and for which the two semantics $\models$ and $\rsem$ are equivalent.
    \end{enumerate}
\end{rem}

One useful consequence of admitting $\alpha$-renamings is the following observation:
\begin{lem}\label{lem:closurealph}
  The literal language of a RANA is closed
  under $\alpha$-equivalence.
\end{lem}

Let us note another basic property of RANAs, which we refer to as the \emph{support principle}:
\begin{lem}\label{lem:suppRANA}
  Let $A = \maketuple{Q,\delta,q_0}$ be a (positive / explicit-dual) RANA,
  $q, q' \in Q$, and $a \in \names$.
  \begin{enumerate}%
    \item\label{lem:suppRANA:free} If $\Diamond_a q'$
      or $\Box_a q'$ are in $\delta(q)$,
      then $\supp(q') \cup \set{a} \seq \supp(q)$.
    \item\label{lem:suppRANA:bar} If\,$\Diamond_{\scriptnew{a}}q'$\,%
      or\,$\Box_{\scriptnew{a}}q'$\,are in $\delta(q)$, then $\supp(q'){\seq}\supp(q) \cup
      \set{a}$.%
  \end{enumerate}
\end{lem}

To establish the equivalence of the models of \Cref{defn:rana}, we need two important
technical notions for explicit-dual RANAs: \emph{evaluation dags} and \emph{escape letters}.
\begin{defn}[Evaluation dag]\label{defn:reachtree} Let $A=(Q,\delta,q_0)$ be an explicit-dual RANA. The \emph{evaluation dag} of $A$ is the infinite
    directed acyclic graph (dag) with nodes
    of the form $w \qvDash \varphi$ where
    $w \in \barAs$ and $\varphi\in\dualbool(1 + \names \times Q + \Abstr{Q})$, defined inductively as follows:
      \begin{enumerate}
      \item
        $\varphi \in \set{\varepsilon, \neg\varepsilon, \top,
          \bot}$: The node $w \qvDash \varphi$ is a leaf (that is, it has no successors).
      \item $\varphi = \heartsuit_a q$ for
        $\heartsuit \in \set{\Diamond, \Box}$: If~$w$ has the form
        $w = av$ for $a\in\names$, $v\in\barAs$, then
        $v \qvDash \delta(q)$ is the only successor of
        $w \qvDash \varphi$; otherwise, $w \qvDash \varphi$ is
        a leaf.
      \item $\varphi = \heartsuit_{\scriptnew{a}} q$ for
        $\heartsuit \in \set{\Diamond, \Box}$: If $w$ has the form
        $w=\newletter b v$ for $b\in\names$, $v\in\barAs$, then the
        successors of $w \qvDash \varphi$ are the nodes of the form
        $v' \qvDash \delta(q')$ where
        $w=\newletter b v \alphaequiv \newletter{c}v'$ and
        $\braket{a}q = \braket{c}q'$. Otherwise,
        $w \qvDash \varphi$ is a leaf.
      \item$\varphi = \varphi_1 \ast \varphi_2$ for
        $\ast\in\set{\vee,\wedge}$: The node $w\qvDash \phi$ has the two successors
        $w\qvDash\varphi_i$ for $i \in \set{1,2}$.
      \end{enumerate}
    \end{defn}

\noindent The evaluation dag keeps track of the dependency of evaluations: A path from $w\qvDash \phi$ to $w'\qvDash \phi'$ indicates that the evaluation of $w\models \phi$ may eventually require the evaluation of $w'\models \phi'$.
Although we do not currently phrase acceptance in
    game-theoretic terms, the evaluation dag is effectively the
    arena in which an acceptance game would be played.

\begin{defn}[Escape letter]
Let $A=(Q,\delta,q_0)$ be an explicit-dual RANA. Given 
      a bar string $w\in\barAs$ and a 
      formula $\varphi\in\dualbool(1 + \names \times Q + \Abstr{Q})$,
      a free name $a \in \FN(w)$ is an 
      \emph{escape letter for~$w$ at $\varphi$} if from the node
      $w\qvDash\varphi$ in the evaluation dag, 
      a leaf of the form $av \qvDash \Box_\alpha q'$ (in
      which case $a\neq\alpha$) or $av \qvDash \neg\varepsilon$ is reachable. A free name $a 
      \in \FN(w)$ is an \emph{escape letter for $w$ at a state
      $q \in Q$} if it is one at $\delta(q)$.
\end{defn}

Intuitively, an {escape letter} is a letter~$a$ at which the processing of the input word ends
immediately: when checking whether $av \vDash \Box_\alpha q'$ with $a\neq\alpha$ holds for a
suffix $av$ of the input, $\Box_\alpha q'$ is vacuously satisfied because the next letter~$a$
to be read is distinct from~$\alpha$. 
Similarly, $\neg\varepsilon$ is vacuously satisfied for a suffix $av$ of the input.
We will later use these escape letters to \enquote{replace} box modalities by a disjunction
of diamond modalities under preservation of accepted words. This also explains why these
escape letters do not include those arising from diamond modalities, that are those letters
$a$ for which a leaf of the form $av \qvDash \Diamond_\alpha q'$ is reached.

\begin{lem}\label{lem:boundEscapes}
  Let $A = \maketuple{Q, \delta, q_0}$ be an explicit-dual RANA. 
  Then the set of escape letters for~$w\in \barAs$ at $\varphi \in
  \dualbool(1 + \names \times Q + \Abstr{Q})$ is contained in $\supp(\varphi) \cup \set{a}$
  for some $a \in \names$. 
\end{lem}
\begin{IEEEproof}
  Suppose that there is at least one escape letter for $w$ at $\phi$ (otherwise the statement is trivial). Let $a$ be the escape letter that occurs last in $w$. Fix a path 
  from $w\qvDash\varphi$ to a leaf witnessing $a$ 
  as an escape letter. Then for every escape letter $b\neq a$, the path passes a non-leaf node
  $bv\qvDash \heartsuit_bq'$ where
  $\heartsuit\in\{\Diamond,\Box\}$. Since~$b$ is free in~$w$, the support principle(\Cref{lem:suppRANA}) shows that
  $b \in \supp(\varphi)$. Thus the set of escape letters is contained in $\supp(\phi)\cup \{a\}$.
\end{IEEEproof}

\begin{example}\label{rem:boundEscapes:Example}
  Note, that the additional name $a$ in the bound for escape letters
  (\Cref{lem:boundEscapes}) depends on the bar string~$w$ as seen in the
  following small example. Take the formula \mbox{$\varphi = \Box_d q_\bot$} in some
  explicit-dual RANA and two bar strings $w_1 = b$ and $w_2 = c$. Then the
  set of escape letters for~$w_1$ at~$\varphi$ is contained in $\set{d, b}$,
  while the set of escape letters for~$w_2$ at~$\varphi$ is contained in $\set{d, c}$.
  Indeed, every escape letter must occur at some position in $w$.
\end{example}

We are prepared to prove that the three types of RANAs are expressively equivalent:

\begin{theo}\label{thm:equivTypes} Ordinary,
  positive, and explicit-dual RANA are equivalent under
  literal language semantics, hence also under bar language, global and local
  freshness semantics.
\end{theo}

Since every positive RANA is a RANA, it remains to show that every RANA can be transformed into an equivalent explicit-dual RANA, and every explicit-dual RANA can be transformed into an equivalent positive RANA. The first transformation is straightforward:

\begin{prop}\label{lem:constr01}
For every RANA $A$, there exists an explicit-dual RANA $\dual{A}$ that accepts the same
  literal language, has twice as many orbits and the same degree as $A$.
\end{prop}
\begin{IEEEproof}[Proof sketch]
  The explicit-dual RANA $\dual{A}$ has two states $q$, $\negstate{q}$
  for each state $q$ of $A$. The state $q$ accepts the same literal
  language as in $A$, while $\negstate{q}$ accepts the complement of
  this language.  The transition formulae of $\dual{A}$ emerge
  from those of~$A$ by familiar dualities, e.g.~replacing
  $\neg\Diamond_\alpha q$ with~$\Box_\alpha \negstate{q}$.
\end{IEEEproof}

The second transformation is more involved.
Using escape letters, we can restrict the number of ways a box modality can be
satisfied: For instance, a bar string can satisfy $\Box_a q$ either by starting
with~$a$ and continuing with a word accepted by~$q$ (thus satisfying
$\Diamond_a q$), by being empty (thus satisfying~$\epsilon$) or by
starting with a letter~$\beta\in\barNames$ distinct from~$a$ (thus
satisfying $\Diamond_\beta q_\top$ where~$q_\top$ accepts every word).
The bound on escape letters (\Cref{lem:boundEscapes}) allows us to keep this
implicit disjunction finite.

\begin{theo}\label{lem:constr02}
For every explicit-dual RANA $A$ of degree $k$ and with $n$ orbits, there exists a positive RANA~$\pos{A}$ that accepts the same literal language, has degree $2k + 1$, and at most $n \cdot (k + 2) \cdot (2k + 1)^{2k + 1} +1$ orbits. 
\end{theo}
\begin{IEEEproof}[Proof sketch]
  Let $A=(Q,\delta,q_0)$. The positive RANA~$\pos{A}$ has pairs $(q,S)$ as states, where $q\in Q$ and $S\seq \At$ is the current set of escape letters. There are only orbit-finitely many states since the number of escape letters for any
  bar string at any state $q$ is bounded by the size of the support of that state with one additional
  letter (\Cref{lem:boundEscapes}). Therefore, we only need to store at most $k + 1$ letters
  simultaneously. Additionally, $\pos{A}$ has a single equivariant state~$q_{\top}$
  with transition function $\pos\delta(q_{\top}) = \top$ which is the \enquote{totally
  accepting} state. Given a state $\maketuple{q, S}$ in $\pos{A}$,
  the transition formula is defined by modification of $\delta(q)$ in $A$. We sketch the
  different possibilities: Every box modal atom $\Box_a q'$ is replaced by the disjunction
  $\varepsilon \vee \Diamond_a \maketuple{q',S} \vee \bigvee_{\sigma \in S \setminus \set{a}}
  \Diamond_\sigma q_{\top} \vee \Diamond_{\scriptnew{a}} q_{\top}$, whereas diamond modalities
  for plain names are not changed. Whenever a bar modal atom $\Diamond_{\scriptnew{a}}$ or
  $\Box_{\scriptnew{a}}$ is encountered, the transition formula allows a nondeterministic
  choice of adding~$a$ to the currently stored set $S$ of escape letters, while possibly
  removing previously stored names to ensure adherence to the bound $k + 1$. 
  This nondeterminism ensures that any set of escape letters can be reached. Moreover, when encountering~$\Box_{\scriptnew{a}}$, we add both $\varepsilon$
  and $\bigvee_{\sigma \in S} \Diamond_\sigma q_{\top}$ as disjuncts to the newly generated
  formula. It is
then not difficult to verify that a state $q$ in $A$ accepts a bar string $w$ iff $\maketuple{q, S_{q,w}}$
  does in~$\pos{A}$, where $S_{q,w}$ is the set of escape letters for $w$ at $q$. Thus, setting the initial state to $\maketuple{q_0, \emptyset}$ (and observing that closed bar strings have no escape letters), we see that $A$ and $A^+$ accept the same literal languages.
\end{IEEEproof}

An immediate consequence of the above equivalence is:

\begin{theo}[Closure properties] \label{thm:closure} Bar and literal
  languages acceptable by (positive/ explicit-dual) RANAs are closed
  under intersection, union, and complement. More precisely, given RANAs $A_i$ with $n_i$ orbits and degree $k_i$ ($i=1,2$), there exist RANAs with $n_1+n_2+1$ orbits and degree $\max\{k_1,k_2\}$ accepting $L_0(A_1)\cup L_0(A_2)$ and $L_0(A_1)\cap L_0(A_2)$, as well as a RANA with $2\cdot n_1$ orbits and degree $k_1$ accepting $(L_0(A_1))^\complement$.
\end{theo}

\section{Correspondence between RANAs and \muBar}\label{sec:logic}

Applications for alternating automata have been found with constructing equivalent
automata for logical formulae like with LTL and alternating Büchi automata~\cite{mss88,var94}, enabling usage of automata theoretic methods
for logical problems.
Our motivation for introducing alternation to nominal automata with name allocation is to
make translations between a nominal fixed-point logic with name allocation (namely~\muBar~\cite{hms21})
and such nominal automata more succinct. We show that for every \muBar formula, an equivalent
RANA can be constructed with a number of orbits linear in the formula size.

We recap the syntax and semantics of \muBar from Hausmann et al.~\cite{hms21}:

\begin{defn}%
  We fix a countably infinite set $\VAR$ of fixed-point variables. The set $\BarForm$ of bar formulae
  is generated by the grammar $\varphi, \psi ::= \varepsilon\,\vert\,\neg\varepsilon\,\vert\,\varphi\vee\psi
  \,\vert\,\varphi\wedge\psi\,\vert\,\heartsuit_\sigma \varphi\,\vert\,X\,\vert\,\mu X.\varphi$ where
  $\heartsuit \in \set{\Diamond, \Box}$, $\sigma \in \barNames$ and $X \in \VAR$. We put
  $\top := \varepsilon \vee \neg\varepsilon$ and $\bot := \varepsilon \wedge \neg\varepsilon$ and
  refer to $\Diamond_\sigma$ and $\Box_\sigma$ as \emph{$\sigma$-modalities}.

  A name $a$ is \emph{free} in a formula $\varphi$ if $\varphi$ contains an $a$-modality at a position
  not in the scope of any $\newletter{a}$-modality; that is, $\newletter{a}$-modalities bind the name $a$.
  We write $\FN(\varphi)$ for the set of \emph{free names} in $\varphi$ and $\BN(\varphi)$ for
  the set of \emph{bound names} in $\varphi$, i.e.~names~$a$ such that $\varphi$ contains a $\newletter{a}$-modality.
  The \emph{degree} of a formula~$\varphi$ is the number of its free and bound names: $\deg(\varphi) = \card{\FN(\varphi) \cup \BN(\varphi)}$.
  Additionally, we require that all fixed points $\mu X. \varphi$ are \emph{guarded}, that is, all free occurrences of~$X$ lie within the scope of some $\sigma$-modality in $\varphi$.
  A bar formula is \emph{closed}, if every free occurrence of a fixed-point variable $X$ lies within the
  scope of a fixed-point expression~$\mu X. \psi$. We denote by $\phi[\nicefrac{\psi}{X}]$ the
  formula obtained from $\phi$ by replacing every free occurrence of the fixed-point variable $X$ in $\phi$ with $\psi$
  in the usual manner (capture-avoiding).
\end{defn}

Formulae are evaluated over \emph{finite} bar strings. This means that least and greatest fixed points coincide; therefore the syntax only features least fixed points.
$\BarForm$ can be regarded as a nominal set by \enquote{annotating} every fixed-point variable $X$
(with enclosing fixed-point expression~$\mu X.\, \varphi$) with the set $A = \FN(\mu X. \varphi)$ of free names
in the enclosing expression. The $\Perm(\names)$-action then replaces names in the usual manner:
$\pi \cdot \varphi$ is obtained from $\varphi$ by replacing plain names $a \in \names$ with
$\pi(a)$, bar names $\newletter{a} \in \barNames$ with $\newletter{\pi(a)}$ and annotations
$\maketuple{X, A}$ with $\maketuple{X, \pi \cdot A}$. Additionally, we assume all formulae
to be \emph{clean}, that is, fixed-point variables are bound at most once.

\begin{defn}%
  Bar formulae $\varphi$ are interpreted as bar languages. We define
  satisfaction $w \models \varphi$ of a closed formula $\varphi$ by a bar string $w \in \barAs$ recursively
  by the usual clauses for the Boolean connectives, and 
  \begin{IEEEeqnarray}{lCl}
    w \vDash \varepsilon & \iff & w = \varepsilon; \IEEEnonumber\\
    w \vDash \neg\varepsilon & \iff & w \neq \varepsilon; \IEEEnonumber\\
    w \vDash \mu X. \varphi & \iff & w \vDash \varphi[\nicefrac{\mu X. \varphi}{X}]; \IEEEnonumber\\
    w \vDash \Diamond_a \varphi & \iff & \exists v \in \barAs.\ w = av \text{ and } v \vDash \varphi; \IEEEnonumber\\
    w \vDash \Diamond_{\scriptnew{a}} \varphi & \iff &
    \begin{IEEEeqnarraybox}[\relax][t][0.4\linewidth]{l} \exists v, v' \in \barAs, b, c \in \names, \psi \in \BarForm.\\
      w = \newletter{b}v \alphaequiv \newletter{c}v',\ \braket{a}\varphi = \braket{c}\psi \\
      \text{and } v' \vDash \psi;
    \end{IEEEeqnarraybox} \IEEEnonumber\\%
    w \vDash \Box_a \varphi & \iff & \forall v \in \barAs.\ w = av \implies v \vDash \varphi; \IEEEnonumber \\
    w \vDash \Box_{\scriptnew{a}} \varphi & \iff & 
    \begin{IEEEeqnarraybox}[\relax][t][0.4\linewidth]{l}%
      \forall b \in \names, v \in \barAs.\ w = \newletter{b}v \implies\\
      \big(\exists c \in \names, v' \in \barAs, \psi \in \BarForm.\\
      w \alphaequiv \newletter{c}v',\ \braket{a}\varphi = \braket{c}\psi\text{ and }\\
      v' \vDash \psi\big).
    \end{IEEEeqnarraybox} \label{eq:muBar:val-BARBOX}\IEEEyesnumber
  \end{IEEEeqnarray}
  Given a closed formula $\varphi$, its \emph{literal language} $\sem{\varphi}_0$ consists of
  all closed bar strings $w \in \bars(\emptyset)$ such that $w \models \varphi$, and
  the \emph{bar language of $\varphi$} is 
  $\sem{\varphi} = \quotient{\sem{\varphi}_0}{\alphaequiv}$.
\end{defn}

\begin{rem} \label{rem:sem-equivalence}
  We note that the semantics of
  $\Box$-modalities~\eqref{eq:muBar:val-BARBOX} given here is equivalent to the one 
  defined by Hausmann et al.~\cite{hms21}. This is mainly due to the possibility of choosing
  fresh names and equivariance of satisfaction.
\end{rem}

\removeThmBraces
\begin{example}[{{\cite[Ex.~4.8]{hms21}}}] \label{ex:barmutls}
  We give a few examples of languages expressible in $\muBar$. 
  \begin{enumerate}
    \item The bar language $\sem{\Diamond_{\scriptnew{a}}\Box_a\varepsilon}$ is the language of all closed bar strings that
      start with a bar name $\newletter{a}$ and stop after the second letter if that letter exists and is the plain name $a$. Note that $\newletter{a}$, $\newletter{a}a$, $\newletter{a}\newletter{b}ab \in \sem{\Diamond_{\scriptnew{a}}\Box_a\varepsilon}$ but $\newletter{a}aa \notin \sem{\Diamond_{\scriptnew{a}}\Box_a\varepsilon}$.
    \item The formula
      \[\Diamond_{\scriptnew{a}}(\Diamond_{\scriptnew{b}}\mu X.(\varepsilon \vee \Diamond_a(\varepsilon \vee \Diamond_a X \vee \Diamond_b X)) \wedge \Diamond_{\scriptnew{b}}\mu Y.(\Diamond_a Y \vee \Diamond_b \top))\]
      denotes the literal language of~\cref{ex:barmu-rana}.
    \item\label{ex:barmutls:3} Lastly, the bar language of \[\varphi = \Diamond_{\scriptnew{a}}\Diamond_{\scriptnew{b}}\mu X. ((\Diamond_{\scriptnew{b}}X) \vee \Diamond_a\Diamond_b\top)\]
      consists of all closed bar strings that start with a bar name $\newletter{a}$, at some later point contain a
      substring $\newletter{b}ab$ and have only bar names distinct from $\newletter{a}$ in between.
  \end{enumerate}
\end{example}
\resetCurThmBraces

\begin{defn}
  We define the \emph{closure} $\cl(\varphi)$ of a closed bar formula $\varphi \in
  \BarForm$ as usual to be the least
  set of formulae containing $\varphi$ such that
  \begin{align*}
    \psi \ast \chi \in \cl(\varphi) &\text{${}\implies \psi, \chi \in \cl(\varphi)$ for $\ast \in \set{\vee, \wedge}$};\\
    \heartsuit_\sigma \psi \in \cl(\varphi) &{}\implies \psi \in \cl(\varphi)\ \text{for $\heartsuit \in  \set{\Diamond, \Box}$ and $\sigma \in \barNames$};\\
    \mu X. \psi \in \cl(\varphi) &{}\implies \psi[\nicefrac{\mu X.\psi}{X}] \in \cl(\varphi).  
  \end{align*}
  As common for fixed-point logics, the \emph{size} $\card{\varphi}$ of a bar formula~$\varphi$ is the size of its closure set.
\end{defn}

\begin{construction}\label{constr:formaut}
  Given a closed bar formula $\varphi \in \BarForm$ with $\FN(\varphi) = \emptyset$, we
  construct a RANA $A_\varphi = \maketuple{Q_\varphi,\delta_\varphi,\varphi}$.
  \begin{itemize}
    \item $Q_\varphi := \setw{\pi \cdot \psi}{\pi \in \Perm(\names), \psi \in \cl(\varphi)}$. Note that
      $Q_\varphi$ has $\card{\cl(\varphi)} = \card{\varphi}$ orbits and  degree at most
      $\deg(\varphi)$.
    \item $\delta_\varphi\colon Q_\varphi \to \dualbool(1 + \names \times Q_\varphi + \Abstr{Q_\varphi})$
      is defined as follows.
      \begin{align*}
        \delta_\varphi(\pi \cdot \iota)
        &= \iota\ \text{for $\iota \in \set{\varepsilon, \neg\varepsilon}$}; \\
        \delta_\varphi(\pi \cdot \heartsuit_\sigma \chi)
        &= \heartsuit_{\pi\cdot\sigma} \pi \cdot \chi\
        \text{for $\heartsuit \in \set{\Diamond, \Box}$, $\sigma \in \barNames$}; \\
        \delta_\varphi(\pi \cdot \mu X.\chi)
        &= \delta_\varphi(\pi \cdot (\chi[\nicefrac{\mu X.\chi}{X}])); \\
        \delta_\varphi(\pi \cdot (\psi \ast \chi))
        &= \delta_\varphi(\pi \cdot \psi) \ast \delta_\varphi(\pi \cdot \chi)\ \text{for $\ast \in \set{\vee, \wedge}$}.
    \end{align*}
  \end{itemize}
\end{construction}

\begin{rem}\label{rem:formautconstr}
  Note that not all states of~\cref{constr:formaut} are necessarily reachable. Indeed, unfolded fixed-points and
  individual con-/disjuncts are examples for unreachable states.
\end{rem}

\begin{theo}\label{theo:corrFormAut}
  For every closed bar formula $\varphi \in \BarForm$ with no free names, the RANA $A_\varphi$ of~\Cref{constr:formaut}
  accepts the literal language of $\varphi$: $L_0(A_\varphi) = \sem{\varphi}_0$.
\end{theo}
\begin{IEEEproof}[Proof sketch]
  Show that for all $\psi \in Q_\varphi$,
  $w \vDash \psi$ iff $w \vDash \delta_\varphi(\psi)$. The proof
  is by induction over the triple $\maketuple{\card{w}, \textsf{u}(\psi),
  \card{\psi}}$, where $\textsf{u}(\psi)$ denotes the number of \emph{unguarded fixed-point operators} in
  $\varphi$.
\end{IEEEproof}

\begin{example}
  Consider the bar formula $\varphi$ of~\cref{ex:barmutls} \cref{ex:barmutls:3}. Its closure $\cl(\varphi)$
  consists of the following formulae:
  \begin{IEEEeqnarray*}{lClClCl}
    \varphi_0 & := & \mathrlap{\Diamond_{\scriptnew{a}}\Diamond_{\scriptnew{b}}\mu X. ((\Diamond_{\scriptnew{b}}X) \vee \Diamond_a\Diamond_b\top) = \varphi} \\
    \varphi_1 & := & \Diamond_{\scriptnew{b}}\mu X. ((\Diamond_{\scriptnew{b}}X) \vee \Diamond_a\Diamond_b\top) & \qquad & \varphi_4 & := & \Diamond_a\Diamond_b\top  \\
    \varphi_2 & := & \mu X. ((\Diamond_{\scriptnew{b}}X) \vee \Diamond_a\Diamond_b\top) & \qquad & \varphi_5 & := & \Diamond_b\top \\
    \varphi_3 & := & (\Diamond_{\scriptnew{b}}\varphi_2) \vee \Diamond_a\Diamond_b\top & \qquad & \varphi_6 & := & \top
  \end{IEEEeqnarray*}
  Then, the (positive) RANA $A_\varphi$ of~\cref{constr:formaut} with state set $Q_\varphi$ has the following transitions:
  \begin{IEEEeqnarray*}{lClClCl}
    \delta_\varphi(\pi \cdot \varphi_0) & = & \Diamond_{\scriptnew{\pi(a)}} \pi\cdot\varphi_1 & \qquad & \delta_\varphi(\pi\cdot\varphi_4) & = & \Diamond_{\pi(a)} \pi\cdot\varphi_5 \\
    \delta_\varphi(\pi\cdot\varphi_1) & = & \Diamond_{\scriptnew{\pi(b)}} \pi\cdot\varphi_2 & \qquad & \delta_\varphi(\pi\cdot\varphi_5) & = & \Diamond_{\pi(b)} \pi\cdot\varphi_6 \\
    \delta_\varphi(\pi\cdot\varphi_2) & = & \delta_\varphi(\pi\cdot\varphi_3) & \qquad & \delta_\varphi(\pi\cdot\varphi_6) & = &\top \\
    \delta_\varphi(\pi\cdot\varphi_3) & = & \mathrlap{\Diamond_{\scriptnew{\pi(b)}}\pi\cdot\varphi_2 \vee \Diamond_{\pi(a)}\pi\cdot\varphi_5} 
  \end{IEEEeqnarray*}
  Note that the orbits of $\varphi_3$ as well as $\varphi_4$ are unreachable. (cf.~\cref{rem:formautconstr})
  The orbit of $\varphi_1$, however, is still reachable despite its occurrence, since it
  also appears elsewhere in $\varphi$.
\end{example}

\begin{rem}
For the converse direction, a translation taking ERNNAs (Definition~\ref{defn:rana}, Item ($4$))
to \muBar formulae is given in~\cite{hms21}.
A similar construction, omitted here, can be used to transform
RANAs to \muBar formulae. We point
out that both conversions incur exponential blowup
if formula size is measured by the size of the syntax tree.	
\end{rem}

\section{Name-Dropping Modification and Finite Automata}\label{subsec:name-dropping}

We have seen in~\Cref{ex:restrictedsemantics} that the possibility
of $\alpha$-renaming bar strings when encountering bar modalities is necessary for
$\alpha$-invariance of negation.
In the following, we show that every bar language accepted by a positive RANA is also
accepted by a \emph{name-dropping} positive RANA that does not need any implicit renamings.
Additionally, we show that such name-dropping RANAs admit a finite representation in terms
of \emph{bar alternating finite automata}, which are subsequently employed for our decidability
results in~\Cref{sec:decincl}.

\paragraph*{Name-Dropping Modification} We recall that implicit renamings of bar strings
are needed whenever the renaming of a modal atom is \enquote{blocked}, i.e.~has no fitting representative.
To mitigate this, we allow the automaton to forget specific names in the support of a state. This is inspired by a similar construction by Urbat et
al.~\cite{uhms21} for nominal Büchi automata with name allocation. We will use the key fact that every RANA is equivalent to one with a strong
nominal state set.
A nominal set $X$ is \emph{strong}~\cite{Tzevelekos07} if, for all $x\in X$
and $\pi\in \Perm(\At)$, one has $\pi\cdot x = x$ if and only if $\pi$
fixes every element of $\supp(x)$.  (The `if' direction holds in every
nominal set.)  For instance, $\At^{\#n}$, $\At^n$ and
$\Ats$ are strong nominal sets.  Up to isomorphism, orbit-finite strong nominal
sets are precisely coproducts $\coprod_{i=1}^n \At^{\# n_i}$
where $n_i\in \Nat$ and $n$ is the number of orbits of
the nominal set; see e.g.~\cite[Cor.~B.27]{mil-urb-19}.

\begin{theo}\label{thm:representation}
  For every RANA with $n$ orbits and degree $k$, there exists a RANA with $n$ orbits and degree $k$
  that accepts the same literal language and has a strong nominal state set.
\end{theo}

A RANA with states $Q = \coprod_{i=1}^n \names^{\# n_i}$ can be interpreted as a register automaton with a finite set $C=\{c_1,\ldots,c_n\}$ of control states where $c_i$ is equipped with $n_i$ registers.
In \Cref{constr:name-dropping} below, we will modify such an automaton, preserving the accepted literal language, to become lossy in the sense that after each transition some of the
register contents may nondeterministically be erased. Our construction uses the following nominal sets:

\begin{defn}\label{defn:dollar}
  For $n\in\Nat$, we write $\mathbf{n}=\{0,\dots,n-1\}$. We denote by
  $\parnom{n}$ the nominal set of all \emph{partial injective maps}
  from $\mathbf{n}$ to $\names$, with the pointwise group action.  The
  support of~$r$ is $\setw{r(x)}{x \in \dom(r)}$, where $\dom(r)$ is
  the \emph{domain} of $r$, i.e.~the set of all $x \in \mathbf{n}$ for
  which $r(x)$ is defined. The nominal set
  $\parnom{n}$ has $2^{n}$ orbits, one for each possible
  domain. A partial injective map
  $\overline{r} \in \parnom{n}$ \emph{extends} $r \in \parnom{n}$ if
  $\dom(r) \seq \dom(\overline{r})$ and~$r$ and~$\overline{r}$
  coincide on $\dom(r)$, that is, $r(x) = \overline{r}(x)$ for all
  $x \in \dom(r)$.  Given $A \seq \supp(r)$, we write $\restr{r}{A}$
  for the \emph{restriction of $r$ to $A$}. Note that $r$ extends
  $\restr{r}{A}$ and that $\supp(\restr{r}{A}) = A$. 
\end{defn}

The idea of the name-dropping construction is to use partial injective maps (representing the contents of possibly empty registers) as states, which allow to nondeterministically erase stored names.
For transitions, we will replace every occurring modal atom with a disjunction of modal atoms, each of which corresponds to a possible erasure. The details are as follows:

\begin{construction}[Name-dropping modification]\label{constr:name-dropping}
  Let $A = \maketuple{Q, \delta, q_0}$ be a positive RANA 
  with states $Q = \coprod_{i=1}^n \names^{\# n_i}$. The
  \emph{name-dropping modification} of $A$ is the positive RANA $A_{\nd} = \maketuple{Q_{\nd},
  \delta_{\nd}, q_0}$ with states $Q_{\nd} = \coprod_{i=1}^n \parnom{n_i}$ and the following transitions:
  \begin{enumerate}
    \item For each $m\in \Nat$ and $r\in \parnom{m}$ we define the map
      \[f_{\mathsf{r}}\colon \posbool(1 + \names \times Q + \Abstr{Q}) \to
      \posbool(1 + \names \times Q_{\nd} + \Abstr{Q_{\nd}})\]
      recursively as follows: $f_{\mathsf{r}}(\varepsilon) = \varepsilon$, while
      \[\textstyle  f_\mathsf{r}(\Diamond_a \maketuple{j, \overline{s}}) = \bigvee_{s}
      \Diamond_a \maketuple{j, s},\] where the disjunction ranges over
      all~$s$ such that~$\overline{s}$ extends~$s$ and
      $\supp(s) \cup \set{a} \seq \supp(r)$.
      For bar modal atoms,
      $f_{\mathsf{r}}(\Diamond_{\scriptnew{a}} \maketuple{j, \overline{s}}) =
      \bigvee_s \Diamond_{\scriptnew{a}} \maketuple{j, s}$, where the
      disjunction ranges over all~$s$ such that~$\overline{s}$
      extends~$s$ and $\supp(s) \seq \supp(r) \cup \set{a}$.  Finally,
      $f_{\mathsf{r}}(\varphi_1 \ast \varphi_2) = f_{\mathsf{r}}(\varphi_1) \ast
      f_{\mathsf{r}}(\varphi_2)$ for $\ast \in \set{\vee, \wedge}$.
\item For $\maketuple{i, r} \in Q_{\nd}$, choose an extension $\overline{r}$
      of $r$ such that
      $\maketuple{i, \overline{r}} \in Q$, and put
      $\delta_{\nd}\maketuple{i,r}=f_{\mathsf{r}}(\delta\maketuple{i,\overline{r}})$. We show in
      the appendix that~$\delta_{\nd}(i,r)$ is
      independent of the choice of~$\overline{r}$.
  \end{enumerate} 
Note that if $A$ has degree $k$ and $n$ orbits, then $A_{\nd}$ has degree~$k$ and
  at most $n \cdot 2^k$ orbits.
\end{construction}

\begin{theo}\label{thm:relAAnd}
For $A$ and $A_{\nd}$ as in \Cref{constr:name-dropping}, we have $L_0(A)=L_0(A_{\nd})$.
\end{theo}
\begin{IEEEproof}[Proof sketch]
  The inclusion \enquote{$\seq$} clearly holds, since for every $q\in Q$, the transition formula $\delta(q)$ is weaker than $\delta_{\nd}(q)$.
  The inclusion \enquote{$\qes$} is more involved. It uses that the transition formula
  of every state $q$ of $A_{\nd}$ comes from that of a state of $A$ extending $q$. This connection allows us to prove by  induction on words (with an inner induction on transition formulae) that whenever a state of $A_{\nd}$ accepts a word, so does the extended state of $A$.
\end{IEEEproof}

One of the key properties of name-dropping automata is that we may
restrict the states in a run as long as the free names of
the input bar string are not removed. %

\begin{lem}\label{lem:restrRun}
  In the notation of \Cref{constr:name-dropping}, the following
  implication holds for all
  $\varphi \in \posbool(1 + \names \times Q + \Abstr{Q})$, 
  $r,r' \in \parnom{{n_i}}$ and $w\in \barAs$:
  If $r$ extends $r'$, $\supp(r) \cap \FN(w) \seq \supp(r')$ and
  $w \vDash f_r(\varphi)$, then $w \vDash f_{r'}(\varphi)$.
\end{lem}

We proceed to substantiate the main motivation behind name-dropping
RANAs, namely that they do not need to rename bar strings when
reading bar letters.  We formally define a restricted semantics
in this sense, detailing what we have already seen
in~\Cref{ex:restrictedsemantics}:

\begin{defn} \label{rem:implrenW} Let $A = \maketuple{Q, \delta, q_0}$
  be a positive RANA. We inductively define the \emph{restricted
    satisfaction relation} $w\rsem\varphi$ for $w \in \barAs$ and
  $\varphi \in \posbool(1 + \names \times Q + \Abstr{Q})$ just as
  $\vDash$ for~$\varepsilon$ and diamond modalities, see \eqref{eq:val-EPS}
  and \eqref{eq:val-PLAIN},
  except for bar modalities, see \eqref{eq:val-BOUND}, where we put
  \begin{IEEEeqnarray*}{lCl}
    w \rsem \Diamond_{\scriptnew{a}} q & \iff &
    \begin{IEEEeqnarraybox}[\relax][t][0.65\linewidth]{l} \exists v, v' \in \barAs, b \in \names, q' \in Q. \\
      w = \newletter{b}v,\ \braket{a}q = \braket{b}q' \text{ and } v' \rsem \delta(q').
    \end{IEEEeqnarraybox} %
  \end{IEEEeqnarray*}
\end{defn}
\noindent Thus, the only difference between $\rsem$ and $\vDash$ is that  $\vDash$ allows $\alpha$-renaming $w$ in the clause for $\Diamond_{\scriptnew{a}} q'$.
The name-dropping modification makes $\rsem$ and $\vDash$ coincide:

\begin{prop}\label{lem:ndrsem}
  Let $A$ and $A_{\nd}$ be as in \Cref{constr:name-dropping}. For
  every $w \in \barAs$ and $q \in Q_{\nd}$, we have that
  $w \vDash \delta_{\nd}(q)$ iff $w \rsem \delta_{\nd}(q)$.
\end{prop}
\begin{IEEEproof}[Proof sketch]
  The proof proceeds by an inner induction on the word length and an outer
  induction on transition formulae. 
  By construction, bar modal atoms only occur in disjunctions $\bigvee_\ell \Diamond_{\scriptnew{b}} \maketuple{j, s_\ell}$.
  Using~\Cref{lem:restrRun} and standard facts about abstraction sets, one can show that if a bar string satisfies any of the disjuncts, for which the renaming might be blocked, then it is also satisfies another disjunct for which the renaming is not blocked.
\end{IEEEproof}

\paragraph*{Reduction to finite automata} Following earlier work on
(extended) RNNAs~\cite{skmw17,hms21}, we next establish equivalence of
RANAs to a semantic modification of a classical finite-alphabet
automata model, in our case to alternating finite automata
(AFA)~\cite{cks81}. Roughly speaking, we let AFAs accept bar strings
over a finite subset of~$\names$ and quotient modulo
$\alpha$-equivalence. In the following, we say that a set $Q \seq X$ \emph{satisfies} a positive Boolean formula
$\varphi \in \posbool(X)$ if $\varphi$ evaluates to $\top$ when all
atoms $x \in Q$ are set to $\top$ and all atoms $y \in X \setminus Q$ are
set to $\bot$.

\begin{defn}[Bar AFA]\label{defn:bafa}
  \begin{enumerate}
  \item A \emph{bar alternating finite automaton} (bar AFA) $A = \maketuple{Q, \barNames_0, \delta, q_0, F}$ consists of a finite set $Q$ of \emph{states}, a finite \emph{alphabet} $\barNames_0 \seq \barNames$, a \emph{transition function} $\delta\colon Q \times \barNames_0 \to \posbool(Q)$, an \emph{initial state} $q_0 \in Q$ and a set $F \seq Q$ of \emph{final states}.

  \item Given a bar string $w \in \barNames_0^*$ and a state $q \in Q$, a \emph{run dag} for $w$ and $q$ is a rooted dag $\runtree{q}{}{w}$ with nodes $\barNames_0^{\leqslant n} \times (Q + \set{\top})$ and root $\maketuple{w, q}$, satisfying the following properties: Every node $(\varepsilon,-)$ and $(-,\top)$ is a leaf. A node $\maketuple{\alpha v, q'}$ is a leaf if
    $\delta(q', \alpha) \equiv \bot$, and has a single child $\maketuple{v, \top}$ if \mbox{$\delta(q',\alpha)\equiv \top$}. Otherwise, there exists a satisfying
    set $X$ of $\delta(q', \alpha)$ such that precisely the nodes $\maketuple{v, q''}$ (for $q'' \in X$)
    are children of $(\alpha v,q')$.
    
  \item A run dag $\runtree{q}{}{w}$ is \emph{accepting} if all leaves are of the form $(v,\top)$ or $(\varepsilon,q')$ where $q'\in F$. A bar string $w \in \barNames_0^*$ is
    \emph{accepted by $\top$ from $q \in Q$} if there is an accepting run dag $\runtree{q}{}{w}$
    where all leaves are of the form $(v,\top)$. The word $w$ is \emph{accepted in whole from $q \in Q$}, if
    there is an accepting run dag $\runtree{q}{}{w}$ with a leaf $\maketuple{\varepsilon, q'}$ where
    $q' \in F$. %

  \item Given a bar strings $v,w \in \barNames_0^*$, we say that $v$ is an \emph{accepted pre-word} of $w$ if $v$ is a prefix of $w$ with $w = vu$, say, and $w$ is accepted by $\top$ via a run-dag $\runtree{q}{}{w}$ which has a leaf $(u,\top)$ but no leaf $(u',\top)$ where $u'$ is a proper suffix of $u$.

  \item The \emph{pre-language} $\Lpre(A)$ of a bar AFA is defined by
    \begin{IEEEeqnarray*}{lCl}
      \Lpre(A) & = & \setw{\maketuple{w, 1}}{ w' \alphaequiv w,\ w' \text{ is acc.~in whole}}
      \\
      & \cup &  \big\{ \maketuple{v, \top} \ :\ v'\alphaequiv v,\ \text{$v'$ is a pre-word of}
      \\
      &&\ \text{some $w \in \barAs_0$ accepted by $\top$}\big\}. %
    \end{IEEEeqnarray*}
  The \emph{literal language accepted by $A$} then is defined by
  \begin{IEEEeqnarray*}{lCl}
    L_0(A) & = & \big( \setw{w}{\maketuple{w, 1} \in \Lpre(A)} \\
    & \cup & \setw{vu}{\maketuple{v, \top} \in \Lpre(A),\ u \in \barAs} \big) \cap \bars(\emptyset). 
  \end{IEEEeqnarray*}
  Lastly, the \emph{bar language accepted by $A$} is given by $L_0(A)/{\alphaequiv}$. We
  note that the literal language is already closed under $\alpha$-equivalence by definition.
\end{enumerate}
\end{defn}

\begin{rem}\label{ex:cntSimple}
  We give a short reason why this definition of acceptance seems more involved than
  expected from classical AFAs, even though the definition of acceptance in RANAs is not.
  Consider the RANA
  $A = \maketuple{Q, \delta, q_0}$ with $Q = \set{q_0, q_1, q_{\top}} + \set{s, t} \times \names$
  and $\delta$ defined by
  \begin{equation*}
    \begin{IEEEeqnarraybox}[\relax][c]{lCl"lCl}
      \delta(q_0) & = & \Diamond_{\scriptnew{a}} s(a) &
      \delta(q_1) & = & \Diamond_{\scriptnew{a}} s(a) \vee \Diamond_{\scriptnew{a}} t(a), \\
      \delta(q_{\top}) & = & \top, & 
      \delta(t(a)) & = & \Diamond_a q_{\top} \vee \Diamond_{\scriptnew{b}} t(a), \\
      \delta(s(a)) & = & \Diamond_a q_1.
    \end{IEEEeqnarraybox}
  \end{equation*}

  Bar AFAs have a finite alphabet $\barNames_0 \seq \barNames$, which implies that, with a classical definition of acceptance, we can only have a bounded number of different names occuring before a~$\top$-state is reached.
  The above RANA accepts (amongst others) all bar strings of the form
  $w = \newletter{a}a(\newletter{b_i}b_i)_{i = 1}^{m}\newletter{c}\newletter{d}cab_1 \cdots b_m$
  for any $m$, where all letters are distinct. It therefore accepts strings with an unbounded number of names. However, the prefix of $w$ that is processed before reaching the $\top$-state is given by $v = \newletter{a}a(\newletter{b_i}b_i)_{i = 1}^{m}\newletter{c}\newletter{d}c$,
  which is $\alpha$-equivalent to a bar string with only two \emph{distinct} letters. The above
  definition of accepted language via pre-words addresses this
  situation by taking all bar strings $v$ which are $\alpha$-equivalent to pre-words of words accepted by $\top$ in the pre-language of a bar AFA, and then postcomposing $v$ with any other bar string to obtain the literal language.
  Note, that this RANA is also an ERNNA (as no conjunction occurs in any transition formula). We remark that a
  similar problem arose in the definition of extended bar NFAs~\cite{hms21}, which can be resolved in a similar manner.
\end{rem}

We now show how to turn a RANA into an bar language equivalent bar AFA. We consider (w.l.o.g.) name-dropping RANAs, which
allows us work with restricted satisfaction (\Cref{lem:ndrsem}), thus avoid implicit $\alpha$-renamings. This simplifies the construction and its correctness proof.

\begin{construction}[RANA to bar AFA] \label{constr:barafa}
 Given the name-dropping modification $A_{\nd}$ of a RANA, where $A_{\nd}$
  has degree $k$ and $n$ orbits, we construct a bar AFA $A_0$ as follows. If $\delta_{\nd}(q_0) \equiv
  \top$, put $A_0 = \maketuple{\set{q_0}, \set{\newletter{\ast}}, \delta_0, q_0, \set{q_0}}$
  with $\delta_0(q_0, \newletter{\ast}) = \top$. If $\delta_{\nd}(q_0) \not\equiv
  \top$, put $A_0 = \maketuple{Q_0, \barNames_0 \cup \set{\newletter{\ast}}, \delta_0, q_0, F_0}$ with the following data:
  \begin{enumerate}
    \item Let $\names_0$ be a fixed $k$-element subset of $\names$, $\ast \in \names\setminus \names_0$ and
      put $\barNames_0 := \names_0 \cup \setw{\newletter{a}}{a \in \names_0}$.
    \item    
      $Q_0 := \setw{q \in Q_{\nd}}{\supp(q) \seq \names_0 \,\wedge\, \delta_{\nd}(q)
      \not\equiv \top}$.
     \item $F_0 := \setw{q \in Q_0}{\varepsilon\vDash\delta_{\nd}(q)}$.
    \item\label{constr:barafa:3} Define the function
      \[f\colon (\barNames_0 \cup \set{\newletter{\ast}}) \times \posbool(1 + \names \times Q_{\nd} + \Abstr{Q_{\nd}})
      \to \posbool(Q_{\nd})\] recursively by ${f(\alpha, \varepsilon) = \bot}$, $f(\alpha, \top) = \top$,
      $f(\alpha, \bot) = \bot$, $f(\alpha, \Diamond_\beta q) = z$, where $z$ is given as follows 
      \[
        \begin{array}{@{}c|l@{}}
          z %
          & \text{if \dots}
          \\
          \hline
          \bot & \text{$\Diamond_\beta q \neq \Diamond_\alpha q'$ for all $q'$,}\\
          \bot & \text{$\alpha = \newletter{\ast}$, $\Diamond_\beta q = \Diamond_\alpha q'$
            for some $q'$ s.th.~$\ast \in \supp(q')$,}\\
          \top & \text{$\Diamond_\beta q = \Diamond_\alpha q'$ for some $q'$ and $\delta_{\nd}(q') \equiv \top$,}\\
          q' & \text{$\Diamond_\beta q = \Diamond_\alpha q'$ for some $q'$ and $\delta_{\nd}(q') \not\equiv \top$;}
        \end{array}
     \takeout{ %
      f(\alpha, \Diamond_\beta q)
      =
      \begin{cases}
        \bot & \text{if $\Diamond_\beta q \neq \Diamond_\alpha q'$ for all $q'$,} \\
        \bot & \text{if $\alpha = \newletter{\ast}$, $\Diamond_\beta q = \Diamond_\alpha q'$ for some $q'$} \\
             & \text{with $\ast \in \supp(q')$,}\\
        \top & \text{if $\Diamond_\beta q = \Diamond_\alpha q'$ for some $q'$} \\
             & \text{and $\delta_{\nd}(q') \equiv \top$,} \\
          q' & \text{if $\Diamond_\beta q = \Diamond_\alpha q'$ for some $q'$}\\
             & \text{and $\delta_{\nd}(q') \not\equiv \top$;}
           \end{cases}
           }%
      \]
      and $f(\alpha, \varphi \ast \psi) = f(\alpha, \varphi) \ast f(\alpha, \psi)$
      for $\ast \in \set{\vee,\wedge}$.
    \item Define the transition function $\delta_0$  by  $\delta_0(q, \alpha) =
      f(\alpha, \delta_{\nd}(q))$ for $q \in Q_0$ and $\alpha \in \barNames_0 \cup \set{\newletter{\ast}}$. Note that
      every variable $q'$ in $f(\alpha, \delta_{\nd}(q))$ is an element of $Q_0$ because of the support principle
      (\Cref{lem:suppRANA}).
  \end{enumerate}%
\end{construction}

\noindent The key observation towards the correctness of the construction is that every bar string accepted by $A_{\nd}$ has an accepted prefix that is $\alpha$-equivalent to a bar string over $\barNames_0 \cup \set{\newletter{\ast}}$. The additional bar name \enquote{$\newletter{\ast}$} is needed to capture bar strings such as $\newletter{a}a_1 \cdots a_k$ where $\newletter{a}$ is not matched by an occurrence of~$a$.

\begin{rem}\label{R:barafa}
  \begin{enumerate}%
  \item Note that every state contained in the formula $f(\alpha, \delta_{\nd}(q))$ lies in $Q_0$ due to the support principle (\Cref{lem:suppRANA}). 
  \item\label{R:barafa:2} For a name-dropping RANA $A_{\nd}$ with degree $k$ and $n$ orbits, the bar AFA $A_0$ has an alphabet of size $2\cdot k + 1$ and at most $n \cdot k!$ states: each orbit has support size $\ell \leqslant k$ and contributes as many elements to $Q_0$ as there are injections
      from the ordinal set $\ell$ to $\names_0$, at most the number of injections from $k$ to $\names_0$, which is $k!$. Thus, $|Q_0| \leqslant n \cdot k^k = n \cdot 2^{k \cdot \log k}$.  
    \item We remark about the exclusion of \enquote{$\top$-states}, i.e.~states $q \in Q$ where
      $\supp(q) \seq \names_0$ but $\delta_{\nd}(q) \equiv \top$. The proposed automata model does not allow
      a single transition formula for the complete state, but only a transition formula per state and
      input letter. In the worst case, as the counterexample from~\Cref{ex:cntSimple} shows,
      at a $\top$-state no additional letter can be read without increasing the alphabet simultaneously.
      To counteract this, we replace any such occurrences of $\top$-states with $\top$
      at the respective positions in the transition formula. This is also the reason for the separate
      treatment when $\delta_{\nd}(q_0) \equiv \top$.
  \end{enumerate}
\end{rem}

\begin{theo} \label{thm:bafaEquiv}
For $A_{\nd}$ and $A_0$ as in \Cref{constr:barafa}, we have $L_\alpha(A_0) = L_\alpha(A_{\nd})$.
\end{theo}

\begin{rem}\label{rem:bafa2rana}
Conversely, every bar AFA can be transformed into an equivalent name-dropping RANA (see~\Cref{app:secND}). %
Hence, RANAs and bar AFAs are expressively equivalent.
\end{rem}

\section{De-Alternation of RANAs}\label{sec:dealt}

We next present our first main result: The de-alternation of RANAs to
ERNNAs, that is, RANAs whose transition formulae only use the Boolean connectives $\bot$, $\top$, $\vee$ (\Cref{defn:rana}).
Without loss of generality, we will work under the following
assumptions.

\begin{assumption}\label{ass:ranalookslike}
  We fix a positive RANA $A$ with a strong nominal state set and continue to denote its name-dropping modification by $A_{\nd}$ (\Cref{constr:name-dropping}).
  In addition, we assume that every transition formula $\delta_{\nd}(q)$ for $q \in Q_{\nd}$ is in reduced DNF (\Cref{rem:ranadefn}) and that all disjuncts are
  modal atoms of the same kind, that is, of the form $\varepsilon$ or $\bigwedge_j \Diamond_{\alpha} q_j$ for the same $\alpha$.
\end{assumption}

Our de-alternation procedure uses a powerset construction to keep track of all states that the original automaton currently can be in,
just like for the de-alternation of classical alternating finite automata (AFA)~\cite{fjy90}.
However, in order to preserve orbit-finiteness, our construction needs to restrict the number of states that are tracked simultaneously (see \Cref{rem:powerset} below).
For this, we note that if we fix a single bar string to be read, and if two states $q, q' \in Q_{\nd}$ from the same orbit but with different
supports can be reached by reading a prefix of the bar string, then one of the states can be support-restricted to $\supp(q) \cap \supp(q')$.

To formalize this idea, we need to introduce some notation. For $\maketuple{i, r} \in Q_{\nd}$ and $J\seq \At$, put
$\restr{\maketuple{i, r}}{J}=\maketuple{i, \restr{r}{J}}$ (\Cref{defn:dollar}).
Given a finite set $\Phi$ of finite subsets of~$Q_{\nd}$, the set
$\rest(\Phi)$ to which $\Phi$ \emph{restricts} is obtained by
repeatedly replacing $S \in \Phi$ with all the sets
$S \setminus \set{q} \cup \set{\restr{q}{J}}$ and
$S\setminus \set{q'} \cup \set{\restr{q'}{J}}$, where $q, q' \in S$ is
a pair of states that lies in the same orbit of $Q_{\nd}$ with
$\supp(q)\neq \supp(q')$, and $J=\supp(q)\cap \supp(q')$. (A similar
support restriction technique has been applied to nominal fixed-point logics~\cite{hms21}.)

\begin{lem}\label{lem:rest-termination}
  The restriction process defined above terminates after finitely many steps. 
\end{lem}
\noindent
Note that, by definition, no $S \in \rest(\Phi)$ contains two states $q, q'$ that lie in same orbit of $Q_{\nd}$ and satisfy $\supp(q) \neq \supp(q')$.
This implies that each $S \in \rest(\Phi)$ contains at most $n \cdot k!$ elements, where $n$ is the number of orbits of $A$ and $k$ its degree.

\begin{lem}\label{lem:correst}
  Let $S \seq Q_{\nd}$ be a finite set of states. Then a bar string $w \in \barAs$
  is accepted by \emph{all} $q \in S$ iff $w$ is accepted by \emph{all} $q \in S'$
  for some $S' \in \rest(\{S\})$.
\end{lem}
\begin{IEEEproof}[Proof sketch]
  One only needs to show that two states~$q$ and~$q'$ accept a bar string iff $\restr{q}{J}$
  and $q'$ or $q$ and $\restr{q'}{J}$ do for $J = \supp(q) \cap \supp(q')$. This
  is readily verified by use of~\Cref{lem:restrRun} in combination with~\Cref{lem:ndrsem}. The
  statement follows then directly from the definition of $\rest(\{S\})$ and an iterative
  argument.
\end{IEEEproof}
The above lemma is the key ingredient to the de-alternation of a RANA to an  ERNNA:

\begin{construction}[RANA to ERNNA]\label{constr:dealt}
  Given a RANA $A_{\nd}$ as in \Cref{ass:ranalookslike}
  with $n$ orbits and degree $k$, we define the ERNNA $A_E=(Q_E,\delta_E,\{q_0\})$ as follows:
  \begin{enumerate}
    \item The nominal set $Q_E$ of states is the set of all finite subsets 
    $\{q_1,\ldots,q_j\}\seq Q_{\nd}$ with $j \leqslant n \cdot k!$, $\delta_{\nd}(q_i) \not\equiv \top$
    for all $i$ and
    \[ 
      \card{\setw{q_\ell}{\ell \leqslant j, \exists \pi.\ q_\ell = \pi \cdot q_i}} \leqslant k!
      \qquad\text{for all $i \leqslant j$}.
    \]
    The
      group action on $Q_E$ acts as expected for subsets of nominal sets.
    \item The equivariant transition function $\delta_E \colon Q_E \to \posbool(1 + \names \times Q_E + \Abstr{Q_E})$ is defined
      as follows: Put $\delta_E(\emptyset) = \top$.
      For $S = \set{q_1, \dots, q_j} \in Q_E$ with $j \geqslant 1$:
      \begin{enumerate}
        \item If \emph{all} $\delta_{\nd}(q_i)$'s contain an $\varepsilon$-disjunct, add one to
          $\delta_E(S)$. %
        \item For all $\alpha \in \barNames$, if \emph{every} $\delta_{\nd}(q_i)$ contains an $\alpha$-disjunct $C_i = \bigwedge_{\ell \in I_i} \Diamond_{\alpha} q^i_\ell$, then for every choice $C_1, \ldots, C_j$ of $\alpha$-disjuncts, one from every $q_i$, the following $\alpha$-disjuncts are added:
          put \(\Delta_\alpha^S := \setw{q^i_\ell}{1 \leqslant i \leqslant j, \ell \in I_i}\) and add an $\alpha$-disjunct $\Diamond_{\alpha} (S' \setminus \setw{q \in S'}{\delta_{\nd}(q) \equiv \top})$ for every $S' \in \rest\left(\set{\Delta_\alpha^S}\right)$.
        \item If none of the above two cases applies, put $\delta_E(S) = \bot$.
      \end{enumerate}
  \end{enumerate}
\end{construction}

\noindent The ERNNA $A_E$ has a degree of at most $n \cdot k \cdot k! \leqslant
n \cdot 2^{(k + 1)\log(k)}$, since each element of~$Q_E$ contains at
most $n \cdot k!$ elements whose least supports are, in the worst
case, mutually disjoint and of size $k$.
To obtain an upper bound on the number of orbits of $Q_E$, we record %
\begin{lem}\label{L:Bell}
  Given a nominal set with $n$ orbits and degree~$k$, the nominal set of all subsets of size at most $\ell$ has at most $\ell \cdot n^\ell \cdot 2^{k\cdot \ell \cdot \log(k \cdot\ell)}$ orbits.
\end{lem}

Applying the lemma to $\ell = n \cdot k!$ shows that $Q_E$ has at most
$
  n \cdot k! \cdot  n^{n \cdot k!} \cdot 2^{k \cdot n \cdot k! \cdot\log(k \cdot n \cdot k!)}$ orbits, which is exponential in~$n$ and doubly exponential in $k$, using $k! \leqslant k^k \leqslant 2^{k\cdot \log k}$.

  The bound on the sets of states in $A_E$ is necessary since we do not only need to keep track
  of states in different orbits but also states in the same orbits having the same support.
  All other states of a single orbit get removed by the state restriction process.  It is easy
  to verify that~\Cref{constr:dealt} yields an ERNNA.  That $\delta_E$ is well-defined follows
  from the bounds of sets obtained in the state restriction process.  (Namely, a finite subset
  of $Q_{\nd}$ such that any two elements in the same orbit have the same support has
  cardinality at most $n\cdot k!$.)  The correctness of the construction in the sense that
  $A_E$ accepts the same literal language as the RANA $A_{\nd}$ is readily verified by
  induction on the word length using the correctness of the state restriction process
  (\Cref{lem:correst}). Hence, we obtain

\begin{theo}\label{thm:corrdealt}
  RANAs can be de-alternated to ERNNAs with exponential blowup of the degree
  and doubly exponential blowup of the number of orbits.
\end{theo}

\noindent In fact, the latter blowup is parametrized singly exponential with the
degree as parameter.

\begin{rem}[Need for $\top$-states]%
A `full' de-alternation from RANAs to RNNAs (that is, RANAs whose transition formulae only use the Boolean connectives $\bot$, $\vee$) under preservation of the literal language is not possible. Consider the RANA $A = \maketuple{\set{q_0},\delta,q_0}$ with
    $\delta(q_0) = \top$, which accepts the universal bar language, i.e.~every closed bar string.
    There exists no RNNA accepting that language. Indeed, suppose that such an RNNA~$A$ exists.
    Then, for each natural number $n$, it accepts the closed bar string
    $w_n = \newletter{a_1}a_1\newletter{a_2}a_1a_2\cdots\newletter{a_n}a_1\cdots a_n$.
    Hence $A$ has a state $q$ accepting the suffix $a_1 \cdots a_n$. The support principle for RNNAs
    (cf.~\cite[Lemma~5.4]{skmw17}) entails that supports can only grow along bar transitions. This
    implies that $a_1, \dots, a_n$ must be contained in $\supp(q)$. Choosing $n>\deg(A)$ yields a contradiction.

    Under global freshness semantics, a similar argument shows that a translation from RANAs
    to RNNAs is impossible. However, a translation is possible under local freshness semantics,
    where ERNNAs and RNNAs are equivalent~\cite[Rem.~5.2]{hms21}. 
\end{rem}

\begin{rem}\label{rem:powerset}%
To see that naïve de-alternation via an unrestricted powerset construction would not work,
consider the positive RANA $A = \maketuple{Q, \delta, q_0}$ with the set of states $Q = \set{q_0} +
  \set{q_1} \times \names$ and transitions $\delta(q_0) = \Diamond_{\newletter a} q_0 \wedge \Diamond_{\newletter a} q_1(a)$ and $\delta(q_1(a)) = \Diamond_{\newletter b} q_1(a)$.
  On input $w = \newletter{a_1}\cdots\newletter{a_n}$, the naïve powerset construction reaches the set $\{q_0,q_1(a_1),\ldots, q_1(a_n)\}$; hence it needs to store unboundedly many states, contradicting the required orbit-finiteness of
  the state set.
  \Cref{constr:dealt} avoids this by reducing the above $(n+1)$-element set to a two-element set.
\end{rem}

\section{Decidability of Inclusion} \label{sec:decincl}

We now show our second main result:
Language inclusion of RANAs is decidable under the bar language, the global and the local freshness semantics.
To derive the corresponding complexity bounds, we will use the equivalence to bar AFAs (\Cref{defn:bafa}).

\begin{rem}
  For algorithms deciding properties of RANAs, a finite representation
  of the underlying nominal set of states and transitions is
  required.
  We assume without loss of generality that RANAs have a strong nominal
  state set, which clearly have a finite representation. Since the transition
  function is equivariant, we can represent it via finite representations of
  finitely many formulae, namely one for every orbit.
\end{rem}

\begin{rem} \label{rem:compConstr}
  We summarize the complexities of constructions in previous sections.
  Given a RANA with $n$ orbits and degree $k$, its name-dropping modification (of an equivalent positive RANA) has at most $(2n(k + 2)+1) \cdot 2^{(2k + 1) \cdot \log(4k + 2)}$ orbits, which is linear in $n$ and exponential in~$k$, and a degree of $2k + 1$ (\Cref{lem:constr01,lem:constr02,constr:name-dropping}).
  After de-alternation (\Cref{constr:dealt}), the resulting ERNNA has a degree that is linear in $n$ and exponential in~$k$, and  a number of orbits that is exponential in~$n$ and doubly exponential in $k$.
\end{rem}

We first consider the \emph{non-emptiness problem}, which asks whether the accepted language of a RANA $A = \maketuple{Q, \delta, q_0}$ is non-empty. We will reduce this problem to the non-emptiness problem for bar AFAs. We first introduce a different, \emph{finite}, semantics for bar AFAs:

\begin{defn}[Finite semantics]
  Given a bar AFA $A = \maketuple{Q, \barNames_0, \delta, q_0, F}$, its \emph{AFA-language}
  $L_{\textsf{\scriptsize AFA}}(A)$ consists of those bar strings $w \in \barAs_0$ that are accepted by $A$ when
  interpreted as a classical AFA~\cite{cks81} over the finite alphabet $\barNames_0$.
\end{defn}

\begin{lem}\label{lem:afaequivalence}
  For every bar AFA $A$, we have $L_0(A) = \emptyset$ iff
  $L_{\textsf{\scriptsize AFA}}(A) \cap \bars(\emptyset) = \emptyset$.
\end{lem}
\begin{IEEEproof}
  There are two difference between the AFA-semantics and the bar language semantics (\Cref{defn:bafa}):
  Firstly, a bar AFA can classically (with the AFA-semantics) accept bar strings $w$ that are not closed, i.e.~have
  free names ($\FN(w) \neq \emptyset$), for which we use the intersection; indeed, $L_{\textsf{\scriptsize AFA}}(A)
  \cap \bars(\emptyset)$ only contains closed bar strings. 
  
  Secondly, the bar language semantics handles $\alpha$-equivalent words and
  states with $\delta(q, \alpha) = \top$ (for some $\alpha \in \barNames_0$) differently.  More concretely, with the
  AFA-semantics, a bar AFA can only accept bar strings over $\barNames_0$, not over $\barNames$, which is possible
  using the bar language semantics. However, every closed bar string accepted with the AFA-semantics is also accepted
  using the bar semantics, yielding $L_{\textsf{\scriptsize AFA}}(A) \cap \bars(\emptyset) \seq L_0(A)$ and in turn the
  \enquote{if} direction.

  For \enquote{only if}, let $L_0(A) \neq \emptyset$ and $w \in L_0(A)$. Note, that $w$ is closed by design.
  Then either there is a word $w' \alphaequiv w$ such that $w'$ is accepted in whole by $A$, or there
  is a prefix $v$ and an $\alpha$-equivalent~$v'$ such that $v'$ is the pre-word of a bar string $w$ that is accepted
  by $\top$. By definition, $w'$ and $v'$ both are words over $\barNames_0$ and closed.
  Therefore $w' \in L_{\textsf{\scriptsize AFA}}(A) \cap \bars(\emptyset)$ in the first case, and
  $v' \in L_{\textsf{\scriptsize AFA}}(A) \cap \bars(\emptyset)$ in the second case, proving
  $L_{\textsf{\scriptsize AFA}}(A) \cap \bars(\emptyset) \neq \emptyset$.
\end{IEEEproof}

\begin{rem}\label{rem:fixAFA}
  The language $L_{\textsf{\scriptsize AFA}}(A) \cap \bars(\emptyset)$ of all closed bar
  strings accepted by $A$ (\Cref{lem:afaequivalence}) is accepted by a classical AFA. To see
  this, note that the intersection is a language over the finite alphabet $\barNames_0$ of the
  bar AFA $A$. Thus, an AFA accepting all closed bar strings over $\barNames_0$ is enough to
  prove the claim: Fixing the finite alphabet $\barNames_0$ of size $k$ and the set
  $S := \setw{a \in \barNames_0}{\newletter{a} \in \barNames_0}$, we define the (classical) AFA
  $A_{\textsf{\scriptsize c}} = \maketuple{\pow(S), \barNames_0, \delta_{\textsf{\scriptsize
        c}}, \emptyset, \pow(S)}$ with the following transitions: For $M \seq S$ and
  $\newletter{a}, b \in \barNames_0$, put
  \[
    \delta_{\textsf{\scriptsize c}}(M, \newletter{a}) = (M \cup \set{a}) \cap S, \quad
    \delta_{\textsf{\scriptsize c}}(M, b)
    =
    \begin{cases}
      M & \text{if $b \in M$,} \\
      \bot & \text{otherwise.}
    \end{cases}
  \]
  Then  
  $A_{\textsf{\scriptsize c}}$ has at most $2^{\nicefrac{k}{2}} < 2^k$ states and accepts all closed bar strings over the finite alphabet $\barNames_0$. 
  With a similar construction as in the proof of~\Cref{thm:closure}, we get a classical AFA $A'$ accepting
  the intersection language with at most $n + 2^{\nicefrac{k}{2}} + 1 \leqslant n + 2^k$ states over the finite
  alphabet $\barNames_0$.
\end{rem}

Decidability of non-emptiness now follows from the corresponding result for classical AFAs~\cite[Thm.~3.1]{jr91}
and the equivalence between bar AFAs and RANAs (\Cref{lem:afaequivalence}):

\begin{theo}[Decidability of non-emptiness]
  \label{thm:decneRANA}
  \begin{enumerate}%
    \item Non-emptiness of bar AFAs is decidable
      in space polynomial in the number of states and the size of the alphabet.
    \item Non-emptiness of RANAs with $n$ orbits and degree $k$ 
      is decidable in \EXPSPACE, more precisely, using space polynomial in $n$
      and exponential in $k$.
  \end{enumerate}
\end{theo}

We turn to the language inclusion problem for RANAs, which asks for two given RANAs $A_1$ and $A_2$ whether $L_\alpha(A_1) \seq L_\alpha(A_2)$ holds. Just like for classical AFAs, inclusion can be reduced to non-emptiness, using that $L_\alpha(A_1) \seq L_\alpha(A_2)$ iff $L_\alpha(A_1)\cap ({L_\alpha(A_2)})^\complement = \emptyset$.

\begin{theo}[Decidability of inclusion] \label{thm:decincl}
  The inclusion problem for RANAs
  is decidable in \EXPSPACE; more precisely, 
  in space polynomial in the number of orbits and exponential in the degrees of both RANAs.
\end{theo}

\noindent %
This result also yields inclusion checking under
global freshness semantics with the same complexity, since the global
freshness operator $N$ in~\eqref{eq:ND} preserves and reflects
language inclusion, as mentioned at the end
of~\Cref{sec:prelims}. As an immediate consequence of~\Cref{theo:corrFormAut,thm:decneRANA,thm:decincl},
we obtain both an upper bound on model checking bar formulae $\varphi$ over  RANAs $A$ (understood as the
question whether $L_\alpha(A) \seq \sem{\varphi}$), as well as an upper bound on satisfiability checking bar formulae
$\varphi$ (understood as checking $\sem{\varphi} \neq \emptyset$).
\begin{cor}
  The model checking problem for \muBar{} over RANAs and the satisfiability problem for \muBar{} are in
  \EXPSPACE.
\end{cor}
\noindent
More precisely, for model checking, the space complexity is polynomial in the number of orbits of
the automaton and the size of the formula and exponential in the degrees of both the automaton
and the formula; this improves on a previous result~\cite[Thm.~7.2]{hms21} by generalizing models
from RNNAs to RANAs and improving the complexity bound by one exponential
level. For satisfiability, the space complexity is polynomial in the size of the formula and
exponential in its degree, whence we recover the previous result from op.~cit.

Lastly, with a slightly different approach, we are
also able to decide the inclusion problem under local freshness
semantics.

\begin{theo}[Decidability of inclusion under local freshness]
  The inclusion problem for the local freshness semantics of RANAs is decidable
  in 2\EXPSPACE; more precisely, in space exponential in the number
  of orbits and doubly exponential in the degrees of both RANAs.
\end{theo}
\begin{IEEEproof}
  We convert both RANAs to positive RANAs
  (\Cref{lem:constr01,lem:constr02})
  and form their name-dropping modifications
  (\Cref{constr:name-dropping}).  We use~\Cref{constr:dealt} to
  de-alternate both modifications into ERNNAs, which, under local
  freshness semantics, are equivalent to RNNAs with the same number of
  orbits and the same degree as the ERNNAs (\cite[Rem.~5.2]{hms21}). The
  claim then follows since %
  inclusion of RNNAs under local
  freshness can be checked in space bounded polynomially in the numbers
  of orbits and exponentially in the degrees~\cite[Cor.~7.4]{skmw17}. By
  \Cref{rem:compConstr}, this results in space singly
  exponential in the number of orbits of both RANAs and doubly
  exponential in their degrees.
\end{IEEEproof}

\section{Conclusion}\label{sec:concl}

We have introduced \emph{regular alternating nominal automata}
  (RANAs), an alternating nominal automata model with name
allocation. RANAs generalize both their nondeterministic restriction,
\emph{regular nondeterministic nominal automata} (\mbox{RNNAs})~\cite{skmw17}, and the extension of RNNAs with deadlocked
universal states, i.e.~the \emph{extended RNNAs }used in work on
nominal fixed-point logics~\cite{hms21}.  We have shown that every
RANA can be converted into a positive one having no negation in its
transition formulae. Building on this, we have shown that
RANAs can be de-alternated to extended RNNAs, incurring (only)
doubly exponential blowup.  We have also shown that the emptiness and
inclusion problems of the RANA model are decidable in
\EXPSPACE under global freshness semantics, and in
2\EXPSPACE under local freshness. Lastly, we improved the
model-checking problem for the linear-time nominal fixed-point logic
\muBar~by one exponential level.

In future work we will extend our approach to infinite data words, aiming for an
alternating extension of \emph{Büchi RNNA}~\cite{uhms21}. This will provide an
automata-theoretic basis for specification logics over infinite data words, with a perspective to
obtain elementary decidability of model checking and satisfiability checking.

\section*{Acknowledgments}
We are grateful to Thorsten Wißmann for discussions on the upper bound of orbits in the sets of subsets of a nominal~set.

\bibliographystyle{IEEEtranS}
\bibliography{IEEEabrv,refs}

\clearpage
\appendices
\onecolumn
\begingroup
\crefalias{section}{appendix}

\section{Details for~\Cref{sec:rana}} \label{app:secRANA}

\begin{rem}\label{R:Bell}
  We collect a number of facts on the size of various orbit-finite sets.
  Below we make use of the \emph{Bell number}~$B_n$, which is the number of equivalence relations on a set with $n$ elements.
  Note that $B_n$ is less than $n^n = 2^{n \cdot \log n}$. We also have $2^n \leq B_{n+1}$, since
  every subset (together with its complement) yields a partition of cardinality two. 
  \begin{enumerate}
  \item The nominal set $\names^{\# n}$ has a single orbit and degree $n$.
  \item\label{R:Bell:2} A power $\names^n$ of the set of all names has degree $n$ and $B_n$ orbits; indeed, the orbit of a tuple $(a_1, \ldots, a_n)$ is uniquely determined by the equivalence relation on the set $\set{1, \ldots,n}$ relating indices of coordinates that have the same name: $i$ and $j$ are related if $a_i = a_j$.
  \item\label{R:Bell:3} The disjoint union of a pair of nominal sets with $n_i$ orbits and degree $k_i$ ($i = 1, 2$) has $n_1 + n_2$ orbits and degree $\max\set{k_1,k_2}$.
  \item\label{R:Bell:4} The product of a pair $X_1, X_2$ of nominal sets that have $n_i$ orbits and degree $k_i$, $i = 1, 2$, has degree $k_1 + k_2$ and at most $n_1 \cdot  n_2 \cdot (k_1+k_2)^{k_1 + k_2}$ orbits.
    For the former, note that the support of an ordered pair $(x,y)$ is $\supp(x) \cup \supp(y)$. For the latter, note first that $X_i$ is a quotient of $n_i \bullet \names^{\# k_i}$, the $n_i$-fold coproduct of $\names^{\# k_i}$, for $i = 1, 2$. We write $\norb(X)$ for the number of orbits of the orbit-finite nominal set $X$ and estimate:%
    \begin{IEEEeqnarray*}{lClClCl}
      \norb(X_1 \times X_2)
      &\leqslant&
      \norb\big(n_1 \bullet \names^{\# k_1} \times n_2 \bullet \names^{\# k_2}\big)
      &\leqslant&
      \norb\big(n_1 \bullet \names^{k_1} \times n_2 \bullet \names^{k_2}\big)
      &=&
      \norb\big((n_1\cdot n_2) \bullet \names^{k_1+k_2}\big) \\
      &=&
      n_1 \cdot n_2 \cdot \norb(\names^{k_1+k_2})
      &=&
      n_1 \cdot n_2 \cdot B_{k_1 + k_2}
      &\leqslant&
      n_1 \cdot n_2 \cdot (k_1 + k_2)^{k_1 + k_2}.
    \end{IEEEeqnarray*}

  \item\label{R:Bell:5} For an orbit-finite nominal set $X$ that has $n$ orbits and degree $k$, the nominal set of all subsets of size at most $\ell$ has degree $k \cdot \ell$ and at most $\ell \cdot n^\ell \cdot (k\cdot \ell)^{k \cdot\ell}$ orbits.
    Indeed, we use again that $X$ is a quotient of $n \bullet
    \names^{\# k}$. The number of orbits of the set of all subsets of $X$ of a fixed size $\leqslant j$ is bounded by the number of orbits of the product $X^j$, which is bounded above by 
    $n^j \cdot (k\cdot j)^{k \cdot j}$ using~\Cref{R:Bell:4}.
  \end{enumerate}
\end{rem}

\begin{rem}\label{rem:ernna}
Given an ERRNA $A=(Q,\delta,q_0)$ we write $q\xto{\alpha} q'$ if $\Diamond_\alpha q'$ is a disjunct of $\delta(q)$, and a state $q$ is called \emph{final} if $\delta(q)$ contains the disjunct $\epsilon$. Acceptance of words then can be phrased in terms of accepting runs, as expected for nondeterministic automata~\cite{skmw17,hms21}: the word $w\in \barAs$ is accepted iff either (i) there exists a \emph{complete accepting run}, i.e.~a $w$-labelled path $q_0\xto{w} q_f$ ending in a final state, or (ii) there exists a \emph{partial accepting run}, i.e.~a $w'$-labelled path $q_0\xto{w'} q_\top$ ending in a $\top$-state for some prefix $w'$ of $w$.
\end{rem}

\paragraph*{Details for \Cref{ex:succinct}}

We begin by explaining how the positive RANA works: Informally, for bar strings of the form $uvvw$ the RANA nondeterministically guesses the plain names occuring in $v$ (second big disjunction in $\delta\maketuple{0,S}$).
Since the input is a closed bar string, all plain names occuring in $v$ form a subset of the bar names occuring in
the prefix $u$. When at position $i$ in the first occurrence of $v$, we \enquote{store} the plain name $\sigma$
at that position in the state $\maketuple{N, S, \sigma}$. Here, the index is decreased by one in every step, regardless
of the input name, which can be a name in $S \cup \set{\sigma}$ (see $\delta\maketuple{j,S,\sigma}$ for $j > 1$).
Precisely $N$ positions after position $i$ in the first occurrence, $\sigma$ has to reoccur in the input bar string
(see $\delta\maketuple{1,S,\sigma}$). Due to the universal branching (conjunction) with states $\maketuple{i,S}$
for $0 \leqslant i < N$, we immediately see that after the first copy of $v = \sigma_1\cdots\sigma_N$ is read, each
of the states $\maketuple{j,S,\sigma_j}$ for $1 \leqslant j \leqslant N$ has to accept the remaining input bar string.
This happens iff the remaining input starts with a second copy of $v$.

More formally, we have the following:
\begin{itemize}
  \item States $\maketuple{j, S, \sigma}$ for $1 \leqslant j \leqslant N$ accept precisely bar strings of the form 
    $v'\sigma w$ where $v' \in (S \cup \set{\sigma})^{j - 1}$ and $w \in \barAs$. This follows immediately from
    induction over $j$.
  \item Similarly, states $\maketuple{i, S}$ for $1 \leqslant i < N$ accept precisely bar strings of the form
    $\sigma_{i+1}\cdots\sigma_Nv'\sigma_{i+1}\cdots\sigma_N w$ for $v' \in S^{i}$ and $w \in \barAs$.
    We show this via a \enquote{backward} induction over $i$: The base case $i = N - 1$ follows directly, we have
    \[ w\vDash\bigvee_{\sigma \in S} (\Diamond_\sigma\maketuple{N, S}\wedge\Diamond_\sigma\maketuple{N, S, \sigma}) \]
    iff $w$ starts with $\sigma$ ($\delta\maketuple{N, S} = \top$) and is of the form $w = \sigma v\sigma w'$
    for $v \in S^{N - 1}$ and $w' \in \barAs$ (see first point).
    For arbitrary $i$, we see that
    \[ w\vDash\bigvee_{\sigma \in S} (\Diamond_\sigma\maketuple{i + 1, S}\wedge
      \Diamond_\sigma\maketuple{N, S, \sigma}) \] 
    holds iff $w$ is of the form
    $w = \sigma\sigma_{i+2}\cdots\sigma_Nv'\sigma_{i+2}\cdots\sigma_Nw'$ for $v' \in S^{i+1}$ and $w' \in \barAs$
    (by induction hypothesis) and of the form $w = \sigma v''\sigma w''$ for $v'' \in S^{N-1}$ and $w'' \in \barAs$
    (by the first point). Combining both, we get $w$ of the required form.
  \item Lastly, for states $\maketuple{0,S}$ we have two kinds of bar strings that are accepted: Firstly, those
    of the form $\sigma_1\cdots\sigma_n\sigma_1\cdots\sigma_Nw'$ for $w' \in \barAs$ and $\sigma_i \in S$ (via
    the first big disjunction), and secondly, those of the form 
    $\alpha_1\cdots\alpha_m\sigma_1\cdots\sigma_N\sigma_1\cdots\sigma_Nw'$ with $m \in \Nat$, $\alpha_i \in
    \barNames\setminus\names$, $\sigma_i \in S \cup \setw{\ub(\alpha_j)}{1 \leqslant j \leqslant m}$ and
    $w' \in \barAs$ (via the second big disjunction).
\end{itemize}
Thus, the accepted language of the RANA corresponds precisely to just those closed bar strings of the form
$uvvw$ with $u \in (\barNames\setminus\names)^*$, $v \in \namesN$ and $w \in \barAs$.

Next, we show that the literal language 
\[L_{N} = \text{all closed bar strings
  $uvvw \in \barAs$ such that $w \in \barAs$, $u \in (\barNames\setminus\names)^*$, $v \in \namesN$}\]
is not accepted by any ERNNA with less than $B_N$ orbits, where $B_N\geq 2^{N-1}$ is the $N$th Bell
number (\Cref{R:Bell}\ref{R:Bell:2}). The argument uses a nominal adaptation of the \emph{fooling set} technique~\cite{hk11} for proving lower bounds on the state complexity of nondeterminstic finite automata. Suppose that $A=(Q,\delta,q_0)$ is an ERNNA accepting $L_N$. Using the terminology of \Cref{rem:ernna}, note that every partial accepting run for $uvvw\in L_N$ must fully process the prefix $uvv$; otherwise, the same partial run would be accepting for some proper prefix of $uvv$, but no such prefix lies in $L_N$. 

 Let $v,v'\in \namesN$ be two words in distinct orbits of $\namesN$. Choose arbitrary $u,u' \in (\barNames\setminus\names)^*$ and $w,w'\in \barAs$ such that $uvvw,u'v'v'w'\in L_N$, and let
\[ q_0\xto{uv} q\xto {vw_1} q_1 \qquad\text{and}\qquad q_0 \xto{u'v'} q' \xto{v'w_1'} q_1' \]
be partial or complete accepting runs for the two words. Thus either $w_1=w$ and $q_1$ is
final, or $w_1$ is a prefix of $w$ and $q_1$ is a $\top$-state; similarly for $w_1'$ and
$q_1'$. We claim that the two intermediate states $q$ and $q'$ lie in different orbits. Suppose not, that is, $q'=\pi\cdot q$ for some permutation $\pi$. Since transitions are equivariant, we obtain an accepting run
\[ q_0 = \pi\cdot q_0 \xto{(\pi\cdot u)(\pi\cdot v)} \pi\cdot q = q' \xto{v'w_1'} q_1' \]
for the word $(\pi\cdot u)(\pi\cdot v) v'w'$. Since $\pi\cdot v\neq v'$, this word does not lie in $L_N$, a contradiction. 

This proves that $Q$ has at least as many orbits as $\namesN$, i.e.~at least $B_N$ orbits.

\begin{IEEEproof}[Proof of \Cref{lem:closurealph}]
Let $A=(Q,\delta,q_0)$ be a RANA, and suppose that $w \alphaequiv w'$.
  For every state $q\in Q$ we show that $q$ accepts $w$ iff $q$ accepts $w'$. We may assume w.l.o.g.~that $w = \newletter{a}v$ and $w' = \newletter{b}\makecycle{b,a}\cdot v$ for $b \fresh v$. The proof proceeds by induction on $\varphi = \delta(q)$. The cases $\varphi \in \set{\varepsilon,\top,
    \bot}$ and $\varphi = \Diamond_\sigma q'$ with $\sigma \in \names$ are trivial.
  Suppose that $\varphi = \Diamond_{\scriptnew{c}} q'$ for some $\newletter{c} \in \barNames$.
  If $q$ accepts $w$, i.e.~$w\vDash\Diamond_{\scriptnew{c}}q'$, there is some $\newletter{d}v' \alphaequiv w$ with $\braket{c}q'  = \braket{d}q''$, such that $v' \vDash \delta(q'')$.
 By transitivity of $\alpha$-equivalence, we have $\newletter{d}v' \alphaequiv w'$; therefore, $w'\vDash\Diamond_{\scriptnew{c}}q'$, so $q'$ accepts $w'$.
  The cases $\varphi\in \{\varphi_1\vee \varphi_2, \varphi_1\vee \varphi_2, \neg \phi_1\}$ are immediate by induction.
\end{IEEEproof}

\begin{IEEEproof}[Proof of \Cref{lem:suppRANA}]
    \begin{enumerate}
        \item Suppose that $\Diamond_a q'$ occurs in $\delta(q)$. Since $\delta$ is equivariant, we have $\supp(q') \cup \set{a} = \supp\maketuple{a,q'} = \supp \Diamond_a q' \seq \supp(\delta(q)) \seq \supp(q)$. Similarly for $\Box_a q'$.
        \item Suppose that $\Diamond_{\scriptnew a} q'$ occurs in $\delta(q)$. Then  $\supp(q') \seq \supp(\braket{a}q') \cup \set{a} = \supp \Diamond_{\scriptnew a} q' \cup \{a\} \seq \supp(\delta(q)) \cup \set{a}
            \seq \supp(q) \cup \set{a}$. Similarly for $\Box_{\scriptnew a} q'$. \hfill\IEEEQEDhere
    \end{enumerate}
\end{IEEEproof}

\paragraph*{Details for~\Cref{lem:constr01}}

\begin{construction}[RANA to explicit-dual RANA]\label{constr:neg2dual}
    Given a RANA $A = \maketuple{Q, \delta, q_0}$, we construct an explicit-dual RANA
$\dual{A} = \maketuple{\dual{Q},\dual{\delta},q_0}$ as follows:
    \begin{enumerate}
    \item The nominal set of states is given by $\dual{Q} := Q + \setw{\negstate{q}}{q \in Q}$, with
     $\pi \cdot \negstate{q} =
            \negstate{\left(\pi \cdot q\right)}$. Hence, $\dual{Q}$ just consists of two copies of $Q$.
        \item Define the function $\dual{\trans}_1\colon \negbool(1 + \names \times Q + \Abstr{Q}) \to \dualbool(1 +
            \names \times \dual{Q} + \Abstr{\dual{Q}})$ for formulae in NNF recursively by
\begin{align*}
\dual{\trans}_1(x) &= x & \text{for $x \in \set{\top,\bot,\varepsilon,\neg\varepsilon} + \names \times Q + \Abstr{Q}$};\\
\dual{\trans}_1(\neg\Diamond_a q) &= \Box_a \negstate{q}; & \\
\dual{\trans}_1(\neg\Diamond_{\scriptnew{a}} q) &= \Box_{\scriptnew{a}} \negstate{q};& \\ 
\dual{\trans}_1(\varphi \ast \psi) &= \dual{\trans}_1(\varphi) \ast \dual{\trans}_1(\psi) & \text{for $\ast \in \set{\vee, \wedge}$}.
\end{align*}
        \item Define the function $\dual{\trans}_2\colon \negbool(1 + \names \times Q + \Abstr{Q}) \to \dualbool(1 +
            \names \times \dual{Q} + \Abstr{\dual{Q}})$ recursively by
\begin{align*}
\dual{\trans}_2(\varepsilon) &= \neg \varepsilon &
\dual{\trans}_2(\top) &= \bot &
\dual{\trans}_2(\bot) &= \top \\
\dual{\trans}_2(\Diamond_a q) &= \Box_a \negstate{q} &
\dual{\trans}_2(\Diamond_{\scriptnew{a}} q) &= \Box_{\scriptnew{a}} \negstate{q} & & \\
\dual{\trans}_2(\neg \varphi) &= \varphi &
\dual{\trans}_2(\varphi \vee \psi) &=
              \dual{\trans}_2(\varphi) \wedge \dual{\trans}_2(\psi) &
\dual{\trans}_2(\varphi \wedge \psi) &= \dual{\trans}_2(\varphi) \vee \dual{\trans}_2(\psi). 
\end{align*}
        \item Lastly, the transition function $\dual{\delta}\colon \dual{Q} \to \dualbool(\names \times \dual{Q} + \Abstr{\dual{Q}})$ of $\dual{A}$ is given by
          \[
            \dual{\delta}(q) = \dual{\trans}_1(\delta(q))
            \qquad\text{and}\qquad
            \dual{\delta}(\negstate{q}) = \dual{\trans}_2(\delta(q))
            \qquad
            \text{for every $q \in Q$.}
          \]
    \end{enumerate}
\end{construction}

\begin{rem}
  \begin{enumerate}%
  \item If $A$ has degree $k$ and
    $\ell$ orbits, then $\dual{A}$ has degree $k$ and
    $2 \cdot \ell$ orbits.
  \item The newly introduced states $\negstate{q}$ are intended to
    accept the complement of the languages accepted by the~$q \in Q$. Thus,
    the translation $\dual{\trans}_1$ of transition formulae
    effectively takes a negation normal form of the original
    transition formula, while $\dual{\trans}_2$ takes the negation
    normal form of the negation of the original transition formula, in
    both cases exploiting the logical equivalence between formulae
    $\neg\Diamond_\alpha \varphi$ and $\Box_\alpha \neg\varphi$ for
    $\alpha \in \barNames$.
  \end{enumerate}
\end{rem}

\begin{lem}[Correctness of~\Cref{constr:neg2dual}]\label{lem:exprNegStates}
$A$ and $\dual{A}$ accept the same literal language.
\end{lem}
\begin{IEEEproof}
  We show that for every formula $\varphi \in \negbool(1 + \names \times Q + \Abstr{Q})$ and $w\in \barAs$:
\begin{equation}\label{eq:claim-ad} w \vDash \dual{\trans}_1(\varphi) \text{ in~$\dual{A}$} \quad\text{iff}\quad w \vDash \varphi \text{ in~$A$} \quad\text{iff}\quad w \not\vDash \dual{\trans}_2(\varphi) \text{  in~$\dual{A}$}.\end{equation}
Applying this to closed bar strings $w$ and $\phi=\delta(q_0)$ proves $L_0(A)=L_0(\dual{A})$. The proof of the above claim
  proceeds by a straightforward outer induction on the word length and an inner induction on the size of the formula $\phi$ in negation normal form.
  For the base case ($w = \varepsilon$), the cases $\varphi \in \set{\top, \bot, \varepsilon, \neg\varepsilon}$ are
  easily seen. For the cases $\varphi = \Diamond_\alpha q'$ and $\varphi = \neg\Diamond_\alpha q'$, it is enough to
  notice that $\varepsilon \not\vDash \Diamond_\alpha q'$, but $\varepsilon \vDash \neg\Diamond_\alpha q'$ and
  $\varepsilon \vDash \Box_\alpha q'$. The inductive cases for $\varphi = \varphi_1 \ast \varphi_2$ with $\ast \in \set{\vee,\wedge}$ are then just simply the inner induction hypothesis.
  Therefore, let $w = \alpha v$: Again, the cases $\varphi \in \set{\top, \bot, \varepsilon, \neg\varepsilon}$ are
  readily verified. Similar for $\varphi = \Diamond_a q'$ and $\varphi = \neg\Diamond_a q'$ with $a \in \names$.
  For $\varphi = \Diamond_{\scriptnew{a}} q'$ we have that $w \vDash \varphi$ holds iff there is a $v' \in \barAs$,
  $b \in \names$ and $q'' \in Q$ with $w \alphaequiv \newletter{b}v'$, $\braket{a}q' = \braket{b}q''$ and
  $v' \vDash \delta(q'')$. Per the outer induction hypothesis, the latter is equivalent to $v' \not\vDash \dual{\trans}_2(\delta(q'')) = \dual{\delta}(\negstate{q''})$. However, this holds iff $w \not\vDash \Box_{\scriptnew{a}} q'$.
  Similarly, $\varphi = \neg\Diamond_{\scriptnew{a}} q'$ is verified. Again, $\varphi = \varphi_1 \ast\varphi_2$ with
  $\ast \in \set{\vee,\wedge}$ follows directly from the inner induction hypothesis.
\end{IEEEproof}

\paragraph*{Details for~\Cref{lem:constr02}}

In this transformation, we need to eliminate all box modalities in an
explicit-dual RANA to get a positive one. The main idea behind the
construction is to store possible escape letters of a bar string, the
number of which is bounded by the degree of the RANA plus one
(\Cref{lem:boundEscapes}). Using those escape letters, we can
restrict the number of ways a box modality can be satisfied: For
instance, a bar string can satisfy $\Box_a q$ either by starting
with~$a$ and continuing with a word accepted by~$q$ (thus satisfying
$\Diamond_a q$), by being empty (thus satisfying~$\epsilon$) or by
starting with a letter~$\beta\in\barNames$ distinct from~$a$ (thus
satisfying $\Diamond_\beta q_\top$ where~$q_\top$ is totally
accepting); the bound on escape letters allows us to keep the
disjunction implicit in these considerations finite. Formal details
are as follows:

\begin{construction}[Explicit-dual RANA to positive RANA]\label{constr:dual2pos}
    Given an explicit-dual RANA $A = \maketuple{Q, \delta, q_0}$ 
    with degree $k$, we define a positive RANA $\pos{A} = \maketuple{\pos{Q},\pos{\delta},\maketuple{q_0,\emptyset}}$ as follows:
    \begin{enumerate}
    \item The orbit-finite nominal set of states is given by
      $\pos{Q} := Q \times \pown{k + 1}(\names) + \set{q_{\top}}$,
      where
      $\pown{k + 1}(\names) = \setw{S \seq \names}{\card{S} \leqslant
        k + 1}$. Here, the element $q_{\top}$ is equivariant, and the
      group action of $\pown{k + 1}(\names)$ is given by
      $\pi\cdot S = \setw{\pi\cdot s}{s\in S}$. Intuitively, an
      element $\maketuple{q, S}$ of $\pos{Q}$ denotes the state
      $q \in Q$ together with a set~$S$ of possible escape letters.
        \item \label{posconstr:impl} We define the auxiliary function
          \[\pos{\trans}_S\colon \pown{k + 1}(\names) \times 
          \dualbool(1 + \names \times Q + \Abstr{Q}) \to \posbool(1 + \names 
          \times \pos{Q} + \Abstr{\pos{Q}})\] recursively
          as follows for $S \in \pown{k + 1}$ and $a\in \At$:
\begin{align*}
\pos{\trans}(S, x) &= x \quad & \text{for  $x \in \set{\varepsilon,\top,\bot}$};\\
\pos{\trans}(S, \neg{\varepsilon}) & = \bigvee_{a \in S} \Diamond_{a} q_{\top} \vee
          \Diamond_{\scriptnew{a}} q_{\top}; & \\
\pos{\trans}(S, \Diamond_a q')
          &= \Diamond_a \maketuple{q',S}; & \\
\pos{\trans}(S, \Box_a q') &= \varepsilon \vee \Diamond_a
          \maketuple{q',S} \vee  \bigvee_{\sigma \in S \setminus \set{a}} \Diamond_\sigma q_{\top} \vee
          \Diamond_{\scriptnew{a}} q_{\top}; & \\
\pos{\trans}(S, \Diamond_{\scriptnew{a}} q') &= \bigvee_{\substack{S' \subseteq
          S \cup \set{a}\\\card{S'} \leqslant k + 1}} \Diamond_{\scriptnew{a}}\maketuple{q',S'} & \text{where $a \notin S$};\\
\pos{\trans}(S, \Box_{\scriptnew{a}} q') &= \varepsilon \vee \bigvee_{\substack{S' \subseteq S
          \cup \set{a}\\\card{S'} \leqslant k + 1}} \Diamond_{\scriptnew{a}}\maketuple{q',S'} \vee
          \bigvee_{\sigma \in S} \Diamond_\sigma q_{\top} & \text{where $a \notin S$}; \\
\pos{\trans}(S, \varphi \ast \psi)
          & = \pos{\trans}(S, \varphi) \ast \pos{\trans}(S, \psi) & \text{for $\ast \in \set{\vee, \wedge}$}.
\end{align*}
        \item The transition function $\pos{\delta}\colon \pos{Q} \to \posbool(1 + \names \times
          \pos{Q} + \Abstr{\pos{Q}})$ is given by
          \[ \pos{\delta}\maketuple{q,S} = \pos{\trans}(S, \delta(q)) \quad\text{and}\quad \pos{\delta}({q_{\top}}) = \top.\]
    \end{enumerate}
\end{construction}

\begin{rem}
  \begin{enumerate}%
    \item Given an explicit-dual RANA $A$ with degree $k$ and $n$ orbits, the positive
      RANA $\pos{A}$ has degree $2k + 1$ and at most $n \cdot (k + 2) \cdot (2k + 1)^{2k + 1}+1$ orbits.
      This follows from \Cref{R:Bell} items~\ref{R:Bell:3} and~\ref{R:Bell:4}. 
      
    \item In Step~\ref{posconstr:impl}, we replace any box modal atom in a transition
      formula with an explicit finite disjunction of diamond modal atoms as we considered
      above. In a state $\maketuple{q, S}$, the set $S$ is to be seen as all possible
      escape letters that are previously bound names and that can be used to \enquote{escape}
      from a box modality. Whenever a bar modal atom $\Diamond_{\scriptnew{a}}$ or
      $\Box_{\scriptnew{a}}$ is encountered, $\pos{A}$ may nondeterministically choose
      to add the letter $a$ to the currently stored set~$S$ of escape letters, possibly
      removing other names from~$S$ to ensure that $S$ does not exceed size $k+1$. 
      The requirement that $a \notin S$ for bar modal atoms ensures that the possibility
      of continuing with the escape letter $a$ remains. This
      nondeterminism also ensures that any set of escape letters can be reached.
  \end{enumerate}
\end{rem}

\begin{lem}\label{lem:relStatesPos}
  Given $A$ and $\pos{A}$ as in~\Cref{constr:dual2pos} and a
  bar string $w \in \barAs$, if the state $\maketuple{q,S} \in \pos{Q}$
  accepts $w$, then so does every state $\maketuple{q,S'} \in \pos{Q}$ with $S \seq S'$.
\end{lem}
\begin{IEEEproof}
  By construction of $\pos\delta$, we immediately notice that
  if $S \seq S'$ then the transition formula $\pos\delta\maketuple{q,S'}$
  contains more disjuncts than $\pos\delta\maketuple{q,S}$,
  thereby weakening the formula and allowing more bar strings to be accepted.
\end{IEEEproof}

\begin{lem}\label{lem:helperDual2Pos}
 Given $A$ and $\pos{A}$ as in~\Cref{constr:dual2pos}, $w \in \barAs$ and $q\in Q$, we have that
\[ \text{$q$ accepts~$w$ in $A$}\qquad\text{iff}\qquad
  \text{$\maketuple{q,S_{w,q}}$ accepts~$w$ in~$\pos{A}$},\]
  where $S_{w,q} \seq \supp(q) \cup \set{b_{w,q}}$ is the set of escape
  letters of $w$ for $q$.
\end{lem}
\begin{IEEEproof}
  We prove more generally for all 
  $\varphi \in \dualbool(1 + \names \times Q + \Abstr{Q})$ with
  $\card{\supp(\varphi)} \leqslant k$:
\[ \text{$w \vDash \varphi$}\qquad\text{iff}\qquad
  \text{$w \vDash \pos{\trans}({S_{w,\varphi}}, \varphi)$},\] 
where $S_{w,\varphi}$ is the set of escape letters for $w$ at $\varphi$. The statement of the lemma then follows by taking $\phi=\delta(q)$. We prove the above claim via an outer induction on~$w$ and an inner induction
  on~$\varphi$. The \enquote{if} direction is straightforward, essentially because the
  formulae $\pos{\trans}({S_{w,\varphi}}, \varphi)$ is, by construction,
  stronger than those of the corresponding $q \in Q$. Indeed, the construction
  {restricts} the implicit infinite disjunction of every
  box modality to a finite one.

  We proceed to prove \enquote{only if}. The outer induction base
  $w=\epsilon$ is straightforward. So let $w = \alpha v \in \barAs$ where
  $\alpha \in \barNames$, and let $w\vDash\varphi$. The inner induction
  steps for Boolean operators and $\epsilon$ are trivial. The remaining
  cases are as follows.

  $\varphi = \neg\varepsilon$: If $\alpha \in \names$, then $\alpha$
  is an escape letter, i.e.~$\alpha\in S_{w,\varphi}$, so $\Diamond_\alpha
  q_\top$ is a disjunct of $\pos{\trans}({S_{w,\varphi}}, \varphi)$.
  Otherwise, $\alpha$ is a bar letter, meaning $\Diamond_\alpha q_\top$
  is again a disjunct of $\pos{\trans}({S_{w,\varphi}}, \varphi)$.
  Since $w=\alpha v$, it follows that $w \vDash \pos{\trans}({S_{w,\varphi}}, \varphi)$.

  $\varphi = \Diamond_a q'$: By hypothesis and the semantics
  of~$\Diamond_a$, we have $\alpha=a$ and $v\vDash\delta(q')$,
  whence $v\vDash\pos{\trans}({S_{v,\delta(q')}}, \delta(q'))$
  by the outer induction. Since
  $\pos\delta(q',S_{v,\delta(q')})=\pos{\trans}({S_{v,\delta(q')}}, \delta(q'))$
  by definition, this means that $\maketuple{q',S_{v,\delta(q')}}$
  accepts~$v$ in~$\pos A$. Now $v\qvDash\delta(q')$ is a successor of
  $w\qvDash\varphi$ in the evaluation dag, and $\FN(v)\seq\FN(w)$,
  so we have $S_{v,\delta(q')}\seq S_{w,\varphi}$ by the
  definition of escape letters. By \Cref{lem:relStatesPos}, it
  follows that $\maketuple{q',S_{w,\varphi}}$ accepts~$v$ in~$\pos A$.
  Since $\pos\trans({S_{w,\varphi}}, \varphi)=\Diamond_a(q',S_{w,\varphi})$, this
  implies that $w\vDash\pos\trans({S_{w,\varphi}}, \varphi)$ as required.

  $\varphi = \Diamond_{\scriptnew{a}} q'$: For $w$ to satisfy
  $\varphi$, $\alpha$ needs to be equal to some $\newletter{b}$,
  $w \alphaequiv \newletter{c}v'$ for some $c \in \names$,
  $\braket{c}q'' = \braket{a}q'$ and $q''$ needs to accept
  $v'$. Additionally, by the standard freshness principle~\cite{Pitts2013},
  we assume w.l.o.g.~that $a$ is not an escape letter for $w$ at
  $\varphi$. By the outer induction hypothesis, we see that $v'$
  satisfies $\pos{\trans}({S_{v',\delta(q'')}}, \delta(q''))$.
  We continue by case distinction: If $c \notin S_{v',\delta(q'')}$,
  we have $S_{v',\delta(q'')} = S_{w,\varphi}$ because $v' \qvDash \delta(q'')$
  is a successor of $w \qvDash \varphi$ in the evaluation dag and
  $\FN(v') \setminus \set{c} = \FN(w)$. Therefore, $\maketuple{q'',S_{w,\varphi}}$ must accept $v'$.
  Since $\braket{c}q'' = \braket{a}q'$ and $c \notin S_{w,\varphi}$,
  we have $\braket{c}\maketuple{q'',S_{w,\varphi}} = \braket{a}\maketuple{q',S_{w,\varphi}}$.
  Hence, $w \vDash \pos\trans({S_{w,\varphi}}, \varphi)$ as required.
  However, if $c$ is an escape letter for $v'$ at $q''$,
  i.e.~$c \in S_{v',\delta(q'')}$, it cannot be an escape letter for $w$
  at $\varphi$ (since $c\not\in\FN(w)$). Moreover, $S_{v',\delta(q'')} = S_{w,\varphi} \cup \set{c}$,
  and given the bound on escape letters, we have that $S_{v',\delta(q'')}$
  has at most $k + 1$ elements. Therefore, $\Diamond_{\scriptnew{a}} \maketuple{q',S_{v',\delta(q'')}}$
  is a disjunct in $\pos\trans({S_{w,\varphi}}, \varphi)$, and
  $w \vDash \pos\trans({S_{w,\varphi}}, \varphi)$ as required.

  For $\varphi = \Box_a q'$, we have three possibilities for $w$ to be
  accepted: Firstly, if $\alpha = a$ and $v$ satisfies $\delta(q')$, the
  statement follows from the same argument as in the case of $\varphi = \Diamond_a q'$.
  Secondly, if $\alpha \notin \names$, we are done by construction of $\pos{\trans}({S_{w,\varphi}}, \varphi)$.
  Lastly, if $\names \ni \alpha \neq a$, then~$\alpha$ must be an escape letter
  whence an element of $S_{w,\varphi}$. Again, we are done by construction.

  For $\varphi = \Box_{\scriptnew{a}} q'$, we have two possiblities for
  $w$ to be accepted: Firstly, if $w$ is also accepted by $\varphi' =
  \Diamond_{\scriptnew{a}} q'$ we conclude using the same
  argument as in that case.
  Lastly, if $\alpha \in \names$, then $\alpha$ has to be a free
  escape letter, i.e.~is an element of $S_{w,\varphi}$, and we are done
  by construction.
\end{IEEEproof}

\begin{IEEEproof}[Proof of~\Cref{lem:constr02}]
Immediate from~\Cref{lem:helperDual2Pos}; just notice that $S_{w,q}=\emptyset$ for closed bar strings $w\in \barAs$. 
\end{IEEEproof}

\begin{IEEEproof}[Proof of~\Cref{thm:closure}]
  Closure under intersection and union follows from a standard construction for alternating automata. Given RANAs $A_i = \maketuple{Q_i, \delta_i, s_i}$
  for $i = 1, 2$ and $\ast\in\set{\vee,\wedge}$, the RANA $A_{\ast} = \maketuple{Q_1 + Q_2 + \set{s_\ast},
  \delta_1 + \delta_2 + \delta_s, s_\ast}$ 
  with $\delta_s(s_\ast) = \delta_1(s_1) \ast \delta_2(s_2)$ accepts $L_0(A_1) \ast L_0(A_2)$.

Closure under complement follows from the observation that for a RANA $A=(Q,\delta,s)$, the explicit-dual RANA $A^\complement=(\dual{Q},\dual{\delta},\negstate{s})$ accepts the literal language $L_0(A^\complement)=\bars(\emptyset)\setminus L_0(A)$, see \eqref{eq:claim-ad} in the proof of \Cref{lem:exprNegStates}. Alternatively, it is not difficult to directly construct an (ordinary) RANA with states $Q+Q$ accepting that language.
\end{IEEEproof}

\section{Details for~\Cref{sec:logic}\label{app:secLOG}}

\begin{IEEEproof}[Details for~\Cref{rem:sem-equivalence}]
  We show that for a bar formula $\varphi \in \BarForm$, $a \in \names$ and
  a bar string $w \in \barAs$,
  \[ 
    \left(\forall b, c \in \names, v, c' \in \barAs, \psi \in \BarForm.\ \left(w = \newletter{b}v \alphaequiv \newletter{c}v' \wedge \braket{a}\varphi = \braket{c}\psi\right) \implies v' \vDash \psi\right) \iff w \vDash \Box_{\scriptnew{a}} \varphi.
  \]
  \enquote{$\Rightarrow$}: Suppose that for all $b, c \in \names, v, v' \in \barAs$ and $\psi \in \BarForm$, we have
  $w = \newletter{b}v \alphaequiv \newletter{c}v'$ and $\braket{a}\varphi = \braket{c}\psi$
  implying $v' \vDash \psi$. If $w \neq \newletter{b}v$ for any $b \in \names$ and $v \in \barAs$, then the right hand
  side is trivially satisfied. Otherwise, for $c \in \names$ fresh for $a$, $b$, $\varphi$ and $v$, and
  $v' = \makecycle{b,c} \cdot v$, we have $\newletter{c}v' \alphaequiv \newletter{b}v$ and $\braket{a}\varphi =
  \braket{c}\left(\makecycle{a,c} \cdot \varphi\right)$, implying $v' \vDash \psi$. Taking the same choice of
  $c$, $v'$ and $\psi$, the right hand side is also satisfied.

  \noindent\enquote{$\Leftarrow$}: Suppose the right hand side holds. If $w \neq \newletter{b}v$ for any $b \in \names$
  and $v \in \barAs$, the left hand side is also trivially satisfied. Suppose $w = \newletter{b}v$, then
  there are some $c \in \names$, $v' \in \barAs$ and $\psi \in \BarForm$, such that $w \alphaequiv \newletter{c}v'$,
  $\braket{a}\varphi = \braket{c}\psi$ and $v' \vDash \psi$. Suppose there is some other $c \neq d \in \names$,
  $v'' \in \barAs$ and $\psi' \in \BarForm$ with $w \alphaequiv \newletter{d}v''$ and $\braket{a}\varphi =
  \braket{d}\psi'$.
  Then $c$ is fresh for $\psi'$ and $\psi = \makecycle{d,c} \cdot \psi'$ (by $\braket{d}\psi' = \braket{c}\psi$). 
  Additionally, $v' = \makecycle{d,c} \cdot v''$ and $v' \vDash \psi$ holds iff $\makecycle{d,c}\cdot v'' \vDash
  \makecycle{d,c} \cdot \psi'$. By equivariance of the satisfaction relation, this holds iff $v'' \vDash \psi'$.
\end{IEEEproof}

\begin{IEEEproof}[Proof of~\Cref{theo:corrFormAut}]
  We define (or \emph{extend $\delta_\varphi$ to}) an equivariant function $f\colon \BarForm \to \dualbool(1 +
  \names \times \BarForm + \Abstr{\BarForm})$ as follows:
  \begin{itemize}
    \item $f(\varepsilon) = \varepsilon$;
    \item $f(\neg\varepsilon) = \neg\varepsilon$;
    \item $f(\heartsuit_\sigma \varphi) = \heartsuit_\sigma \varphi$ for $\heartsuit \in \set{\Diamond, \Box}$ and
      $\sigma \in \barNames$;
    \item $f(\varphi \ast \psi) = f(\varphi) \ast f(\psi)$ for $\ast \in \set{\vee, \wedge}$; and
    \item $f(\mu X. \varphi) = f(\varphi[\nicefrac{\mu X.\varphi}{X}])$.
  \end{itemize}
  Both equivariance and restriction to $\delta_\varphi$ are easily verified. We show per induction over the triple
  $\maketuple{\card{w}, \textsf{u}(\varphi), \card{\varphi}}$, ordered by lexicographic ordering, that for any
  $\varphi \in \BarForm$ and any bar string $w \in \barAs$, we have $w \models \varphi$ iff $w \models f(\varphi)$.
  Here, $\textsf{u}(\varphi)$ denotes the number of \emph{unguarded} fixed-point operators in $\varphi$,
  i.e.~subformulae $\mu X. \psi$ of $\varphi$ that do not occur under some modality.
  The base case for $\varphi \in \set{\varepsilon, \neg\varepsilon}$ are clear.
  For $\varphi = \psi \ast \chi$ (with $\ast \in \set{\vee, \wedge})$, we have
  \[ w \models \psi \ast \chi \iff w \models \psi \text{ and/or } w \models \chi \iff w \models f(\psi)
  \text{ and/or } w \models f(\chi) \iff w \models f(\psi \ast \chi). \]
  Here, the second equivalence holds by the induction hypothesis, since the third component is now smaller than
  with $\psi \ast \chi$.
  For $\varphi = \heartsuit_\sigma \psi$ (with $\heartsuit \in \set{\Diamond, \Box}$ and
  $\sigma \in \barNames$), we notice that the semantics of $\Diamond$- and $\Box$-modalities are equivalent
  between transition and bar formulae. The statement then follows directly from the induction
  hypothesis, since the first component gets smaller when going along the modality, even though the second
  component might be larger.
  Lastly, for $\varphi = \mu X. \psi$, we see that
  \[ w \models \mu X. \psi \iff w \models \psi[\nicefrac{\mu X.\psi}{X}] \iff 
    w \models f(\psi[\nicefrac{\mu X.\psi}{X}]) \iff w \models f(\mu X. \psi), \]
  where the second equivalence holds by the induction hypothesis, since the second component gets smaller
  by assumption on the formula that all fixed-point variables are guarded.
  Since $\delta_\varphi$ is then $f$ (co-)restricted to $Q_\varphi$, i.e.~the (nominalized) closure of $\varphi$, and
  well-defined, this equivalence also holds for all formulae in $Q_\varphi$, including $\varphi$ itself.
\end{IEEEproof}

\section{Details for~\Cref{subsec:name-dropping}}\label{app:secND}

\begin{rem}\label{rem:strong}
 For every orbit-finite nominal set $X$, there
exists a strong orbit-finite nominal set $Y$ and a quotient (i.e., surjective
equivariant map) $e\colon Y\epito X$ that is
\emph{$\supp$-nondecreasing}~\cite[Cor.~B.27]{mil-urb-19}. The latter means that $\supp(e(y)) = \supp(y)$ for every
$y \in Y$.  Specifically, one may choose $Y$ to be
$\coprod_{i=1}^n \names^{\# n_i}$, where $n$ is the number of orbits of
$X$ and $n_i$ is the degree of the $i$th orbit. Strong nominal sets are \emph{projective} with respect to
  $\supp$-nondecreasing quotients~\cite[Lemma~B.28]{mil-urb-19}: Given
  a strong nominal set $Y$, an equivariant map $h\colon Y \to Z$ and a
  $\supp$-nondecreasing quotient $e\colon X \epito Z$, there exists an
  equivariant map $g\colon Y \to X$ such that $h = e \circ g$.
\end{rem}

\paragraph*{Proof of~\Cref{thm:representation}}
Given a (positive/explicit-dual) RANA $A = \maketuple{Q, \delta, q_0}$ viewed as a coalgebra
  \[
    Q \xra{\delta} \RANAFunc_x Q = \allbool_x(1 + \names \times Q + \Abstr{Q}) \qquad \text{ for } x \in \set{+,\mathsf{n},\mathsf{d}},
  \]
express the orbit-finite nominal set $Q$ of states as a $\supp$-nondecreasing quotient $e\colon S \epito Q$ of an orbit-finite strong nominal set $S$ (\Cref{rem:strong}).
Note that all type functors $\RANAFunc_x(-)$ preserve $\supp$-nondecreasing quotients, because the functors $\names \times (-)$, $\Abstr{-}$, and $\allbool_x(-)$ all do, and the preservation property is stable under coproduct.
Therefore, the right vertical arrow in the square~\eqref{eq:rana:strong:coalg} is $\supp$-nondecreasing.
By the projectivity of $S$, we obtain an equivariant map $\beta$ such that the square commutes.
  \begin{equation}
      \begin{tikzcd}
          S \ar[two heads]{d}[left, midway]{e} \ar[dashed]{rr}[above, midway]{\beta} &&
          \allbool_x(1 + \names \times S + \Abstr{S}) \ar[two heads]{d}[right, midway]{\RANAFunc_x(e)} \\
          Q \ar{rr}[below, midway]{\delta} && \allbool_x(1 + \names \times Q + \Abstr{Q})
      \end{tikzcd}\label{eq:rana:strong:coalg}
  \end{equation}
Therefore, the map $e$ is a coalgebra morphism from $\maketuple{S, \beta}$ to $\maketuple{Q, \gamma}$.
Take $q_0'$ to be any element of the preimage of $q_0$ under $e$ and define the RANA $S$ by $S = \maketuple{S, \beta, q_0'}$.
Using this, we will prove that the literal languages of $S$ and $A$ are equal.

By commutativity of~\eqref{eq:rana:strong:coalg} we see that for every $q \in S$ the transition formula $\delta(e(q))$ is the formula $\beta(q)$ in which every state $q'\in S$ is replaced by $e(q')$.
Consequently, it is easy to see that for every $w \in \barAs$ the following equivalence holds:
\[ w \vDash \beta(q) \qquad\Longleftrightarrow\qquad w \vDash \delta(e(q)) \]
Since $q_0' = e(q_0)$, the desired result follows.

\paragraph*{Details for~\Cref{constr:name-dropping}}

\begin{lem}
  The automaton $A_{\nd}$ of~\Cref{constr:name-dropping} is a positive RANA.
\end{lem}
\begin{IEEEproof}
  We only have to show that $\delta_{\nd}$ is well-defined and equivariant:

  Let $\maketuple{i, r} \in Q_{\nd}$ and $\overline{r}, \overline{r}'$ be two extensions of $r$ with
  $\maketuple{i, \overline{r}}, \maketuple{i, \overline{r}'} \in Q$. From standard facts about nominal
  sets, it follows that there is a permutation $\pi$ with $\pi\cdot \overline{r} = \overline{r}'$
  that fixes $\supp(r)$. We show that $f_{\mathsf{r}}(\delta\maketuple{i,\overline{r}}) = f_{\mathsf{r}}(\delta\maketuple{i,\overline{r}'})$.
  For plain names, we see that if $\Diamond_a \maketuple{j,\overline{s}}$ is an atom of $\delta\maketuple{i, \overline{r}}$,
  then either $a \in \supp(r)$ or the at is mapped to $\bot$ under $f_{\mathsf{r}}$.
  For $a \in \supp(r)$, we have $f_{\mathsf{r}}(\Diamond_a \maketuple{j,\overline{s}}) = f_{\mathsf{r}}(\Diamond_a \maketuple{j,\pi\cdot\overline{s}})$
  because these two have the same disjuncts $\Diamond_a\maketuple{j,s}$: we see that $\pi$ fixes $s$ since it fixes
  $\supp(s) \cup \set{a} \seq \supp(r)$, and therefore $s = \pi\cdot s$ is also extended by $\pi\cdot\overline s$, and vice versa.

  For bar names, we show $f_{\mathsf{r}}(\Diamond_{\scriptnew{a}} \maketuple{j,\overline{s}}) = f_{\mathsf{r}}(\Diamond_{\scriptnew{\pi(a)}} \maketuple{j,\pi\cdot\overline{s}})$
  by proving that every disjunct of the former is also a disjunct of the latter (the reverse holds using $\pi^{-1}$ in lieu of $\pi$).
  Let $\Diamond_{\scriptnew{a}} \maketuple{j, s}$ be a disjunct of $f_{\mathsf{r}}(\Diamond_{\scriptnew{a}} \maketuple{j,\overline{s}})$
  so that $\supp(s) \seq \supp(r) \cup \set{a}$ and $s$ is extended by $\overline{s}$. Then $\pi\cdot s$ is extended by
  $\pi\cdot\overline{s}$ and, since $\pi$ fixes $\supp(r)$, satisfies $\supp(\pi\cdot s) \seq \supp(r) \cup \set{\pi(a)}$.
  Therefore, $\Diamond_{\scriptnew{\pi(a)}} \maketuple{j, \pi\cdot s}$ is a disjunct of $f_{\mathsf{r}}(\Diamond_{\scriptnew{\pi(a)}}
  \maketuple{j,\pi\cdot\overline{s}})$.
  From the conditions on the support of $s$ and $\pi\cdot s$, we see that $\braket{a}\maketuple{j, s} = \braket{\pi(a)}\maketuple{j, \pi\cdot s}$.
  Therefore, $\delta_{\nd}$ is well-defined.

  Now we verify that $\delta_{\nd}$ is equivariant. Let $\pi \in \Perm(\names)$ be a permutation.
  We show that if $\Diamond_\alpha \maketuple{j,s}$ is an atom of $\delta_{\nd}\maketuple{i,r}$, then
  $\Diamond_{\pi\alpha} \maketuple{j,\pi\cdot s}$ is an atom of $\delta_{\nd}\maketuple{i,\pi\cdot r}$.
  By construction of $\delta_{\nd}$, we know that $\Diamond_\alpha \maketuple{j,\overline{s}}$ is an atom of
  $\delta\maketuple{i, \overline{r}}$, where $\overline{s}$ and $\overline{r}$ extend $s$ and $r$, respectively.
  Hence, by the equivariance of $\delta$, we see that $\Diamond_{\pi\alpha} \maketuple{j, \pi\cdot\overline{s}}$ is an atom of $\delta\maketuple{i, \pi\cdot\overline{r}}$, where $\pi\cdot\overline{r}$ and $\pi\cdot\overline{s}$ extend $\pi\cdot r$
  and $\pi\cdot s$, respectively. We only have to show that the required conditions on the support of $\pi\cdot s$ hold.
  For $\alpha \in \names$, we see that $\supp(\pi\cdot s) \cup \set{\pi\cdot \alpha} = \pi \cdot (\supp(s) \cup \set{\alpha})
  \seq \pi \cdot \supp(r) = \supp(\pi\cdot r)$. For $\alpha=\newletter a \notin \names$, we have $\supp(\pi\cdot s) = \pi \cdot \supp(s) \seq
  \pi \cdot (\supp(r) \cup \set{a}) = \supp(\pi\cdot r) \cup \set{\pi(a)}$.
\end{IEEEproof}

\paragraph*{Proof of~\Cref{thm:relAAnd}}
  \enquote{$\seq$}: Since $Q \seq Q_{\nd}$, the initial states of $A$ and $A_{\nd}$ coincide,
    and for every $q \in Q$ the transition formula $\delta(q)$ is weaker than $\delta_{\nd}(q)$,
    we see that $L_0(A) \seq L_0(A_{\nd})$.

  \enquote{$\qes$}: We prove a slightly stronger statement: for every $\maketuple{i, r}
    \in Q_{\nd}$, every  $w \in \barAs$ and every $\phi\in \allbool_+(1 + \names \times Q + \Abstr{Q})$, we have that
\[ w \vDash f_{\mathsf{r}}(\varphi) \text{ in $A_{\nd}$} \qquad\text{implies}\qquad w \vDash \varphi \text{ in $A$}.\]
Applying this to $\phi=\delta(q_0)$ yields the statement of the theorem.

 We show the above claim via an outer induction on the word length and an inner induction on the formula.
    For $w = \varepsilon$, the statement is obvious, since both the structure as well as the types of atoms are preserved by $f_{\mathsf{r}}$.
    Indeed, $f_{\mathsf{r}}$ only maps single modal atoms $\Diamond_\alpha q$ to  disjunctions of modal atoms $\Diamond_\alpha q'$, both of which are never satisfied by the empty word.
    Now let $w = \alpha v \in \barAs$ and $\maketuple{i, r} \in Q_{\nd}$.
    The only cases of interest are the modal atoms.

    For $\varphi = \Diamond_a \maketuple{i, \overline{s}}$, we have
    $\alpha v \vDash \Diamond_a \maketuple{i, s}$ for some $s$ which is extended by $\overline{s}$ and
    satisfies the support condition $\supp(s) \cup \set{a} \seq \supp(r)$. Expanding the definition of
    $\vDash$, this is equivalent to $\alpha = a$ and $v \vDash \delta_{\nd}\maketuple{i, s} =
    f_s(\delta\maketuple{i, \overline{s}})$. By the outer induction hypothesis, the latter is equivalent to
    $v \vDash \delta\maketuple{i, \overline{s}}$, which implies $w \vDash \varphi$.

    For $\varphi = \Diamond_{\scriptnew{a}} \maketuple{i, \overline{s}}$, we have
    $\alpha v \vDash \Diamond_{\scriptnew{a}} \maketuple{i, s}$ for some $s$ which is extended by $\overline{s}$
    and satisfies the support condition $\supp(s) \seq \supp(r) \cup \set{a}$. By the definition of $\vDash$,
    this is equivalent to $\alpha v \alphaequiv \newletter{b}v'$, $\braket{a}\maketuple{i, s} = \braket{b}%
    \maketuple{i,s'}$, and $v' \vDash \delta_{\nd}\maketuple{i,s'} = f_{s'}(\delta\maketuple{i,\overline{s}'})$.
    Because of standard freshness principles~\cite{Pitts2013}, we can assume that $b$ is not in the support of $\overline{s}$. 
    Moreover, since $\makecycle{a,b} \cdot s = s'$ holds by the equality of equivalence classes above, we can put $\overline{s}' = \makecycle{a,b} \cdot \overline{s}$. 
    Therefore, $\braket{a}\maketuple{i,\overline{s}} = \braket{b}\maketuple{i,\overline{s}'}$ holds and, by the outer induction hypothesis, $v' \vDash \delta\maketuple{i,\overline{s}'}$ does as well. 
    This yields $w \vDash \varphi$.

\paragraph*{Proof of~\Cref{lem:restrRun}}

  We prove this via induction on the word length: The base case ($w=\varepsilon$) is clear by the construction of $\delta_{\nd}$.
  So let $w = \alpha v$ and $\maketuple{i, r} \in Q_{\nd}$ such that $w \vDash \delta_{\nd}\maketuple{i, r}$.
  Additionally, let~$r'$ be a partial injective map that is extended by $r$ and satisfies $\supp(r) \cap \FN(w) \seq \supp(r')$.
  We prove the statement via induction on $\varphi$. The only interesting cases are the modal atoms.

  For $\Diamond_a \maketuple{j, \overline{s}}$,
  we only need to look at the case where $\alpha = a$, since $w$ would not satisfy the formula otherwise. 
  Then, $w \vDash \bigvee_s \Diamond_a \maketuple{j, s}$ where the disjunction ranges over all $s$ which are extended
  by $\overline{s}$ and satisfy the support condition $\supp(s) \cup \set{a} \seq \supp(r)$.
  Let $\Diamond_a \maketuple{j, s}$ be a satisfied disjunct. Then $v \vDash \delta_{\nd}\maketuple{j, s}=f_{s}(\delta(j,\ol{s}))$.
  Let $s'$ be the restriction of $s$ to $\supp(s) \cap \FN(v)$. Then by the induction hypothesis $v \vDash \delta_{\nd}\maketuple{j, s'}=f_{s'}(\delta(j,\ol{s}))$. We have
  We have $\supp(s') \cup \set{a} = (\supp(s)\cap \FN(v))\cup  \set{a} = (\supp(s)\cup \set{a})\cap (\FN(v)\cup \{a\})\seq \supp(r)\cap \FN(w)\seq \supp(r')$. Thus, $\Diamond_a \maketuple{j, s'}$ is a disjunct in $f_{r'}(\varphi)$ and satisfied by $w$.
  
  For bar modal atoms $\Diamond_{\scriptnew{a}} \maketuple{j, \overline{s}}$, we have $w \vDash \bigvee_s
  \Diamond_{\scriptnew{a}} \maketuple{j, s}$ where the disjunction ranges over all $s$ which are extended by $\overline{s}$
  and satisfy the support condition $\supp(s) \seq \supp(r) \cup \set{a}$. Let $\Diamond_{\scriptnew{a}} \maketuple{j, s}$
  be a satisfied disjunct.
  Then $\alpha v \alphaequiv \newletter{b} v'$, and we have some $t$ such that $\braket{a}\maketuple{j, s} =
  \braket{b}\maketuple{j, t}$ and $v' \vDash \delta_{\nd}\maketuple{j, t}$.
  Let $t'$ be the restriction of $t$ to $\supp(t) \cap \FN(v')$.
  By the induction hypothesis, we have $v' \vDash \delta_{\nd}\maketuple{j, t'}$.
  Define $s' := \makecycle{a,b} \cdot t'$. Because of $\braket{a}\maketuple{j, s} = \braket{b}\maketuple{j, t}$,
  we follow that $a \notin \supp(t')$, hence $\braket{a}\maketuple{j, s'} = \braket{b}\maketuple{j, t'}$.
  We see that $s'$ is extended by $s$ and $\overline{s}$, and $\supp(s') = \supp(s) \cap \FN(v)
 \seq (\supp(r)\cup \set{a})\cap (\FN(w)\cup \set{a}) = (\supp(r)\cap \FN(w))\cup \set{a} \seq \supp(r') \cup \set{a}$. Thus, $\Diamond_{\scriptnew{a}} \maketuple{j, s'}$ is a
  disjunct in $f_{r'}(\varphi)$ and satisfied by $w$.

\paragraph*{Proof of~\Cref{lem:ndrsem}}

  We proceed by induction on the word length $w$: The base case ($w = \varepsilon$) is trivial
  since neither the satisfaction of the classical Boolean connectives nor the satisfaction by the
  empty word has changed.

  Let $w = \alpha v$ with $\alpha \in \barNames$ and $v \in \barAs$. For $\alpha \in \names$, the statement follows
  from the induction hypothesis and the fact that the semantics of the $\Diamond_a q'$'s is identical for
  both $\vDash$ and $\rsem$.

  For $\alpha = \newletter{a} \notin\names$,
  the \enquote{if}-direction is easy to see, since $\rsem$ is a restriction of $\vDash$.
  For the \enquote{only if}-direction, let $\maketuple{i, r} \in Q_{\nd}$ be a state in $Q_{\nd}$ and let $\maketuple{i, \overline{r}}
  \in Q$ where $\overline{r}$ extends~$r$. We proceed by induction over $\delta\maketuple{i, \overline{r}}$ and notice that the
  only case of interest is $\Diamond_{\scriptnew{b}} \maketuple{j, \overline{s}}$.
  Suppose 
  \begin{equation}\label{eq:prf:lem:ndrsem}
    \newletter{a} v
    \vDash
    \textstyle\bigvee_s \Diamond_{\scriptnew{b}} \maketuple{j, s},
  \end{equation}
  where $s$ ranges over all partial injective functions extended by $\overline{s}$ and having $\supp(s) \seq \supp(r) \cup \set{b}$.
  Let $\Diamond_{\scriptnew{b}} \maketuple{j, s}$ be any satisfied disjunct.
  Then we have to show that there is a disjunct $\Diamond_{\scriptnew{b}} \maketuple{j, s'}$ that $\alpha v$ satisfies restrictively, i.e.~$\alpha v \rsem \Diamond_{\scriptnew{b}} \maketuple{j, s'}$.
  If $\newletter{a} v \not\rsem \Diamond_{\scriptnew{b}} \maketuple{j, s}$, then $a \neq b$ and
  $a \in \supp(s)$, since otherwise the renaming of the modal atom would not be blocked.
  Since $\newletter{a} v \vDash \Diamond_{\scriptnew{b}} \maketuple{j, s}$, we find an $\alpha$-equivalent
  $\newletter{c}v' \alphaequiv \newletter{a} v$, such that $\braket{b}\maketuple{j, s} = \braket{c}\maketuple{j, t}$
  and $v' \vDash \delta_{\nd}\maketuple{j, t}$. Therefore, $a \notin \FN(v')$. Let $t'$ be the restriction of $t$ to $\supp(t) \cap \FN(v')$.
  By~\Cref{lem:restrRun}, we have $v' \vDash \delta_{\nd}\maketuple{j, t'}$.
  Taking $\braket{b}\maketuple{j, s'} = \braket{c}\maketuple{j, t'}$, we see $\newletter{a} v \vDash \Diamond_{\scriptnew{b}}
  \maketuple{j, s'}$ where $v \vDash \delta_{\nd}\maketuple{j, \makecycle{b, a}\cdot s'}$. By induction hypothesis, we have $v \rsem \delta_{\nd}\maketuple{j,
  \makecycle{b, a}\cdot s'}$, which yields $\newletter{a} v \rsem \Diamond_{\scriptnew{b}}\maketuple{j, s'}$.
  This is indeed a disjunct in~\eqref{eq:prf:lem:ndrsem}, since $s'$ is extended by both $s$ and $\overline{s}$ and satisfies the
  support condition $\supp(s') \seq \supp(s) \seq \supp(r) \cup \set{b}$.

\paragraph*{Restriction of Names}

\begin{lem}\label{lem:restrNames}
  Let $A_{\nd} = \maketuple{Q_{\nd}, \delta_{\nd}, q_0}$ as in \Cref{constr:name-dropping}. Fix a finite set $\names_0\seq \names$ of at least cardinality $\deg(A_{\nd})$ and an additional name
  $\ast \in \names\setminus\names_0$, and put  $\barNames_0 =
  \names_0 \cup \setw{\newletter{a}}{a \in \names_0}$. Let $q \in Q_{\nd}$ with support $\supp(q) \seq
  \names_0$ such that $\delta_{\nd}(q)\neq\top$. Then every bar string $w \in \barAs$ accepted
  by $q$ has a prefix $v \in \barAs$
  such that there is an $\alpha$-equivalent $v' \alphaequiv v$ with $v' \in \left(\barNames_0
  \cup \set{\newletter{\ast}}\right)^*$ which is also accepted by $q$.
\end{lem}
\begin{IEEEproof}
Suppose that $q$ accepts $w$, and let $\runtree{w}{A_{\nd}}{q}$ be an accepting run dag (\Cref{app:secDA}). 
Let $v$ be the prefix of $w$ of length $h$, where $h$ is the length of the longest path from the root. Then $v$ is accepted by $q$ via the same run dag. By the RANA support principle (\Cref{lem:suppRANA}), every name that occurs freely in a suffix of $v$ must be contained in the support of the state from which this suffix is accepted; in particular, every suffix of $v$ contains at most $\deg(A_{\nd})$ free names. This implies that $v$ can be $\alpha$-renamed to bar string $v'\in \left(\barNames_0
  \cup \set{\newletter{\ast}}\right)^*$.
\end{IEEEproof}

\begin{IEEEproof}[Proof of~\Cref{thm:bafaEquiv}]
The case for $\delta_{\nd}(q_0) \equiv \top$ is easily seen since every closed bar string
starts with a bar name and $\barNames \times \set{\top}$ is part of the pre-langauge
of $A_0$. In the case that $\delta_{\nd}(q_0) \not\equiv \top$, we show the desired
equality by proving for all $q \in Q_0$ and $w \in \barAs$ that $w \rsem \delta_{\nd}(q)$
iff there is an accepting run dag for $w$ from $q$. The statement is obvious for the
empty word ($w = \varepsilon$) by the definition of final states in $A_0$. So suppose
$w = \alpha v \in \barAs$. We show both implications: Let $w \rsem \delta_{\nd}(q)$,
and $\varphi_q$ denote the reduced DNF of $\delta_{\nd}(q)$, and $\bigwedge \Diamond_{\alpha}q'$
be a satisfied disjunct with all $q' \in Q_0$, i.e.~$\alpha v \rsem \bigwedge \Diamond_{\alpha}q'$.
Such a disjunct exists by~\Cref{lem:restrNames}, the definition of $A_{\nd}$ and the
support principle~(\Cref{lem:suppRANA}). Furthermore, let $S$ be the set of all those
$\Diamond_{\alpha}q'$'s together with all $\Diamond_{\alpha}q''$'s in $\varphi_q$ where
$\delta_{\nd}(q'') \equiv \top$ and $S'$ the set of all those $q'$'s. We denote by $\restr{\psi}{S}$
for $\psi \in \posbool(1 + \names \times Q_{\nd} + \Abstr{Q_{\nd}})$ the positive formula,
where all atoms $x \in 1 + \names \times Q_{\nd} + \Abstr{Q_{\nd}}$ are replaced by
$\top$ if $x \in S$ and $\bot$ otherwise. Similar for $\psi \in \posbool(Q_0)$ we denote
by $\restr{\psi}{S'}$ the positive formula, where all atoms $q'$ are replaced by $\top$
if $q' \in S'$ and $\bot$ otherwise. We show equality of $\restr{\delta_{\nd}(q)}{S}$
and $\restr{f(\alpha, \delta_{\nd}(q))}{S'}$ if $S' \neq \emptyset$ via induction on the
structure of $\delta_{\nd}(q)$: The only case of interest is $\varphi = \Diamond_{\beta} q''$:
The equality is easily seen for $\beta \neq \alpha$. For $\beta = \alpha$, if
$\delta_{\nd}(q'') \equiv \top$, both $\restr{\varphi}{S}$ and $\restr{f(\alpha, \varphi)}{S'}$
are $\top$. If $\delta_{\nd}(q'') \not\equiv \top$, we have $\varphi \in S$ iff
$q'' \in S'$, wherefore the equality is shown.
We see that $S' = \emptyset$ if $\delta_0(q, \alpha) = \top$. In this case, an accepting run
dag for $w$ from $q$ can easily be built. Otherwise, $S'$ is a satisfying set for $\delta_0(q, \alpha)$.
For all $\Diamond_\alpha q' \in S$ we have $v \rsem \delta_{\nd}(q')$, whence, accepting run dags
for $v$ from all $q'$'s.
Using the satisfying set $S'$ together with those accepting run dags, we can build a run dag for
$w$ from $q$.
For the other direction, if there is an accepting run dag $\runtree{q}{}{w}$ for $w$ from $q$, the
children $\maketuple{v, q'}$ of its root $\maketuple{\alpha v, q}$ can be seen as accepting run dags for $v$
from the $q'$'s. By induction hypothesis, we have $v \rsem \delta_{\nd}(q')$ and $w \rsem \Diamond_{\alpha}q'$.
Because of how $\delta_0(q, \alpha)$ is built and the $q'$'s make up a satisfying set for $\delta_0(q, \alpha)$,
we easily conclude $w \rsem \delta_{\nd}(q)$. Lastly, the statement follows because $q_0 \in Q_0$.
\end{IEEEproof}

\paragraph*{Details for~\Cref{rem:bafa2rana}} We construct a name-dropping RANA from a bar AFA:

\begin{construction}\label{constr:bafa2rana}
  Given a bar AFA $A = \maketuple{Q, \barNames_0, \delta, q_0, F}$, we define the following nominal sets and equivariant functions:
  \begin{enumerate}
    \item Let $\names_0 := \setw{a_i \in \barNames_0}{\newletter{a_i} \in \barNames_0}$, and $\mathbf{n}
      := \card{\names_0}$ be its cardinality.
    \item Define the nominal set $Q_{R} := \coprod_{q \in Q} \parnom{n} + \set{q_\top}$, where
      $q_{\top}$ is equivariant. Let $q_R := \maketuple{q_0, r_\emptyset} \in Q_R$ be an equivariant state
      where $r_\emptyset \in \parnom{n}$ is the unique partial map that is nowhere defined
    \item Define the transition function $\delta_R\colon Q_R \to \posbool(1 + \names \times Q_R + \Abstr{Q_R})$
      as follows: For ${q_{\top}}$, let $\delta_{R}({q_\top}) = \top$, while for
      $\maketuple{q,r} \in Q_R$, the transition formula $\delta_R\maketuple{q,r}$ is defined using the following disjuncts:
      \begin{itemize}
        \item If $q \in F$, the formula has an $\varepsilon$-disjunct.
        \item For $\alpha = a_i \in \names_0$ and $i \in \dom(r)$, if $\delta(q,\alpha) = \top$ we add $\Diamond_{r(i)}
          q_{\top}$ as a disjunct. Otherwise, let $\varphi_{i}$ describe the formula constructed of $\delta(q, r(i))$
          where all atoms $q'$ are replaced by $\bigvee_j \Diamond_{r(i)} \maketuple{q', r_j}$ for all possible
          $r_j \in \parnom{n}$ extended by $r$. Add $\varphi_{i}$ as a disjunct to $\delta_R\maketuple{q,r}$.
        \item For $\alpha = \newletter{a_i} \in \barNames_0$ with $a_i \in \names_0$, if $\delta(q, \alpha) = \top$ we
          add $\Diamond_{\alpha} q_{\top}$ as a disjunct. Otherwise, denote by $\varphi_{i}$ the formula constructed of 
          $\delta(q, r(i))$ where all atoms $q'$ are replaced by $\bigvee_j \Diamond_{\scriptnew{a_i}}
          \maketuple{q', r_j}$ for all possible $r_j \in \parnom{n}$ such that for every $k \in \dom(r_j) \setminus
          \set{i}$ we have $r_j(k) = r(k)$, and if $r_j$ is defined at position $i$, it must be equal to $a_i$.
          Add $\varphi_{i}$ as a disjunct to $\delta_R\maketuple{q,r}$.
        \item For $\alpha = \newletter{a} \in \barNames_0$ with $a \notin \names_0$, if $\delta(q, \alpha) = \top$ we
          add $\Diamond_{\alpha} q_{\top}$ as a disjunct. Otherwise, denote by $\varphi_{\alpha}$ the formula
          constructed of $\delta(q, \alpha)$ where all atoms $q'$ are replaced by $\bigvee_j \Diamond_{\alpha}
          \maketuple{q', r_j}$ for all possible $r_j \in \parnom{n}$ such that $r_j(k) = r(k)$ or $r_j(k) = a$ for
          $k \in \dom(r_j)$. Add $\varphi_\alpha$ as a disjunct to $\delta_R\maketuple{q, r}$.
      \end{itemize}
  \end{enumerate}
  With this, define the positive name-dropping RANA $A_{R} = \maketuple{Q_{R}, \delta_{R}, q_R}$.
\end{construction}

\begin{theo}\label{thm:corrConstrB2R}
  Given a bar AFA $A = \maketuple{Q, \names_0, \delta, q_0, F}$, the positive name-dropping RANA~$A_{R}$ of~\Cref{constr:bafa2rana} accepts the same bar language as $A$.
\end{theo}
\begin{IEEEproof}
  To show that $L_0(A) \seq L_0(A_R)$, we show that if a bar string $w$ is accepted from a state
  $q \in Q$ via an accepting run dag $\runtree{q}{}{w}$ then it also is accepted by $\maketuple{q, r_w}
  \in Q_R$, where $r_w \in \parnom{n}$ is the partial injective map that has $r_w(i) = a_i$ if
  $a_i \in \FN(w)$. We prove this via induction on the word length:
  The statement is clear for the empty word, so suppose $w = \alpha v$ with $\alpha \in \barNames_0$
  and $v \in \barAs$. We see that if $\alpha \in \names$, then $\alpha \in \supp(r_w)$ holds by design.
  We prove the following two cases: Firstly, let the root of the accepting run dag
  $\runtree{q}{}{w}$ have just one child $\maketuple{v, \top}$. By construction of accepting run dags,
  $\delta(q, \alpha) = \top$, so $\Diamond_{\alpha} q_{\top}$ is a disjunct in $\delta_R\maketuple{q, r_w}$.
  Therefore, $w \vDash \delta_R\maketuple{q, r_w}$.
  Secondly, let $X$ denote the set of all states of the root's children $\maketuple{v, q'}$
  in the accepting run dag $\runtree{q}{}{w}$. Then all $q' \in X$ accept $v$, hence, by induction hypothesis
  we have $v \rsem \delta_R\maketuple{q',r_v}$. The correctness follows immediately from the
  construction of $\delta_R$ together with the fact that $X$ is a satisfying set of $\delta(q, \alpha)$,
  and that $\Diamond_\alpha\maketuple{q', r_v}$ is an atom at the correct position in $\delta_R$.
  Indeed, if $\alpha \in \names$, then $\FN(v) \seq \FN(w)$. If $\alpha \notin \names$, then $\FN(v)
  \seq \FN(w) \cup \set{\ub(\alpha)}$. Hence, the atom exists in $\delta_R$.

  For the other direction, let $w \in L_0(A_R)$, then by~\Cref{lem:restrNames} there exists an
  $\alpha$-equivalent $w' \alphaequiv w$, which is also accepted by $A_R$, but only uses letters over any finite
  alphabet $\overline{\Sigma}$ which is built from a finite set of names $\Sigma \seq
  \names$ of at most cardinality $\deg(A_R) := \mathbf{n}$ and one additional bar name $\newletter{\ast}$.
  We see immediately, that $\Sigma$ may be chosen such that $\overline{\Sigma} \setminus \set{\newletter{\ast}}
  \seq \barNames_0$. Additionally, we know that $\barNames_0 \setminus \overline{\Sigma}$ consists only of
  bar names. There are two possibilities here: If $\overline{\Sigma} \setminus \set{\newletter{\ast}} \neq
  \barNames_0$, $\newletter{\ast}$ can be chosen to be an element of $\barNames_0$. If the inequality
  does not hold, any accepted bar string by $A_R$ need not make use of this additional element.
  This is because of how $A_R$ is constructed, since every bar modal atom $\Diamond_{\scriptnew{a}} \maketuple{q',r'}$
  for any state $\maketuple{q, r}$ in $A_R$ cannot have $r = r'$ (only the second bullet point applies).
  Therefore, it is enough to prove that for every $w \in (\overline{\Sigma} \cap \barNames_0)^*$ we have that
  if a state $\maketuple{q, r} \in Q_R$ accepts $w$, so does $q$ in $A$. We continue with induction over
  the word length: For the base case ($w = \varepsilon$), the statement holds by construction. Take
  $w = \alpha v$ with $\alpha \in \overline{\Sigma} \cap \barNames_0$ and $v \in \barAs$: We look at the
  disjunct in $\delta_R\maketuple{q, r}$ which arises for $\alpha$ from $q$ in $A$: If the disjunct is equal
  to $\Diamond_\alpha q_{\top}$, the statement is obvious, since $\delta(q, \alpha) = \top$. Otherwise,
  we easily get a satisfying set of states for $\delta(q, \alpha)$ by construction of the disjunct. 
  Since $\varphi_{\alpha}$ is satisfied by $w$, we have a set of modal atoms $\Diamond_{\alpha}
  \maketuple{q', r'}$ satisfied by $w$. By induction hypothesis, this means that all those
  $q'$'s accept $v$ in $A$. Lastly, because of how we constructed $\delta_{R}$, we see that
  those $q'$ build a satisfying set of $\delta(q, \alpha)$, thereby allowing us to have an
  accepting run dag for $w$ from $q$. Therefore, $A$ accepts $w'$, and by closure under
  $\alpha$-equivalence also $w$.
\end{IEEEproof}

\section{Details for~\Cref{sec:dealt}} \label{app:secDA}

\begin{IEEEproof}[Proof of \Cref{lem:rest-termination}]
Let $p$ be the maximum cardinality of any $S\in \Phi$, and let $A=\bigcup_{S\in \Phi}\bigcup_{q\in S} \supp(q)$. The restriction process only generates subsets of $Q_{\nd}$ of cardinality at most $p$ all of whose elements are supported by $A$. Since the nominal set $Q_{\nd}$ is orbit-finite, it contains only finitely many elements with support $A$, hence there are only finitely many such subsets.  
\end{IEEEproof}

\paragraph*{Proof of~\Cref{lem:correst}} We first introduce run dags to capture acceptance for RANAs:

\begin{defn}[Run DAG]
  Given the name-dropping modification $A_{\nd} = \maketuple{Q_{\nd}, \delta_{\nd}, q_0}$ of a RANA,
  a state $q \in Q_{\nd}$, and a bar string $w = \alpha_1 \cdots \alpha_n \in \barAs$, a
  \emph{run dag} of $w$ from $q$ in $A_{\nd}$ is a (rooted) directed acyclic graph $\runtree{w}{A_{\nd}}{q}$ with nodes  $\barNames^{\leqslant n} \times Q_{\nd}$ and the following properties,
  defined recursively on the length of $w$:
 
The root is the node
  $\maketuple{w, q}$. If $w = \varepsilon$
  or $\delta_{\nd}(q)$ has no $\alpha_1$-disjunct $\bigwedge_{j} \Diamond_{\alpha_1} q_j$,
  then $(w,q)$ has no outgoing edges.
  Otherwise, there exists an
  $\alpha_1$-disjunct $\bigwedge_{j} \Diamond_{\alpha_1} q_j$ and edges from $(w,q)$ to the roots of run dags $\runtree{w'}{A_{\nd}}{q_j}$ of $w' = \alpha_2 \cdots \alpha_n$ from $q_j$ in $A_{\nd}$.%

A run dag $\runtree{w}{A_{\nd}}{q}$ is \emph{accepting} whenever the following holds: If the root $\maketuple{w, q}$ has no children, then $\delta_{\nd}(q)=\top$ or
  $w = \varepsilon$ and $\delta_{\nd}(q)$ has an $\varepsilon$-disjunct. If the root has children, then all children are accepting run trees.
\end{defn}

\begin{lem}
  Given a state $q$ of $A_{\nd}$ and a bar string $w \in \barAs$, $q$ accepts $w$ restrictively, i.e.~$w \rsem \delta_{\nd}(q)$, iff there exists an accepting run dag  $\runtree{w}{A_{\nd}}{q}$ of $w$ from $q$ in $A_{\nd}$.
\end{lem}
\begin{IEEEproof}
  We prove this lemma via induction on the word length: For the base case ($w = \varepsilon$), we see
  that $\varepsilon \rsem \delta_{\nd}(q)$ iff either $\delta_{\nd}(q)$ is $\top$ or has an
  $\varepsilon$-disjunct, which is equivalent to having an accepting run dag $\runtree{\varepsilon}{A_{\nd}}{q}$.
  Suppose then that the statement holds for words of length $n$, and
  let $w = \alpha v \in \barAs$ with $v \in \barAs$ of length $n$. By the definition of the restricted
  semantics (\Cref{rem:implrenW}), we see that $w \rsem \delta_{\nd}(q)$ holds iff either $\delta_{\nd}(q)$
  is $\top$ or has an $\alpha$-disjunct $\bigwedge_{j} \Diamond_\alpha q_j$ where each $q_j$ accepts $v$. In
  the first case we are done, since the root of any run dag $\runtree{w}{A_{\nd}}{q}$ has no children and
  is accepting by definition, so suppose such an $\alpha$-disjunct exists. By the induction hypothesis,
  the acceptance of $v$ by all $q_j$'s is equivalent to having accepting run dags $\runtree{v}{A_{\nd}}{q_j}$
  for each $q_j$. This allows us to build a run dag $\runtree{w}{A_{\nd}}{q}$ which has $\maketuple{w, q}$ as
  its root and the $\runtree{v}{A_{\nd}}{q_j}$ as children, making it accepting.
\end{IEEEproof}

\begin{lem}[Accepting run DAGs]\label{lem:accruntree}%
  A run dag $\runtree{w}{A_{\nd}}{q}$ is accepting iff all leaves $\maketuple{w',q'}$
  either have $w' = \varepsilon$ and an $\varepsilon$-disjunct in $\delta_{\nd}(q')$,
  or have $\delta_{\nd}(q')$ be equal to $\top$.
\end{lem}
\begin{IEEEproof}
  This is trivially shown via structural induction on the run dags: For the base case,
  if the run dag $\runtree{w}{A_{\nd}}{q}$ consists of just one root node $\maketuple{w, q}$
  with no children, then the equivalence follows by the definition of acceptance there.
  Now suppose the statement holds for run dags less than a specific height $k$ and
  let $\runtree{w}{A_{\nd}}{q}$ be a run dag of height $k + 1$, then the run dag is
  accepting iff all its children are which by the induction hypothesis lets $\runtree{w}{A_{\nd}}{q}$
  fulfill the equivalence.
\end{IEEEproof}

\begin{IEEEproof}[Proof of~\Cref{lem:correst}]
  We only need to show that whenever $q$ and $q'$ accept a bar string then so
  do either $\restr{q}{J}$ and $q'$ or $q$ and $\restr{q'}{J}$ for $J = \supp(q) \cap \supp(q')$
  and vice versa. The statement then follows from the definition of $\rest(\{S\})$
  and an iterative argument. So suppose that both $q$ and $q'$ accept $w \in \barAs$ and let $\runtree{w}{A_{\nd}}{q}$ and $\runtree{w}{A_{\nd}}{q'}$ accepting run dags. By symmetry, we may assume that $\mathrm{height}(\runtree{w}{A_{\nd}}{q})\leqslant \mathrm{height}(\runtree{w}{A_{\nd}}{q'})$, where $\mathrm{height}(-)$ denotes the height (i.e.~length of a longest path from the root) of a dag. Then every free name of $w$ processed by $\runtree{w}{A_{\nd}}{q}$ is also processed by $\runtree{w}{A_{\nd}}{q'}$. By repeated application of the RANA support principle (\Cref{lem:suppRANA}), every such name is contained in both $\supp(q)$ and $\supp(q')$, hence in $J=\supp(q)\cap \supp(q')$. Therefore, the accepting run dag $\runtree{w}{A_{\nd}}{q}$ can be transformed into an accepting run dag of $w$ from $\restr{q}{J}$ by replacing the root $(w,q)$ by $(w,\restr{q}{J})$ and propagating the restrictions along the edges. This proves that $\restr{q}{J}$ accepts $w$. 
\end{IEEEproof}

\begin{IEEEproof}[Proof of~\Cref{L:Bell}]
Use \Cref{R:Bell}.\ref{R:Bell:5} and that $m^m = 2^{m \cdot \log m}$.
\end{IEEEproof}

\begin{IEEEproof}[Proof of~\Cref{thm:corrdealt}]
We show equality of the literal languages by the stronger statement that a state
$S = \set{q_1, \dots, q_j} \in Q^E$ (where $j \leqslant n \cdot k!$) accepts a bar string $w \in \barAs$
iff all $q_i$ with $1 \leqslant i \leqslant j$ do in $A_{\nd}$. Equality then follows by taking $S=\{q_0\}$.

We proceed by induction on the word length. In the base case
($w = \varepsilon$), we have $\varepsilon \rsem \delta_E(S)$ iff $\delta_E(S)=\top$ or $\delta_E(S)$ contains an $\varepsilon$-disjunct. By
construction, this occurs iff $S = \emptyset$ or all $q \in S$ have an $\varepsilon$-disjunct, i.e.~accept
$\varepsilon$. For the induction step, assume that $w = \alpha v \in \barAs$. If $S = \emptyset$, then the equivalence holds trivially,
so assume $S\neq \emptyset$.
If $w \rsem \delta_E(S)$ then there is an $\alpha$-disjunct $\Diamond_{\alpha} S'$
in $\delta_E(S)$ and $S'$ accepts $v$. Additionally, $S'$ is a subset of some $S'' \in \rest(\Delta_{\alpha}^S)$
where the difference are states $q'$ with $\delta_{\nd}(q') \equiv \top$. Per induction hypothesis, all $q \in S'$
accept $v$, which implies (\Cref{lem:correst}) that all $q' \in \Delta_{\alpha}^S$ accept $v$.
By construction of $\Delta_{\alpha}^S$, this is equivalent to the fact that all $q \in S$ accept $w$. The proof of the converse is analogous.    
\end{IEEEproof}

\section{Details for \Cref{sec:decincl}}

\paragraph*{Details for~\Cref{rem:compConstr}}

Recall from \Cref{lem:constr01} that for a RANA $A$ that has degree $k$ and $n$ orbits, its equivalent explicit-dual RANA $\dual{A}$ has degree $k$ and $2n$ orbits.
Using the bounds in \Cref{lem:constr02}, the equivalent positive RANA $(\dual{A})^+$ then has degree $2k+1$ and at most $2 \cdot n\cdot (k+2) \cdot (2k+1)^{2k+1} + 1$ orbits.
Finally, the bound in \Cref{constr:name-dropping} yields that the name-dropping modification $(\dual{A})^+_{\nd}$ has degree $2k+1$, and its number of orbits is at most:
\begin{align*}
  \big(2\cdot n\cdot (k+2) \cdot (2k+1)^{2k+1} + 1\big)\cdot 2^{2k+1}
  &=
  2\cdot n\cdot (k+2) \cdot 2^{(2k+1) (\log(2k+1) +1)} + 2^{2k+1}
  \\
  &\leqslant
  (2\cdot n\cdot (k+2) + 1)\cdot 2^{(2k+1) (\log(2k+1) + 1)}
  \\
  &=
  (2\cdot n\cdot (k+2) + 1)\cdot 2^{(2k+1) \log(4k+2)}.
\end{align*}

\begin{IEEEproof}[Proof for~\Cref{thm:decneRANA}]
  For the first statement, see Jiang and Ravikumar~\cite[Thm.~3.1]{jr91} together with~\Cref{lem:afaequivalence} and~\Cref{rem:fixAFA}.
  The second statement follows from the first one, using~\Cref{rem:compConstr} and~the size bound $|Q_0| \leqslant n_{\nd} \cdot 2^{(2k + 1) \cdot \log(2k + 1)}$ from \Cref{R:barafa}.\ref{R:barafa:2}, where $n_{\nd}$ is the number of orbits of the name-dropping modification.%
\end{IEEEproof}

\begin{IEEEproof}[{Proof of \Cref{thm:decincl}}]

Just like for ordinary AFAs, the inclusion problem can be reduced to the non-emptiness problem. Specifically, we obtain the following algorithm that decides for given RANAs $A_i$ with $n_i$ orbits and degree $k_i$, $i = 1,2$, whether {$L_\alpha(A_1) \seq L_\alpha(A_2)$}:
  \begin{enumerate}
    \item Construct the RANA $A_{\cap}$ via the disjoint union of both state sets and
      transition functions and add a new initial state $q_{0,\cap}$ with transition formula
      $\delta_{\cap}(q_{0,\cap}) = \delta_1(q_{0,1}) \wedge \neg\delta_2(q_{0,2})$ where
      $q_{0,i}$ are the initial states of the RANAs $A_i$ for $i = 1,2$. Equivalently, $A_{\cap}$ is the intersection automaton of $A_1$ and the complement of $A_2$ (\Cref{thm:closure}).
      The resulting automaton has $n_1 + n_2 + 1$ orbits and degree $k = \max\set{k_1,k_2}$.
    \item Decide emptiness for $A_{\cap}$, which is possible in
      \EXPSPACE by~\Cref{thm:decneRANA}.
  \end{enumerate}

The RANA $A_\cap$ accepts $L_{\alpha}(A_1) \cap \overline{L_{\alpha}(A_2)}$, which is empty iff $L_{\alpha}(A_1) \seq L_{\alpha}(A_2)$ holds. The complexity follows from that of the constructions used in the algorithm and~\Cref{thm:decneRANA}.
\end{IEEEproof}
\endgroup
\end{document}

%% file: prae-flo-ieee.tex
\usepackage{amsmath,amssymb,amsfonts}
\usepackage{amsxtra,mathrsfs,mathtools,bm,stmaryrd}
\usepackage{graphicx}
\usepackage{textcomp}
\usepackage{amsthm}
\usepackage{thmtools}
\usepackage[export]{adjustbox}
\usepackage{xspace}
\usepackage{dsfont}
\usepackage{tikz}
\usetikzlibrary{arrows,decorations.markings,
  decorations.pathmorphing,decorations.pathreplacing,arrows.meta,automata,fit,calc,positioning,shapes}
\usepackage{tikz-cd}
\tikzcdset{scale cd/.style={every label/.append style={scale=#1},
    cells={nodes={scale=#1}}}}

\usepackage{accents}

\usepackage{ifthen}
\usepackage[notref,notcite]{showkeys}
\usepackage[noadjust]{cite}
\usepackage[babel]{csquotes}
\usepackage{rotating}
\usepackage{blkarray}
\usepackage{calc}
\usepackage{dutchcal}
\usepackage{relsize}
\usepackage{booktabs}
\usepackage{multirow}
\usepackage{nicefrac}
\usepackage{colonequals}
\usepackage{nccmath}

\usepackage{ifdraft}
\ifdraft{
  \overfullrule=5mm %
}{}

\usepackage{hyperref}
\hypersetup{hidelinks,final}
\usepackage[capitalize,nameinlink]{cleveref}

\newtheorem{theo}{Theorem}[section]
\newtheorem{lem}[theo]{Lemma}
\newtheorem{prop}[theo]{Proposition}
\newtheorem{cor}[theo]{Corollary}

\theoremstyle{definition}
\newtheorem{defn}[theo]{Definition}
\newtheorem{example}[theo]{Example}

\newtheorem{rem}[theo]{Remark}
\newtheorem{notation}[theo]{Notation}
\newtheorem{assumption}[theo]{Assumption}

\newtheorem{construction}[theo]{Construction}

\crefname{theo}{Theorem}{Theorems}  
\Crefname{theo}{Theorem}{Theorems}

\newcommand{\resetCurThmBraces}{%
\gdef\curThmBraceOpen{(}%
\gdef\curThmBraceClose{)}}
\resetCurThmBraces
\newcommand{\removeThmBraces}{%
\gdef\curThmBraceOpen{}%
\gdef\curThmBraceClose{}}
\resetCurThmBraces

\usepackage{etoolbox}
\patchcmd{\thmhead}{(#3)}{\curThmBraceOpen #3\curThmBraceClose }{}{}

\usepackage{seqsplit}
\usepackage{xstring}
\newcommand{\defaultshowkeysformat}[1]{%
\StrSubstitute{#1}{ }{\textvisiblespace}[\TEMP]%
\parbox[t]{\marginparwidth}{\raggedright\normalfont\small\ttfamily\(\{\){\color{red!50!black}\expandafter\seqsplit\expandafter{\TEMP}}\(\}\)}%
}

\renewcommand*\showkeyslabelformat[1]{%
\noexpandarg%
\defaultshowkeysformat{#1}%
}

\usepackage[footnote,marginclue,nomargin,draft]{fixme}

\usepackage[inline]{enumitem}
\setlist[enumerate,1]{label=(\arabic*),font=\normalfont,align=left,leftmargin=0pt,labelindent=0pt,listparindent=\parindent,labelwidth=0pt,itemindent=!,topsep=3pt,parsep=0pt,itemsep=3pt,start=1}
\setlist[enumerate,2]{label=(\alph*),font=\normalfont,align=left,leftmargin=0pt,labelindent=0pt,listparindent=\parindent,labelwidth=0pt,itemindent=!,topsep=3pt,parsep=0pt,itemsep=3pt,start=1}
\setlist[itemize]{labelindent=*,leftmargin=*,topsep=5pt,itemsep=3pt}
\setlist[description]{labelindent=*,leftmargin=*,itemindent=-1 em}

\NewCommandCopy{\oldtextsf}{\textsf}
\renewcommand{\textsf}[1]{{\oldtextsf{\small#1}}}

\newcommand\restr[2]{{%
  \left.\kern-\nulldelimiterspace %
  #1 %
  \littletaller %
  \right|{}^{#2} %
}}

\newcommand{\littletaller}{\mathchoice{\vphantom{\big|}}{}{}{}}

\newcommand{\seq}{\subseteq}
\newcommand{\qes}{\supseteq}

\newcommand{\xra}[1]{\xrightarrow{#1}}

\newcommand{\xto}[1]{\xra{#1}}

\DeclareMathOperator{\norb}{\sharp_o}
\newcommand{\muBar}{\mbox{\textsf{Bar-$\mu$TL}}\xspace}

\newcommand{\At}{\mathds{A}}

\newcommand{\Ats}{\At^{\!\raisebox{1pt}{\scriptsize$\star$}}}
\newcommand{\barAs}{\barNames^{\star}}
\newcommand{\ol}{\overline}

\newcommand{\bars}{\mathsf{bs}}

\newcommand{\sem}[1]{\llbracket #1 \rrbracket}

\newcommand{\op}[1]{\operatorname{\mathsf{#1}}}

\newcommand{\cl}{\op{cl}}
\newcommand{\nd}{\op{nd}}
\newcommand{\rest}{\op{rest}}

\newcommand{\card}[1]{\left\vert#1\right\vert}
\let\abs\card

\newcommand{\trans}{\mathsf{t}}

\newcommand{\allbool}{\mathcal{B}}
\newcommand{\posbool}{\allbool_+}
\newcommand{\pos}[1]{{#1}^{+}}
\newcommand{\negsym}{\mathsf{n}}
\newcommand{\negstate}[1]{#1_{\negsym}}
\newcommand{\negbool}{\allbool_{\negsym}}
\newcommand{\dualbool}{\allbool_\mathsf{d}}
\newcommand{\dual}[1]{{#1}^\mathsf{d}}

\newcommand{\runtree}[3]{\mathcal{T}_{#1}^{#2}(#3)}

\newcommand{\rsem}{\vDash^{\textsf{r}}}

\renewcommand{\epsilon}{\varepsilon}

\DeclareMathOperator{\dom}{\textsf{dom}}
\newcommand{\Nom}{\mathbf{Nom}}

\DeclareMathOperator{\supp}{\mathsf{supp}}

\newcommand{\Perm}{\mathsf{Perm}}
\newcommand{\names}{\At}
\newcommand{\namesN}{\At^{\!N}}
\newcommand{\Abstr}[2][\names]{[#1]#2}

\newcommand{\parnom}[1]{\names^{\$\mathbf{#1}}}

\newcommand{\braket}[1]{\langle #1 \rangle}

\newcommand{\pow}{\mathcal{P}}

\newcommand{\pown}[1]{\pow_{\leqslant #1}}
\newcommand{\fresh}{\mathbin{\#}}

\renewcommand{\phi}{\varphi}

\newcommand{\epito}{\twoheadrightarrow}

\newcommand{\set}[1]{\{#1\}}
\newcommand{\setw}[2]{\{#1\,:\,#2\}}

\newcommand{\takeout}[1]{\empty}

\tikzset{shiftarr/.style={
        rounded corners,%
        to path={--([#1]\tikztostart.center)
                     -- ([#1]\tikztotarget.center) \tikztonodes
                     -- (\tikztotarget)},
}}

\tikzset{shiftarrr/.style={
        rounded corners,%
        to path={-- ([#1]\tikztostart.center)
                    |- (\tikztotarget)  \tikztonodes},
}}

\tikzset{roundcornerarr/.style={
        rounded corners,%
        to path={--([#1]\tikztostart.south)
                     |- (\tikztotarget) \tikztonodes},
}}

\tikzset{roundcornerarrr/.style={
        rounded corners,%
        to path={ -| (\tikztotarget) \tikztonodes},
}}

\tikzset{roundcornerarrrr/.style={
        rounded corners,%
        to path={ |- (\tikztotarget) \tikztonodes},
}}

\DeclareMathOperator{\alphaequiv}{\equiv_{\alpha}}

\newcommand{\mybar}[3]{%
  \mathrlap{\hspace{#2}\overline{\scalebox{#1}[1]{\phantom{\ensuremath{#3}}}}}\ensuremath{#3}
}

\newcommand{\FN}{\mathsf{FN}}
\newcommand{\BN}{\mathsf{BN}}

\newcommand{\VAR}{\mathsf{V}}

\newcommand{\Lpre}{L_{\mathsf{pre}}}

\newcommand{\BarForm}{\mathsf{Bar}}

\newcommand{\barNames}{{\mybar{0.7}{1.25pt}{\names}}}

\newcommand{\ub}{{\mathsf{ub}}}

\newcommand{\Nat}{\mathds{N}}

\newcommand{\RANAFunc}{\ensuremath{\mathbcal{A}}}

\usepackage{turnstile}
\newcommand\Turnstile[3][2]{%
  \mathrel{\calcturnstile{#1}%
  \vcenter{\hbox{\vrule\vphantom{$\tsvstyle\models$}}}%
  \vcenter{\offinterlineskip
   \ialign{\hfil\kern1pt$\tsstyle##\vphantom{by}$\hfil\cr
           \relax\if\relax\detokenize{#2}\relax\hphantom{=}\kern-2pt\else#2\fi\cr
           \noalign{\kern1pt\hrule\kern\fontdimen22\tsfont2\hrule\kern1pt}%
           \relax\if\relax\detokenize{#2}\relax\hphantom{-}\kern-2pt\else#3\fi\cr}}
}}
\newcommand\calcturnstile[1]{%
  \ifcase#1\let\tsstyle\textstyle\let\tsvstyle\textstyle\let\tsfont\textfont\or
    \let\tsstyle\textstyle\let\tsvstyle\textstyle\let\tsfont\textfont\or
    \let\tsstyle\scriptstyle\let\tsvstyle\textstyle\let\tsfont\scriptfont\or
    \let\tsstyle\scriptscriptstyle\let\tsvstyle\scriptstyle\let\tsfont\scriptscriptfont\fi}

\newcommand{\qvDash}{\smash{{}\Turnstile{?}{}{}}}

\renewcommand{\vDash}{\models}

\newcommand{\midmid}{\hspace{0.2ex}{\rule[-0.1ex]{0.6pt}{1.65ex}}\hspace{0.2ex}}
\newcommand{\scriptmidmid}{\hspace{0.2ex}{\rule[-0.1ex]{0.6pt}{1.1ex}}\hspace{0.2ex}}

\newcommand{\newletter}[1]{{\midmid}#1}
\newcommand{\scriptnew}[1]{{\scriptmidmid}#1}

\newcommand{\quotient}[2]{#1/{#2}} %

\let\originalleft\left
\let\originalright\right
\renewcommand{\left}{\mathopen{}\mathclose\bgroup\originalleft}
\renewcommand{\right}{\aftergroup\egroup\originalright}

\ExplSyntaxOn
\NewDocumentCommand{\makecycle}{om}{
	\ensuremath{ \left(~\guest_print_list:nn { #2 } {~{\ }~}~\right)\IfValueT{#1}{^{#1}}}
}
\NewDocumentCommand{\maketuple}{om}{%
\ensuremath{\left(~\guest_print_list:nn { #2 } {~{,\:}~}~\right)\IfValueT{#1}{^{#1}}}
}
\NewDocumentCommand{\makemonad}{om}{
	\ensuremath{ \left\langle~\guest_print_list:nn { #2 } {~{,\ }~}~\right\rangle\IfValueT{#1}{^{#1}}}
}
\NewDocumentCommand{\makegrammar}{om}{
	\ensuremath{ ~\guest_print_list:nn { #2 } {~{\ \,\vert\ \,}~}~\IfValueT{#1}{^{#1}}}
}

\seq_new:N \l_guest_list_seq
\cs_new_protected:Nn \guest_print_list:nn
{
	\seq_set_from_clist:Nn \l_guest_list_seq { #1 }
	\seq_use:Nn \l_guest_list_seq { #2 }
}
\ExplSyntaxOff